\documentclass[lettersize,journal]{IEEEtran}
\usepackage{amsmath,amsfonts}
\usepackage{algorithm}
\usepackage{algpseudocode}
\usepackage{array}
\usepackage[caption=false,font=normalsize,labelfont=sf,textfont=sf]{subfig}
\usepackage{textcomp}
\usepackage{stfloats}
\usepackage{url}
\usepackage{verbatim}
\usepackage{graphicx}
\usepackage{cite}
\usepackage{tikz}
\usetikzlibrary{fit}
\usepackage{hyperref}
\usepackage{colortbl}
\usepackage{enumerate}

\usepackage{setspace}
\usepackage{graphicx}
\usepackage{multirow,multicol}
\usepackage[]{amsmath}
\usepackage{amsbsy}
\usepackage{amssymb}
\usepackage{amsfonts}
\usepackage{pifont}
\usepackage{balance}
\usepackage{fancyvrb}

{\begin{list}               
    {$\bullet$ \hfill}{
        \setlength{\leftmargin}{\parindent}
        \setlength{\parsep}{0.04\baselineskip}
        \setlength{\itemsep}{0.5\parsep}
        \setlength{\labelwidth}{\leftmargin}
        \setlength{\labelsep}{0em}}
    }
{\end{list}}

\providecommand{\eref}[1]{Eq. \eqref{#1}}  
\providecommand{\cref}[1]{Chapter~\ref{#1}}

\providecommand{\fref}[1]{Figure~\ref{#1}}
\providecommand{\tref}[1]{Table~\ref{#1}}

\providecommand{\R}{\ensuremath{\mathbb{R}}}

\providecommand{\norm}[1]{\lVert#1\rVert}

\renewcommand{\vec}[1]{\ensuremath{\boldsymbol{#1}}}

\providecommand{\calF}{\mathcal{F}}

\providecommand{\mF}{\mathbf{F}}

\providecommand{\mH}{\mathbf{H}}

\providecommand{\vh}{\mathbf{h}}

\providecommand{\vn}{\mathbf{n}}

\providecommand{\vx}{\mathbf{x}}
\providecommand{\vy}{\mathbf{y}}
\providecommand{\vz}{\mathbf{z}}

\newcommand{\cmark}{\ding{51}}
\newcommand{\xmark}{\ding{53}}

\newcommand{\overbar}[1]{\mkern 1.5mu\overline{\mkern-1.5mu#1\mkern-1.5mu}\mkern 1.5mu}

\providecommand{\vone}{\vec{1}}

\newcommand{\argmin}[1]{\mathop{\underset{#1}{\mbox{argmin}}}}
\newcommand{\argmax}[1]{\mathop{\underset{#1}{\mbox{argmax}}}}

\newcommand\Tstrut{\rule{0pt}{2.6ex}}         
\newcommand\Bstrut{\rule[-0.9ex]{0pt}{0pt}}  

\usepackage{lscape}

\usepackage{tikz}
\usetikzlibrary{calc}
\usepackage{enumitem}

\begin{document}

\title{The Secrets of Non-Blind Poisson Deconvolution}

\author{Abhiram~Gnanasambandam,~\IEEEmembership{Member,~IEEE},
and~Yash~Sanghvi,~\IEEEmembership{Student~Member,~IEEE}, and~Stanley~H.~Chan,~\IEEEmembership{Senior~Member,~IEEE}
\thanks{Y. Sanghvi and S. Chan are with the School of Electrical and Computer
Engineering, Purdue University, West Lafayette, IN 47907, USA. Email: {
\{ysanghvi, stanchan\}}@purdue.edu. A. Gnanasambandam is with Samsung Research America, Plano, TX, but this work is completed during his time as a PhD student at Purdue. Email: abhiram.g94@gmail.com. This work is supported, in part, by the National Science Foundation under grant IIS-2133032, ECCS-2030570, and DMS-2134209.} 
}



\maketitle

\begin{abstract}
Non-blind image deconvolution has been studied for several decades but most of the existing work focuses on blur instead of noise. In photon-limited conditions, however, the excessive amount of shot noise makes traditional deconvolution algorithms fail. In searching for reasons why these methods fail, we present a systematic analysis of the Poisson non-blind deconvolution algorithms reported in the literature, covering both classical and deep learning methods. We compile a list of five ``secrets'' highlighting the do's and don'ts when designing algorithms. Based on this analysis, we build a proof-of-concept method by combining the five secrets. We find that the new method performs on par with some of the latest methods while outperforming some older ones. \end{abstract}


\section{Introduction}
\subsection{From Gaussian to Poisson deconvolution}
Image deconvolution is one of the most fundamental problems in image restoration. When the blur kernel is fixed and given, the problem is known as \emph{non-blind deconvolution}. For spatially invariant blur and additive i.i.d. Gaussian noise, the goal of deconvolution is to recover $\vx \in \R^N$ from the equation
\begin{align}
    \vy = \mH \vx + \vn, \label{eq:gaussian_blur_model}
\end{align}
where $\vn \in \mathbb{R}^N$ is the i.i.d. Gaussian noise, and $\mH \in \R^{N \times N}$ is the blur kernel represented as a convolution matrix \cite{Andrews_1977_Book, Vogel_1998_TV}. The inverse problem associated with \eref{eq:gaussian_blur_model} has been studied for a few decades, with an extensive list of methods, both classical \cite{Chan_2011_ICIP, Dey_2006_Richardson, Afonso_2010_Splitting, Fergus_2006_TOG, Shan_2008_SIGGRAPH, Cho_2011_ICCV, Zoran_2011_ICCV, Danielyan_2012_TIP, Xu_2010_ECCV} and deep-learning based \cite{Dong_2021_LearningSpatially, Dong_2020_DeepWiener_NIPS, Dong_2022_DeepWiener_PAMI, Gong_2020_LearningDeep, Nan_2020_DeepLearning, Pronina_2020_Microscopy, Zhang_2020_USRNet}. 

With such a large volume of prior work, it would appear that the problem is solved. However, as we push the limit of image deconvolution to \emph{low-light} conditions, the problem remains wide open. Moreover, the growth of advanced photon counting image sensors and the need for extreme low light imaging applications \cite{Chan_2016_MDPI, Choi_2018_ICASSP, Elgendy_2021_Demosaic, Gnanasambandam_2020_ECCV, Gnanasambandam_2020_TCI, Chengxi_2021_ICCVW} makes the problem even more interesting than before. As people have shown in \cite{Chi_2020_ECCV}, even an ideal image sensor with zero read noise cannot escape from the photon shot noise. Thus, signal processing at this limit remains critical.

The change from a well-illuminated condition to a low-light condition is not just about switching the Gaussian model to a Poisson model \footnote{The Poisson model we study in this paper is a simplification of the actual image formation process which should involve dark current, read noise, etc.. However, given that the Poisson problem is already difficult enough, we decided to focus on it in this paper.}
\begin{align}
    \vy = \text{Poisson}\{\alpha \mH \vx\}, \label{eq:forward_model_Poisson_matrix}
\end{align}
where $\alpha$ is the average number of photons in the scene \cite{Snyder_1975_Point}. The increased difficulty is not associated with the unbounded-below and the non-differentiable-at-origin property of the Poisson negative-log-likelihood, but the magnitude of the noise exhibited in the data. In a typical low light condition, the mean photon count can be as low as one to ten photons per pixel. At this photon level, the random fluctuation of the signal would cause many algorithms to fail.

\begin{figure}[t]
\centering
\begin{tabular}{ccc}
\hspace{-2ex}\includegraphics[width=0.33\linewidth]{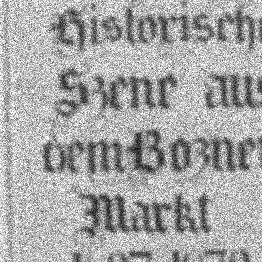}&
\hspace{-2ex}\includegraphics[width=0.33\linewidth]{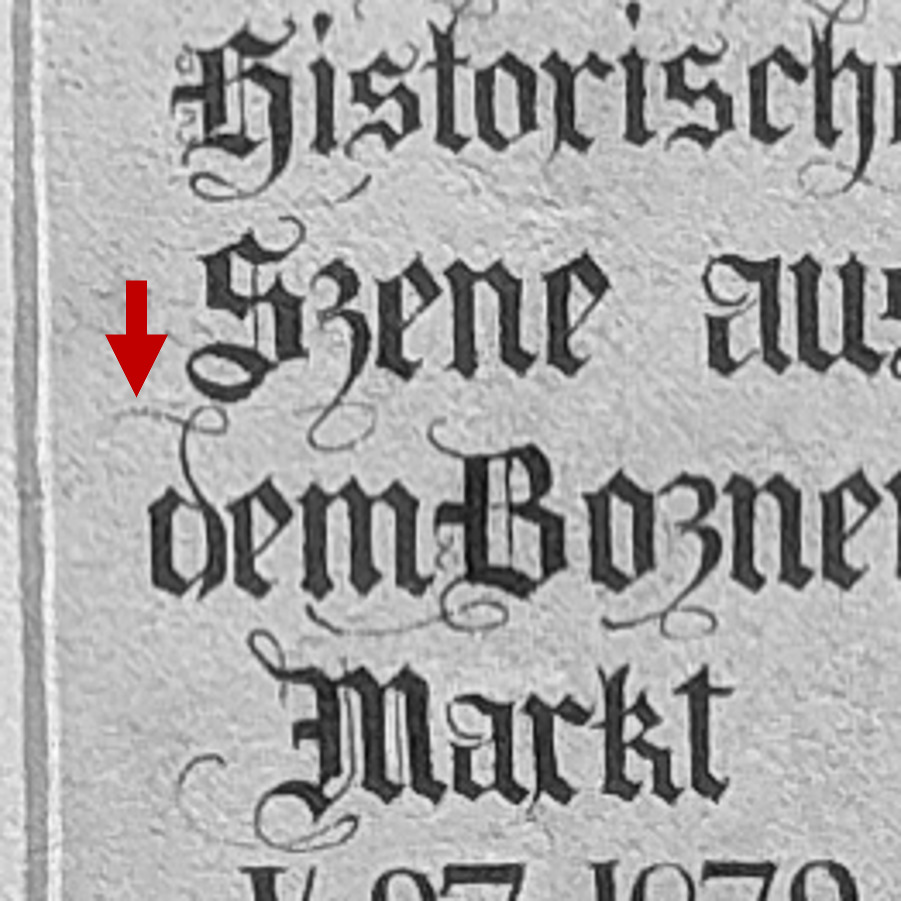}&
\hspace{-2ex}\includegraphics[width=0.33\linewidth]{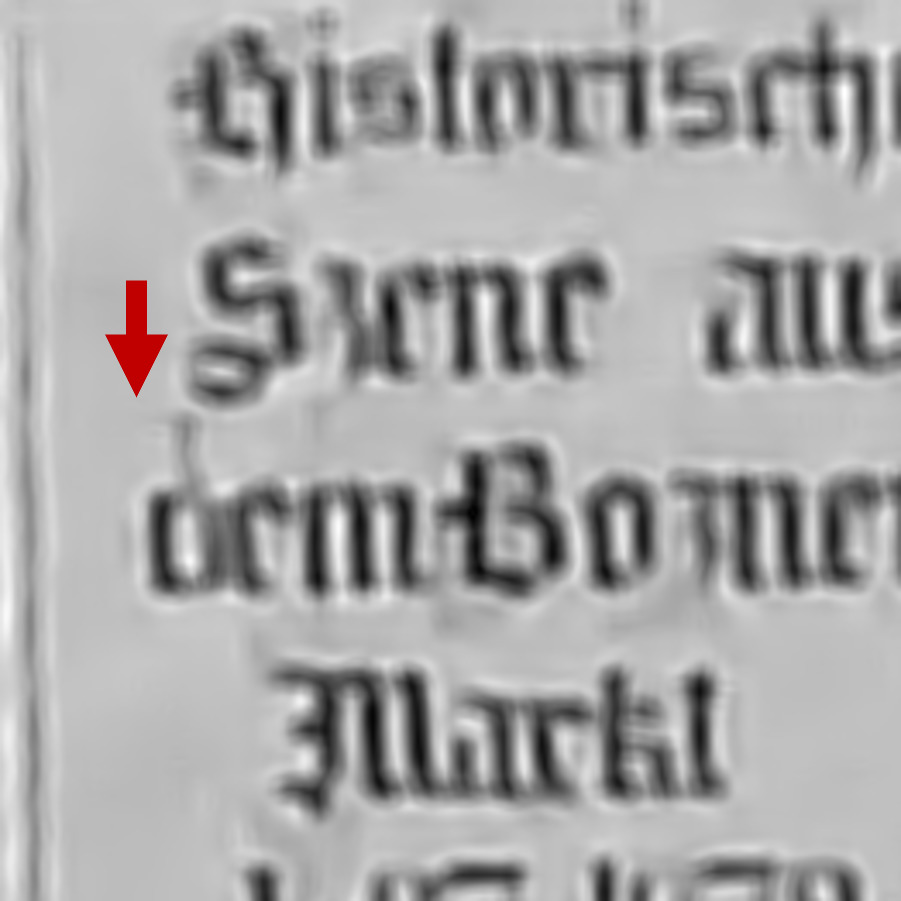}\\
\scriptsize{Noisy-blurry}   & \scriptsize{Ground Truth}    & \scriptsize{PURE-LET \cite{Li_2017_PURELET}} \\
\hspace{-2ex}\includegraphics[width=0.33\linewidth]{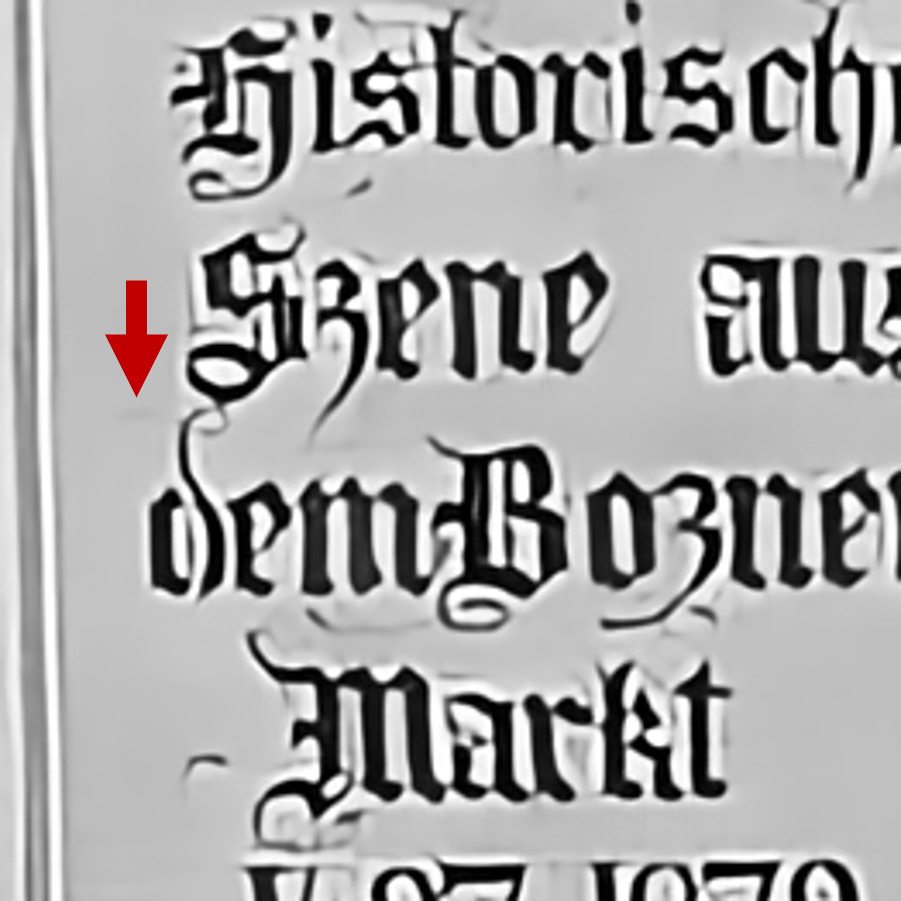}&
\hspace{-2ex}\includegraphics[width=0.33\linewidth]{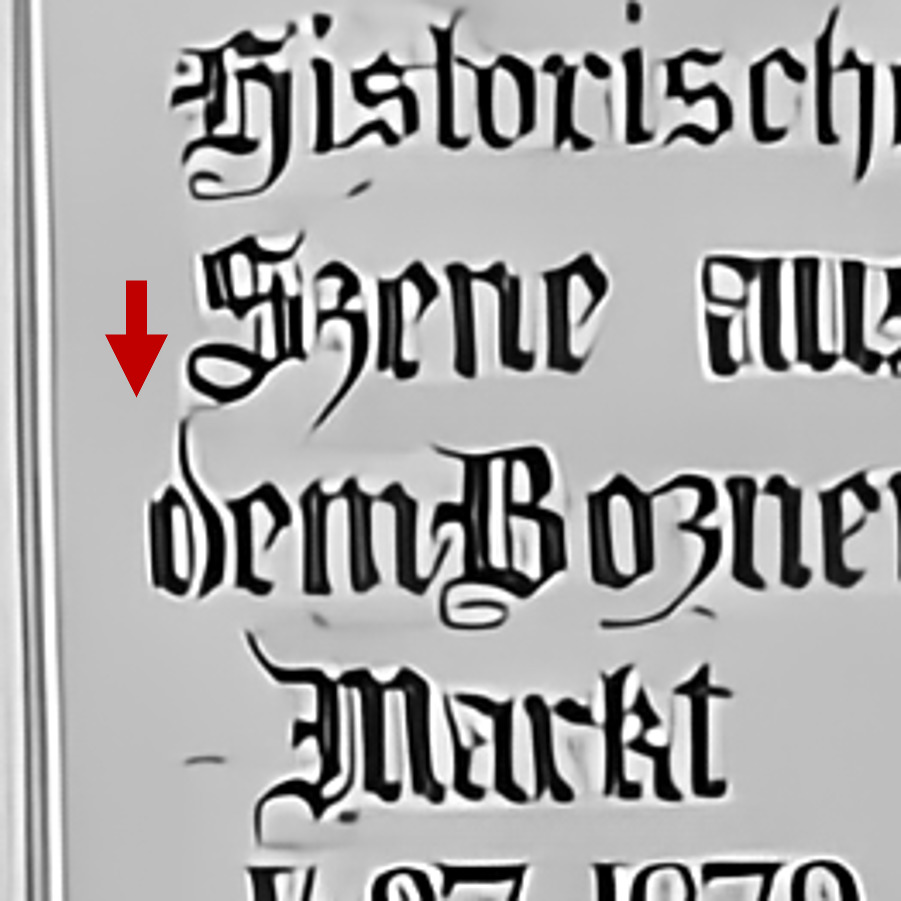}&
\hspace{-2ex}\includegraphics[width=0.33\linewidth]{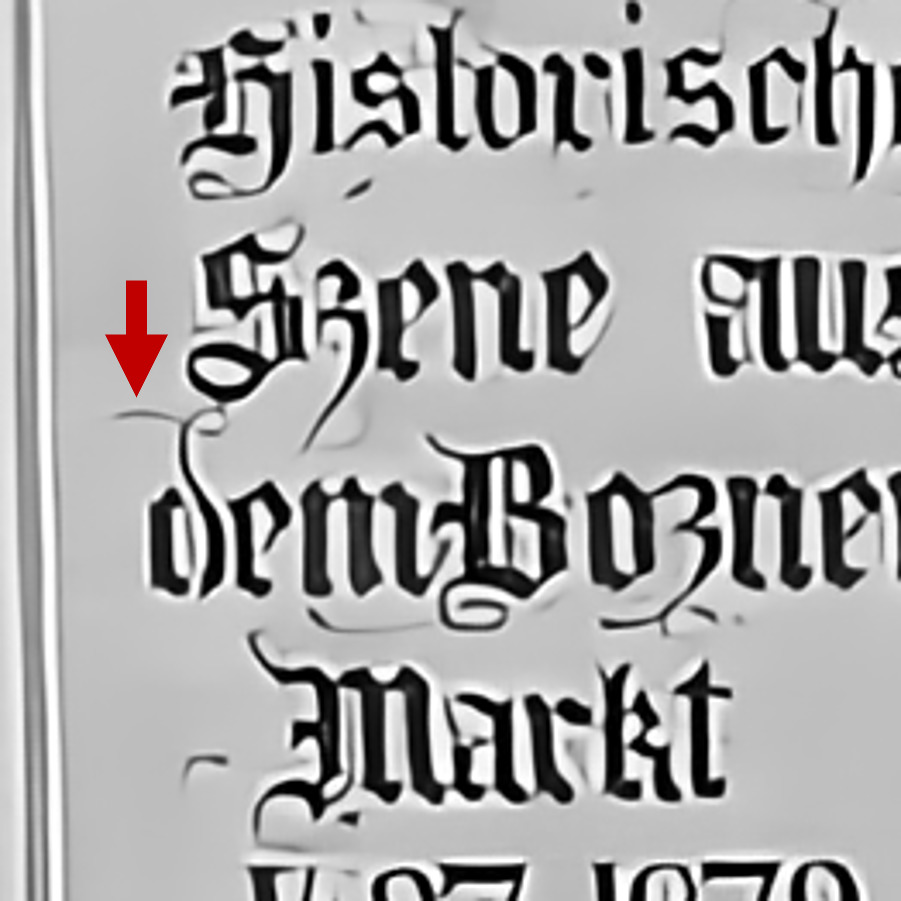}\\
\scriptsize{DWDN \cite{Dong_2022_DeepWiener_PAMI}}  & \scriptsize{USRNet \cite{Zhang_2020_USRNet}} & \scriptsize{Ours}\\
\end{tabular}
\caption{\textbf{Overview}. The goal of this paper is to identify factors that will benefit Poisson image deblurring. Shown in this example are a simulated blurry and noisy image (where the noise is Poisson), and the corresponding image reconstruction results. The proposed method (to be discussed in Section IV) is just a combination of the five factors we identified, \emph{without} introducing any new architectures.}
\label{fig: front page}
\end{figure}

The impact of noise in Poisson deconvolution is noticeable in every step of a deconvolution algorithm. Since there is noise, it becomes much harder for an algorithm to invert the blur (usually in the Fourier space) and remove the noise. Deep learning algorithms also suffer from heavy noise because extracting features from the image becomes more difficult. In fact, Poisson deconvolution has only been discussed in a few deep-learning papers \cite{Sanghvi_2022_TCI, Sanghvi_2022_Iterative, Sanghvi_2023_CVPR, Rond_2016_Poisson}.

\begin{table*}[!t]
\centering
\setlength\doublerulesep{0.5pt}
\caption{We study a comprehensive list of methods as shown in the table below. The gray-colored methods are selected for further analysis to identify the ``secrets'' of Poisson deconvolution.}
\label{tab: List of Methods}
\scalebox{0.8}{
\begin{tabular}{c>{\columncolor[gray]{0.8}}c>{\columncolor[gray]{0.8}}cc>{\columncolor[gray]{0.8}}ccccc>{\columncolor[gray]{0.8}}c>{\columncolor[gray]{0.8}}ccc}
&\multicolumn{3}{c}{Classical Methods}          & \multicolumn{9}{c}{Deep Learning Methods} \\
&\multicolumn{3}{c}{$\overbrace{\hspace{4cm}}$} & \multicolumn{9}{c}{$\overbrace{\hspace{12cm}}$} \\
\hline
& PURE-LET
& VSTP
& Deconvtv
& DWDN
& SVMAP
& KerUnc
& CPCR
& RGDN
& PhDNet
& USRNet
& DPIR
& DWKF \Bstrut\\
& ~\cite{Li_2017_PURELET}
& ~\cite{Azzari_2017_VST}
& ~\cite{Chan_2011_TIP}
& ~\cite{Dong_2022_DeepWiener_PAMI}
& ~\cite{Dong_2021_LearningSpatially}
& ~\cite{Nan_2020_DeepLearning}
& ~\cite{Eboli_2020_EndEnd}
& ~\cite{Gong_2020_LearningDeep}
& ~\cite{Sanghvi_2022_Iterative}
& ~\cite{Zhang_2020_USRNet}
& ~\cite{Zhang_2017_LearningDeep}
& ~\cite{Pronina_2020_Microscopy} \Bstrut\\
\hline\hline
\textbf{Neural Network?}        & \ding{55} & \ding{55} &  \ding{55} & \ding{51}& \ding{51} & \ding{51} & \ding{51} & \ding{51} & \ding{51}& \ding{51}& \ding{51} & \ding{51} \Tstrut\\
\textbf{Decoupling?}            & \ding{51} & \ding{51} & \ding{51} & \ding{51}& \ding{51} & \ding{51} & \ding{51} & \ding{51} & \ding{51}& \ding{51}&\ding{51}&\ding{51}\\
\textbf{Poisson Likelihood?}    & \ding{51} & \ding{51} & \ding{55} & \ding{55}& \ding{55} & \ding{55} & \ding{55} & \ding{55} & \ding{51}& \ding{55}&\ding{55}&\ding{55}\\
\textbf{Iterative?}             & \ding{55} & \ding{51} & \ding{51} & \ding{55}& \ding{51} &\ding{55} & \ding{51}& \ding{51}& \ding{51}& \ding{51}&\ding{51}&\ding{51}\\
\textbf{Learned parameter?}     & \ding{55} & \ding{55} & \ding{55}& \ding{55}& \ding{55} & \ding{55} & \ding{55} & \ding{55} & \ding{51}& \ding{51} &\ding{55} &\ding{55} \\
\textbf{Feature space?}         & \ding{55} & \ding{55} & \ding{55} & \ding{51}& \ding{55} & \ding{55} & \ding{55} & \ding{55} & \ding{55}& \ding{55}&\ding{55}&\ding{55}\Bstrut\\
\hline
\end{tabular}}
\end{table*}

\subsection{Scope and contributions}
Given the success of Gaussian-noise based image deconvolution algorithms, we believe that the lessons learned in the past can shed light on understanding the Poisson problem. To this end, we analyze a large collection of non-blind deconvolution algorithms reported in the literature. We look into the design details of each method and compile a list of do's and don'ts we learned from these methods.

As a preview of our results, we show in \fref{fig: front page} the image reconstruction results of three methods published in the literature: PURE-LET \cite{Li_2017_PURELET} (T-IP, 2017), DWDN \cite{Dong_2022_DeepWiener_PAMI} (T-PAMI, 2022), and USRNet \cite{Zhang_2020_USRNet} (CVPR, 2020). All three methods are fine-tuned using Poisson data. In the same figure, we also report a proof-of-concept method by combining the ``secrets'' we learned in this paper. We stress that this proof-of-concept method is not meant to become a state-of-the-art but rather a confirmation of ideas described in the paper. Interestingly, the performance of this proof-of-concept is quite satisfactory.

So, what are our observations? We found the following five ``secrets'' of non-blind Poisson deconvolution:

\begin{enumerate}[label=(\roman*)]
    \item \textbf{Wiener filter is recommended}. While some networks perform deconvolution and denoising simultaneously, we find that it is better to decouple the deconvolution part using a Wiener filter so that we can leverage the fact that the blur kernel is known. Of course, we assume that the blur is spatially invariant.
    \item \textbf{Iteration is recommended}. Many networks estimate the image in a single shot. We find that iterative algorithms are more effective. For deep neural networks, the iterative algorithms can be implemented via algorithm unrolling.
    \item \textbf{Feature space is recommended}. It is better to perform deconvolution in the feature space than in the spatial domain.
    \item \textbf{Poisson likelihood is not needed}. When handling Poisson noise, there is no need to use customized tools such as variance stabilizing transform or the Poisson likelihood. Any architecture for Gaussian noise also works for Poisson.
    \item \textbf{Learning the hyper-parameters is recommended}. Some algorithms estimate the hyperparameters using an off-the-shelf method or a heuristic rule. We find that end-to-end learning of the hyperparameters helps the performance.
\end{enumerate}

This paper focuses on non-generative methods. Our analysis does not cover generative models (e.g., generative adversarial networks or denoising diffusion probabilistic models) because they belong to a different category of approaches. We do not consider \emph{blind} deconvolution algorithms because we do not estimate the blur kernel.

\section{Analysis of Prior Methods} \label{sec:SOTA}
Given the large number of papers published for non-blind image deconvolution, it would be unrealistic to comment on every single method. The approach we take here is to focus on a  \emph{representative} subset of existing methods. However, the selection of the representative methods would require some work. In what follows, we first list a number of Poisson deconvolution methods. We group them, and discuss their attributes. Afterward, we select the representative methods and discuss their design philosophies.

\subsection{Prior Methods}
To help readers visualize the methods being studied in this paper, we summarize them in Table~\ref{tab: List of Methods}. These methods can be categorized into two main classes:

\textbf{Classical Methods}. By classical methods, we mean methods that do not require learning. These methods are typically developed before the deep-learning era. In this paper, we select three representative methods with code publicly available:
\begin{itemize}
    \item PURE-LET, by Li and Blu \cite{Li_2017_PURELET}, is a non-iterative deblurring algorithm that uses the Poisson unbiased risk estimator (PURE) as a metric to guide the steps in linear expansion thresholding (LET). The thresholding idea used here is similar to several other paper \cite{Cai_2010_FrameletBased,Cheng_2015_TIP,Zhang_2016_TIP}.
    \item VSTP, by Azzari and Foi \cite{Azzari_2017_VST}, uses the variance stabilization transform (VST) to equalize the variance of the Poisson random variable. Then, a deblurring algorithm is applied to handle the blur.
    \item Deconvtv, by Chan et al. \cite{Chan_2011_TIP}, uses total variation for Gaussian noise removal. Its performance is not necessarily the best compared to other total variation solvers such as \cite{Chowdhury_2020_JMIV, Lingenfelter_2009_Poisson, Lefkimmiatis_2013_TIP, Setzer_2010_Bregman, Harmany_2012_SPIRAL, Getreuer_2012_IPOL, Figueiredo_2010_ADMM}, but its code is readily available for experiments.
\end{itemize}
We acknowledge that there are plenty of other classical methods, such as \cite{Xu_2010_ECCV, Sun_2014_ECCV, Krishnan_2009_FastImage, Joshi_2009_ImageDeblurring, Yuan_2008_TOG, Schmidt_2013_CVPR, Bar_2006_IJCV}. These papers made great contributions in improving the prior models of the images so that the deblurring and denoising can be more effective. Some of these methods perform very well whereas some are similar to the three abovementioned methods. For concreteness of this paper and considering the availability of their codes, we decide to focus on the ones we mentioned above.

\textbf{Deep-Learning Methods}. While deep learning based deconvolution algorithms are abundant, many of them are \emph{blind} algorithms. For non-blind methods, we consider nine of them.
\begin{itemize}
    \item Deep Wiener Deconvolution Network (DWDN) \cite{Dong_2022_DeepWiener_PAMI} is a deep neural network that performs Wiener deconvolution in the feature space followed by a decoder. A follow-up method INFWIDE \cite{Zhang_2023_INFWIDE} adds a cross-residual fusion module. In this paper, we focus on DWDN for clarity and simplicity. 
    \item KerUnc \cite{Nan_2020_DeepLearning},  CPCR \cite{Eboli_2020_EndEnd}, USRNet \cite{Zhang_2020_USRNet},  PhDNet \cite{Sanghvi_2022_Iterative}, and \cite{Gong_2020_LearningDeep} are perform fixed iteration unrolling of alternating direction method of multipliers (ADMM), half quadratic splitting or gradient descent methods followed by end-to-end training.
    \item DPIR \cite{Zhang_2017_LearningDeep} uses the plug-and-play (PnP) based ADMM optimization to solve the problem.
    \item DWKF \cite{Pronina_2020_Microscopy} is an iterative method that uses kernel prediction networks for imposing the image priors.  
\end{itemize}

\subsection{Attributes of the Methods}
With more than ten methods listed in Table~\ref{tab: List of Methods}, it would be helpful if we could further categorize them according to their attributes. The attributes we highlight here will be used to inform the do's and don'ts of designing an algorithm.

\begin{itemize}
\item \textbf{Neural network?} This attribute asks if the method uses a neural network - either trained end-to-end \cite{Dong_2020_DeepWiener_NIPS} or as a pretrained block \cite{Zhang_2020_USRNet}. By definition, all classical methods are treated as non-neural network methods in this paper.
\item \textbf{Decoupling?} Decoupling means that a method handles the deblurring step and the denoising step separately. The decoupling can be realized via variable splitting (e.g., in ADMM), or via a two-stage operation (e.g., in PURE-LET). For neural networks, we say that it employs a decoupling strategy if there are modules explicitly performing deblurring and are separated from denoising.
\item \textbf{Poisson likelihood?} If a method explicitly uses the Poisson likelihood in an algorithm, then this attribute is satisfied. Some methods, usually deep neural networks, do not incorporate the Poisson likelihood in its algorithm design, for example \cite{Dong_2022_DeepWiener_PAMI,Eboli_2020_EndEnd}
\item \textbf{Iterative?} Both classical and deep learning methods can be iterative. The iteration can occur in the form of an actual iteration (as in optimization steps) or algorithm unrolling in deep learning methods.
\item \textbf{Learned parameters?} All restoration methods have a set of hyperparameters. If these hyperparameters are picked manually, we say that the parameters are not learned. In contrast, if the hyper-parameters are simultaneously selected by the learning algorithm, then we say that the parameters are learned.
\item \textbf{Feature space?} For some deep learning methods, the deconvolution does not take place in the spatial domain  \cite{Li_2017_PURELET} but in the feature space \cite{Dong_2020_DeepWiener_NIPS, Zhang_2023_INFWIDE}. We check this box to reflect the property.
\end{itemize}

\subsection{Design Principles} \label{sec:Comparing_methods}
We now discuss the design principles of the methods shown in Table~\ref{tab: List of Methods}. To narrow down the discussion to a smaller set of methods, we compared the methods' performance on a testing dataset. The execution of the experiment is described in Section III when we discuss the five secrets of Poisson deconvolution. For the sake of brevity, the detailed numbers are reported in the Supplementary Material. Based on the performance of the methods, we select five leading methods that cover four categories. They are:
\begin{enumerate}[label*=\arabic*)]
    \item Traditional, non-iterative: PURE-LET \cite{Li_2017_PURELET}
    \item Traditional, iterative: VSTP \cite{Azzari_2017_VST}
    \item Neural-network, non-iterative: DWDN \cite{Dong_2022_DeepWiener_PAMI}
    \item Neural-network, iterative: USRNet \cite{Zhang_2020_USRNet}, PhDNet \cite{Sanghvi_2022_TCI}. Two methods were chosen because of their similar performance.
\end{enumerate}

\subsubsection{PURE-LET \cite{Li_2017_PURELET}} The core idea of PURE-LET is to construct multiple initial estimates using the Wiener filter, which is essentially a Fast-Fourier transform (FFT) based deconvolution. Given the blur matrix $\mH$, PURE-LET estimates a set of $K$ initial guesses via
\begin{align}
    \widehat{\vx}^\text{Wiener}_k = \texttt{Wavelet}[\left(\mH^T\mH + \lambda_k \mathbb{I}\right)^{-1} \mH^T \vy],
\end{align}
where $k=\{1,2,\hdots K\}$ denotes the $k$th Wiener estimate, and $\lambda_k$ is the $k$th hyperparameter. The operator \texttt{Wavelet} denotes the wavelet thresholding, which is the method PURE-LET used to clean up the estimates. The estimates are then linearly combined in such a way that they minimize the mean square error, i.e.,
\begin{align}
    \widehat{\vx} = \sum_{k=1}^K a_k \cdot \widehat{\vx}^\text{Wiener}_k,
\end{align}
where $\{a_k \,|\, k = 1,\ldots,K\}$  are the optimal combination weights determined by minimizing the Poisson unbiased risk estimate (PURE).

\begin{figure}[h]
\centering
\includegraphics[width=\linewidth]{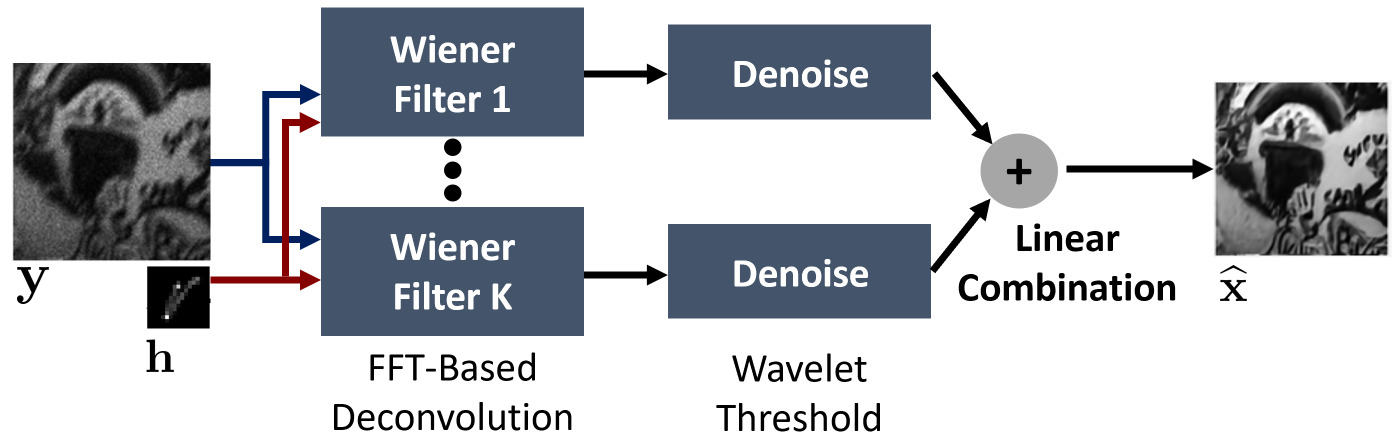}
\caption{PURE-LET \cite{Li_2017_PURELET} constructs a bank of Wiener filters to deblur the image, followed by image denoisers.}
\label{fig: Method: PURELET}
\end{figure}

A conceptual diagram of PURELET is shown in \fref{fig: Method: PURELET}. Referring to Table~\ref{tab: List of Methods}, PURELET employs a decoupling strategy by separating the deconvolution step and the denoising step. The Poisson likelihood is used to compute the risk estimate, but it was not used for the deconvolution step which is a filter bank of Wiener filters.

\subsubsection{DWDN \cite{Dong_2022_DeepWiener_PAMI}} DWDN has many similarities to PURE-LET. Instead of applying the Wiener filter on the images, DWDN applies it to the features:
\begin{align}
    \widehat{\vx}^\text{feature}_k = \left(\mH^T\mH + \lambda_k \mathbb{I}\right)^{-1} \mH^T \calF^\text{feature}_k(\vy),
\end{align}
where $\calF^\text{feature}_k(\cdot)$ is a neural network trained to produce features. The estimated deblurred features $\{\widehat{\vx}^\text{feature}_1, \widehat{\vx}^\text{feature}_2, \hdots, \widehat{\vx}^\text{feature}_K\}$ are then fed to another neural network for refinement $\calF_\text{refine}$ to obtain the final output $\widehat{\vx}$:
\begin{align}
    \widehat{\vx} = \calF_\text{refine}\{\widehat{\vx}^\text{feature}_1, \widehat{\vx}^\text{feature}_2, \hdots, \widehat{\vx}^\text{feature}_K\}.
\end{align}
The feature networks $\{\calF^\text{feature}_k \,|\, k = 1,\ldots,K\}$ and the refinement network $\calF_\text{refine}$ are trained end-to-end. When the mean squared error (MSE) loss is used, DWDN and PURE-LET both aim to find the MMSE estimate.

\begin{figure}[h]
\centering
\includegraphics[width=\linewidth]{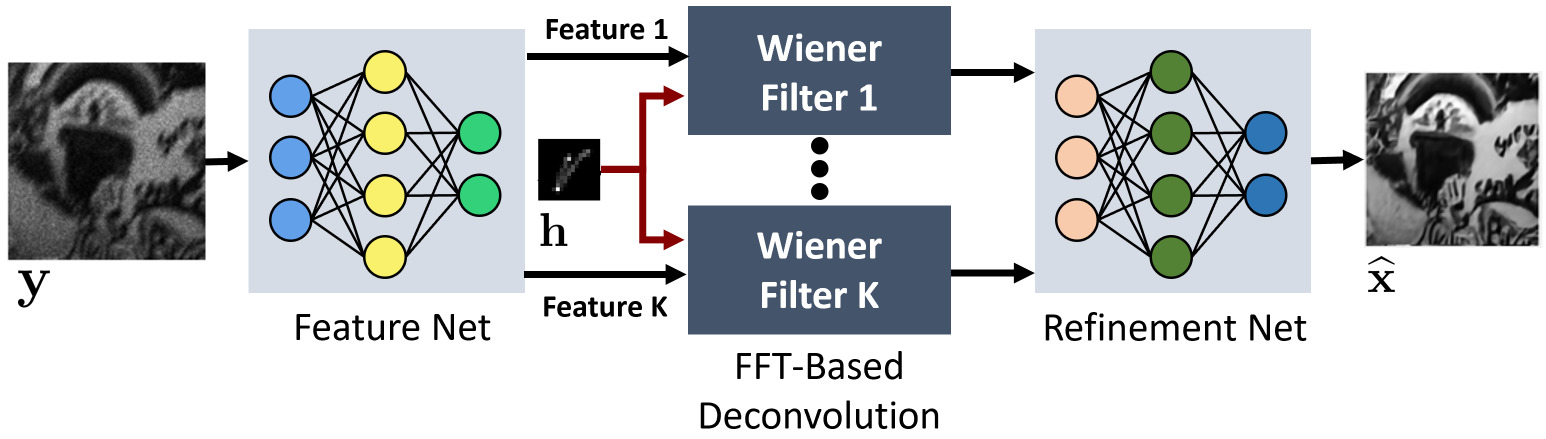}
\caption{ DWDN \cite{Dong_2022_DeepWiener_PAMI}. While it shares similarities with PURELET, it performs Wiener deconvolution in the feature space instead of the image space.}
\label{fig: Method: DWDN}
\end{figure}

A schematic diagram of DWDN is shown in \fref{fig: Method: DWDN}. If we compare DWDN with PURE-LET, we recognize that the overall multi-channel filter bank idea is the same. The only difference is that DWDN performs the deconvolution operations in the feature space. The denoisers are also replaced by neural networks. Moreover, since DWDN does not need to estimate the risk (as in PURE-LET), the Poisson likelihood is not considered.

\subsubsection{VSTP \cite{Azzari_2017_VST}} VSTP extends the idea of PURE-LET to make it iterative. VSTP starts with a single estimate of the deblurred image $\widehat{\vx}^\text{Wiener}$ instead of the multiple estimates used in PURE-LET. However, the overall concept of decoupling the deconvolution and the denoising steps remain the same.

An interesting idea of VSTP is to iteratively update the denoising step so that each denoising step can be ``mild''. To do so, a linear combination of $\widehat{\vx}^\text{Wiener}$ and the denoised estimate from the previous iteration $\widehat{\vx}_{t-1}$ is obtained via
\begin{align}
    \widehat{\vx}^\text{data}_t = \lambda_t \widehat{\vx}_t + (1-\lambda_t) \widehat{\vx}^\text{Wiener}
\end{align}
A variance stabilizing transform (VST) is then used to stabilize the spatially varying noise strength of $\widehat{\vx}^\text{data}_t$, which is then denoised with \texttt{Denoiser},
\begin{align}
    \widehat{\vx}_t = \texttt{Denoiser}\left[\text{VST}\left(\widehat{\vx}^\text{data}_t\right)\right].
\end{align}
The iteration continues until the stopping criteria are met.

In VSTP, the variance stabilizing transform is more of a technical need because the noise is spatially varying. The rationale of using VST is that when the photon level is not too low, VST is able to stabilize the variance so that the spatially varying variance will become invarying.

\begin{figure}[h]
\centering
\includegraphics[width=\linewidth]{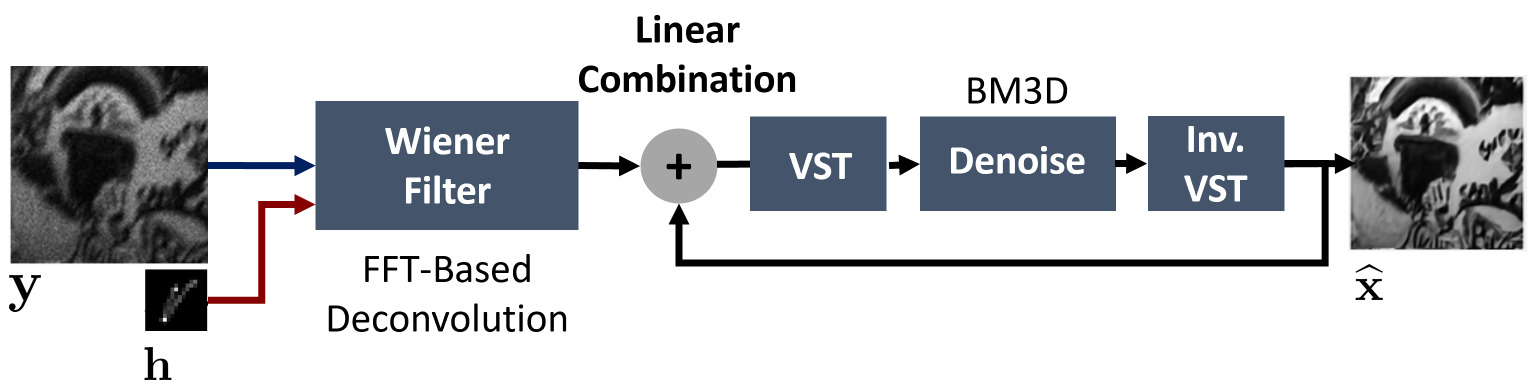}
\caption{VSTP \cite{Azzari_2017_VST} applies variance stabilizing transform and a denoiser for the denoising step. The denoising step is also repeated in an iterative manner to improve the performance.}
\label{fig: Method: VSTP}
\end{figure}

A schematic diagram of VSTP is shown in \fref{fig: Method: VSTP}. In the literature, people sometimes refer the denoising module as transform-denoise \cite{Chan_2016_MDPI}.

\subsubsection{PhDNet \cite{Sanghvi_2022_TCI} and USRNet \cite{Zhang_2020_USRNet}} Both methods are based on maximizing the posterior probability (hence they are a maximum-a-posteriori (MAP) estimator). More specifically, the estimate is obtained by solving the optimization:
\begin{align}
    \widehat{\vx} = \argmax{\vx} \Big[\log \mathbb{P}(\vy|\vx) + \log \mathbb{P}(\vx)\Big],
\end{align}
where $\mathbb{P}(\vy|\vx)$ is the likelihood term and $\mathbb{P}(\vx)$ is the natural image prior. USRNet models the problem by assuming that the noise is Gaussian (without considering the fact that the true noise distribution is Poisson). Thus, in USRNet, the likelihood term is
\begin{align}
    \log \mathbb{P}(\vy|\vx) = -\norm{\vy-\alpha \mH\vx}^2.
\end{align}
PhDNet explicitly takes into consideration of the Poisson noise, which leads to the following likelihood term
\begin{align}
   \log \mathbb{P}(\vy|\vx) = -\alpha \vone^T\mH\vx + \vy^T\log(\alpha \mH \vx),
\end{align}
where $\vone$ is a vector with all ones.

Both methods solve the optimization using an unrolled neural network. Two steps are common for both methods:
\begin{itemize}
    \item The inversion module is similar to a Wiener filter. For iteration $t$, it is given by
        \begin{align}
            \widehat{\vx}^\text{data}_t = \left(\mH^T\mH + \alpha \mathbb{I}\right)^{-1}\left(\mH^T\vy + \alpha \widehat{\vx}_{t-1}\right). \label{eq:data_module}
        \end{align}
    \item The Gaussian denoising module, which can be considered as a refinement step:
        \begin{align}
            \widehat{\vx}_{t} = \calF^\text{refine}(\widehat{\vx}^\text{data}_t) \label{eq:refinement_module}
        \end{align}
\end{itemize}
PhDNet has an additional step in each iteration to deal with the Poisson noise.

\begin{figure}[h]
\centering
\includegraphics[width=\linewidth]{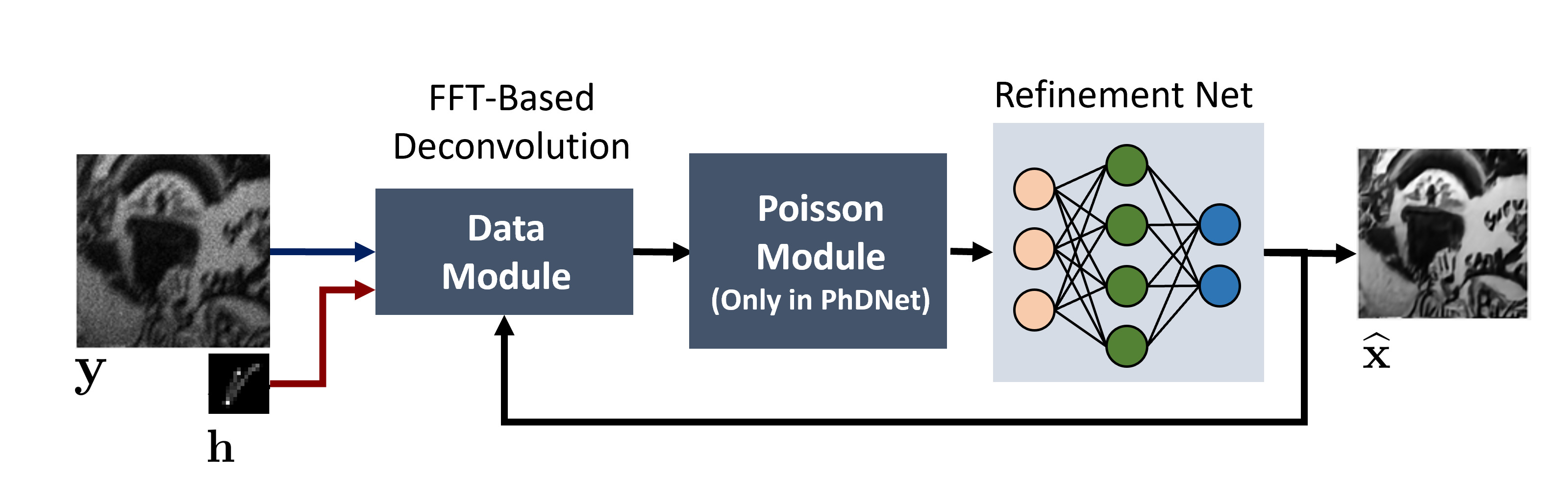}
\caption{USRNet \cite{Zhang_2020_USRNet}, PhDNet \cite{Sanghvi_2022_TCI} is an optimization-based algorithm where the problem is decoupled into deconvolution, Poisson data, and image denoising. The method is iterative; in deep neural networks, the iterations are realized via algorithm unrolling. }
\label{fig: Method: USRNet}
\end{figure}

A schematic diagram of the methods is shown in \fref{fig: Method: USRNet}. On neural networks, the iterations are implemented via algorithm unrolling. That is, we unfold the optimization algorithm into a fixed number of blocks where each block is implemented via a neural network. When looping through this fixed number of blocks, effectively we perform an iterative algorithm. For additional details about algorithm unrolling, we refer the readers to \cite{Monga_2021_Unrolling, Sanghvi_2022_TCI, Sanghvi_2022_Iterative}.

\section{The Secrets} \label{sec:Secrets}
In Section II we analyzed the structures of the prior methods, but this alone does not tell us much about the secrets of Poisson deconvolution. In this section, our goal is to dive into the details by conducting a series of experiments. From the experimental results, we then draw conclusions about the influencing factors for Poisson deconvolution. Some of the discussions are based on the main experimental result Table~\ref{tab:SOTA_comparison}, which are presented in Section V.

\subsection{Experimental setting}
\label{subsec:training}
Our approach to analyzing the performance of the prior methods is based on a series of carefully designed experiments. Since this is an empirical approach, we first state the background experimental settings.

First of all, we consider classical methods and deep learning methods separately, because deep learning methods require training. To make sure that the comparisons are fair, we retrain all the deep learning methods with the exact same training dataset, same training loss, and fine-tune the hyper-parameters to maximize their performances.

For training, we use images from the Flickr2K \cite{Flickr2K} dataset. We generate 500 random kernels based on \cite{Boracchi_2012_Modeling}. These 500 kernels consist of five groups of sizes where each group has 100. The sizes are $9\times9$, $18\times18$, $27\times27$, $36\times36$, and $45\times45$. In addition, we generate 64 Gaussian kernels of varying anisotropy with the blur parameter $\sigma$ between $0.1$ and $5$. Images of size $128\times128$ are cropped randomly from the dataset and then each image is blurred using a random kernel among these 500+64 = 564 kernels. For noise, we assume that the photons per pixel (ppp) is ranged between 1 and 80. \footnote{The average ppp can be adjusted by varying $\alpha$ in \eref{eq:forward_model_Poisson_matrix}}

During training, we use the $\ell_1$ loss between the reconstructed image $\widehat{\vx}$ and the ground truth image $\vx$ to train the networks. The loss function is defined by
\begin{align}
    \mathcal{L}(\widehat{
    \vx}, \vx) =   \|\widehat{\vx} - \vx\|_1, \label{eq:l1}
\end{align}
where $\|\cdot\|_1$ denotes the $\ell_1$ norm. We train all the networks for 500 epochs, with the Adam optimizer. The learning rate is initialized as $10^{-4}$ which gets halved every 100 epochs. The batch size was set to $2$ for all the methods. We do so to ensure a fair comparison because some methods consume more GPU power. The inputs to the networks include the degraded image $\vy$ and the blur kernel $\vh$. Some methods like \cite{Zhang_2020_USRNet,Sanghvi_2022_TCI} take the noise level as inputs. In such cases, the photon level $\alpha$ corresponding to each image was sent as the input.

For testing, we evaluate the methods using synthetically degraded images obtained by blurring 100 images from the BSD300 dataset \cite{martin2001database}. We use 3 different sets of 5 motion kernels of size  $9\times9$ (Small), $27\times27$ (Medium), $45\times45$ (Large) using \cite{Boracchi_2012_Modeling}. Each combination of the image and motion is evaluated at three different photon levels (10, 30, and 50).

\subsection{Secret 1: Using Wiener filters is recommended}
We observe that the five methods discussed in Section~\ref{sec:Comparing_methods} all have a separate Fourier-based deconvolution module - irrespective of whether they are traditional methods or deep learning-based methods. The presence of the Fourier-based deconvolution module hints that a black-box neural network might have some limitations.

The decoupling approach makes sense in classical methods. In these methods, Poisson deconvolution is often posed as MAP-based optimization. Since it is very difficult for a simple optimization step to simultaneously handle blur and noise, it makes sense to decouple them. 

\begin{figure}[!h]
\centering
\begin{tabular}{cc}
    \hspace{-2.0ex}\includegraphics[height=2cm]{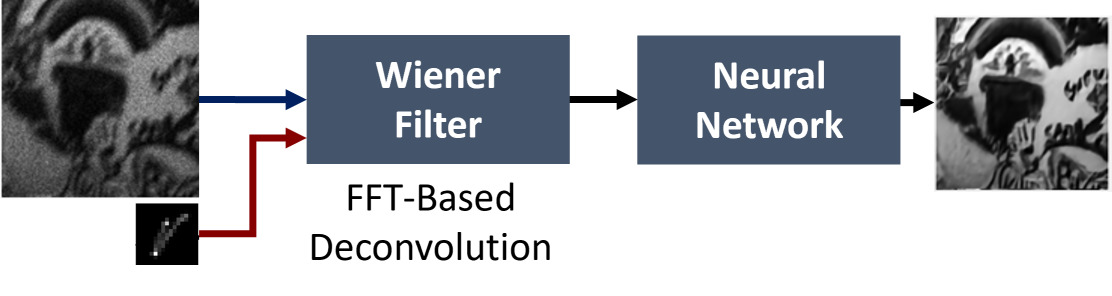}\\
    (a) Neural network w/ Wiener Filter \\
    \includegraphics[height=1.4cm]{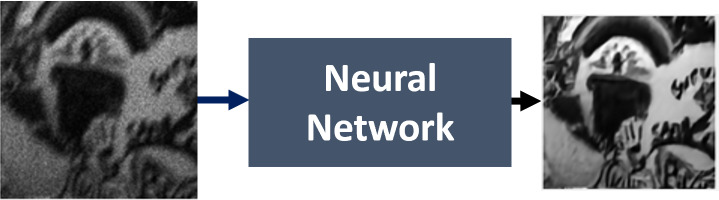}\\
    (b) Neural network w/o Wiener Filter \\
\end{tabular}
\caption{\textbf{How Wiener filters are used}. We consider two neural networks, where in (a) we decouple the inversion step by a Fourier-based deconvolution module which is the Wiener filter, and in (b) we use only a neural network. The added computational complexity of the Wiener filter is minimal because it is a simple inversion in the Fourier space.}
\label{fig:nns_cant_deblur_networks}
\end{figure}

How about deep neural networks? One would expect that since they have a large capacity, they wouldn't need to adopt a decoupling strategy. To examine the need for decoupling, we compare the two configurations as shown in \fref{fig:nns_cant_deblur_networks} - neural networks with and without a Wiener filter.

In this experiment, we use the ResUNet from \cite{Zhang_2020_USRNet} for the task shown in \fref{fig:nns_cant_deblur_networks}(b). We train the network at a particular light level of 10 photons per pixel (ppp). To ensure that there is no domain gap, we train the network for \emph{one} single blur kernel and test it for the exact same blur kernel. For the configuration shown in \fref{fig:nns_cant_deblur_networks}(a), we use a single-iteration USRNet. A single-iteration USRNet is nothing but a deconvolution module followed by a refinement network. We train the network with a large range of photon levels and blur kernels, as described in Section III-A. Our argument is that if a specialized network in \fref{fig:nns_cant_deblur_networks}(b)  cannot beat a generic network in \fref{fig:nns_cant_deblur_networks}(a), then there must be some fundamental limits in the network itself.

\begin{figure}[!h]
\centering
\begin{tabular}{ccc}
    \hspace{-2.0ex}\includegraphics[width=0.28\linewidth]{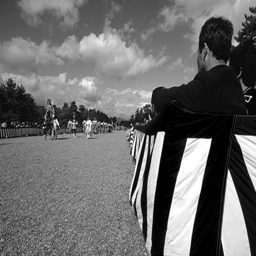}&
    \hspace{-2.0ex}\includegraphics[width=0.28\linewidth]{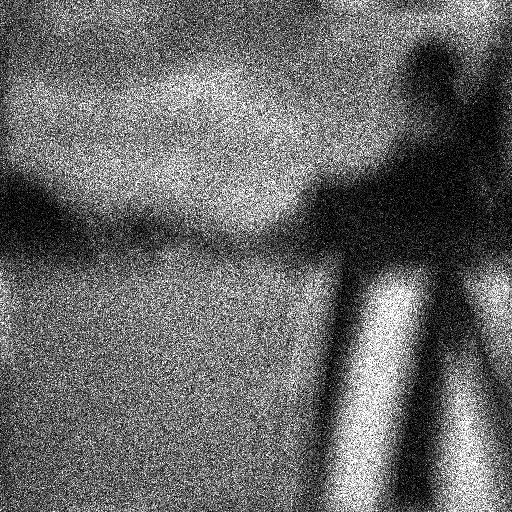}&
    \hspace{-2.0ex}\includegraphics[width=0.28\linewidth]{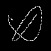}\\
    \hspace{-2.0ex} \small{(a) Ground Truth} & \hspace{-2.0ex} \small{(b) Input} & \hspace{-2.0ex} \small{(c) Blur Kernel}\\
    \hspace{-2.0ex}\includegraphics[width=0.28\linewidth]{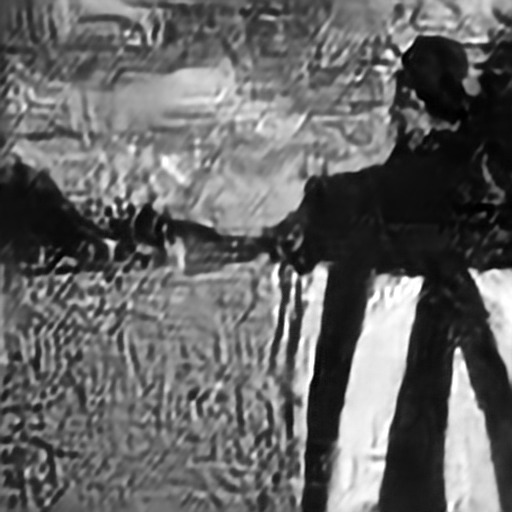}&
    \hspace{-2.0ex}\includegraphics[width=0.28\linewidth]{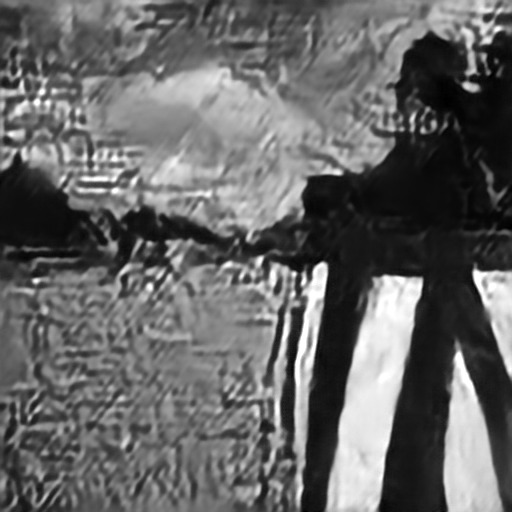}&
    \hspace{-2.0ex}\includegraphics[width=0.28\linewidth]{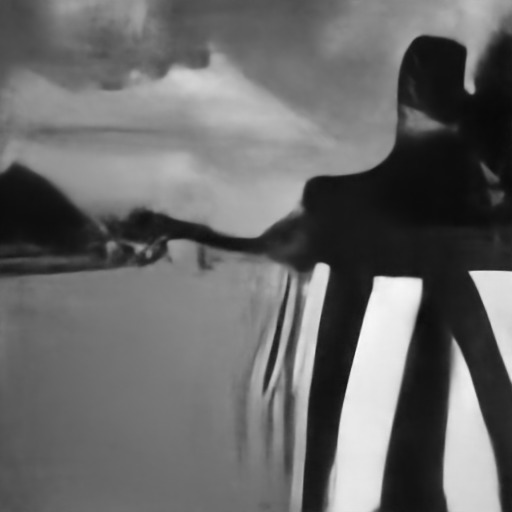}\\
    \hspace{-2.0ex} \small{(d) w/o Wiener} & \hspace{-2.0ex} \small{(e) w/o Wiener} & \hspace{-2.0ex} \small{(f) w/ Wiener}\\
    \hspace{-2.0ex} \small{(general)} & \hspace{-2.0ex} \small{(specific)} & \hspace{-2.0ex} \small{(general)}\\
\end{tabular}
\caption{\textbf{Secret 1: Wiener filter is recommended} (a) Clean. (b) Degraded image. (c) Blur kernel. (d) A deconvolution U-Net trained on a variety of kernels. (e) A deconvolution U-Net trained on the specific blur kernel defined in (b). Note that even if we train the network \emph{specifically for the blur kernel}, the result is still not satisfactory. (f) Deconvolution when the Wiener step is included in the U-Net. }
\label{fig:nns_cant_deblur}
\end{figure}

The results are shown in \fref{fig:nns_cant_deblur}. We observe that the black-box neural network cannot handle blur and noise simultaneously. In contrast, a network with an explicit deconvolution step performs much better. Our conjecture is that since we know the blur kernel, it is better to incorporate this forward model in the solution when deblurring the image.

\subsection{Secret 2: Iterations are recommended}
Iterative methods, regardless if they are traditional or neural-network-based, tend to perform better according to \tref{tab:SOTA_comparison}. Let us explain why this is the case.

Consider USRNet as an example: It is an iterative algorithm where the iterations are given by \eref{eq:data_module} and \eref{eq:refinement_module}. The first step \eref{eq:data_module} is the deconvolution module which produces an estimate $\widehat{\vx}^\text{data}_t$, and the second step \eref{eq:refinement_module} is a neural network denoiser that refines the estimate to generate $\widehat{\vx}_t$. The performance of a denoiser is directly related to the input quality. The noisier the input is, the worse the reconstruction performance will be \cite{Gnanasambandam_2020_ICML}. In a single-shot low-light deconvolution, we need to have a very good estimate from \eref{eq:data_module}, and this needs to be computed directly from the noisy image itself. In iterative schemes, even though the initial estimate $\widehat{\vx}^\text{data}_0$ is not good, the mild refinement steps will gradually improve the image quality because they use both the previous estimate $\widehat{\vx}^\text{data}_{t-1}$ and the current estimate.

\begin{figure}[!h]
\centering
\includegraphics[width=0.8\linewidth]{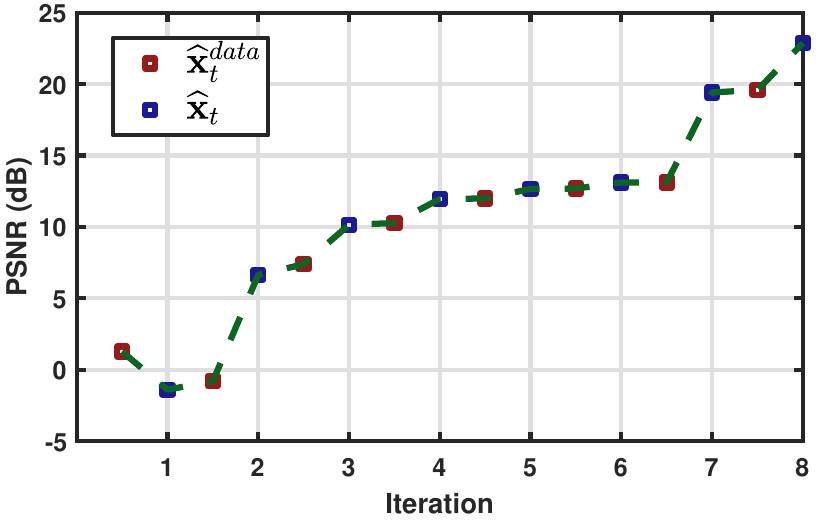}
    \caption{\textbf{Secret 2: Iteration is recommended.} We plot the PSNR of the estimates at each iteration of USRNet. We can notice that ignoring the first iteration, the plot aligns with our claim - A better deconvolution leads to a better refinement, which turn again leads to a better deconvolution. }
    \label{fig:plot_PSNRl_itr1}
\end{figure}

\fref{fig:plot_PSNRl_itr1} shows a typical per-iteration PSNR of an iterative scheme USRNet. Putting aside the initial estimate (which shows a downward PSNR trend), the performance generally goes up as the number of iterations increases. Specifically, we see that after each pair of $\widehat{\vx}_t$ and $\widehat{\vx}_t^{\text{data}}$, the performance improves. The exact dynamics of the PSNR is difficult to track because it is image-dependent. However, the trend confirms our hypothesis that iterations are helpful.

\begin{figure}[!h]
\centering
\includegraphics[width=0.8\linewidth]{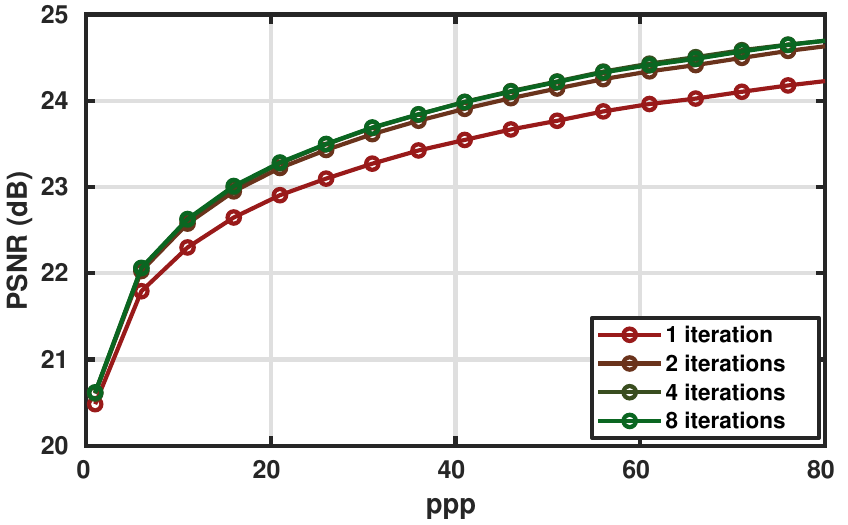}
    \caption{\textbf{Number of iterations}. We retrain USRNet with different numbers of iterations. We can see that we obtain a performance boost by increasing the number of iterations.  }
    \label{fig:plot_PSNRl_itr2}
\end{figure}

For unrolled algorithms, the number of iterations is realized by the number of blocks. A natural question is the number of such iterative blocks --- will more blocks improves the overall deconvolution result? \fref{fig:plot_PSNRl_itr2} shows the results of four USRNets trained at different number of iterative blocks. It is clear from the result that more iterations leads to a better final performance, although there is a diminishing return after several blocks.

\begin{figure}[!h]
\centering
\begin{tabular}{c}
    \includegraphics[height=2cm]{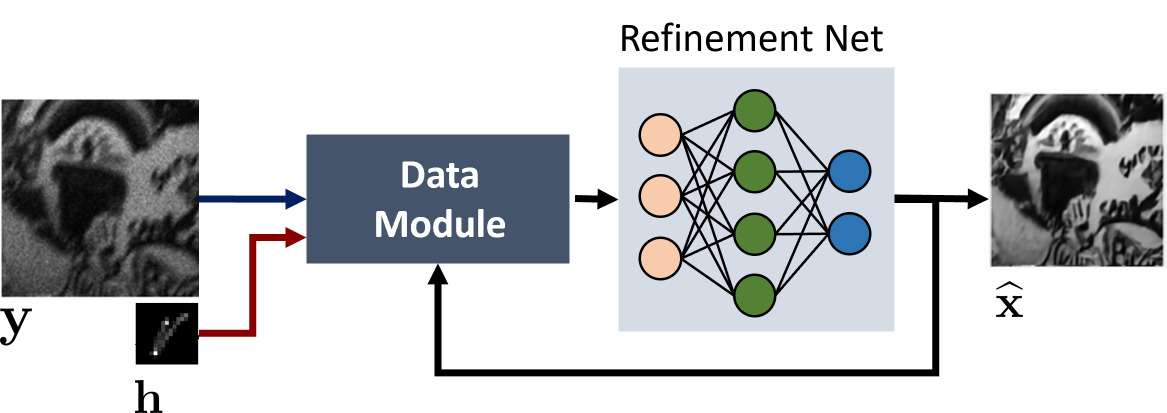}\\
    (a) USRNet\\
    \includegraphics[height=2cm]{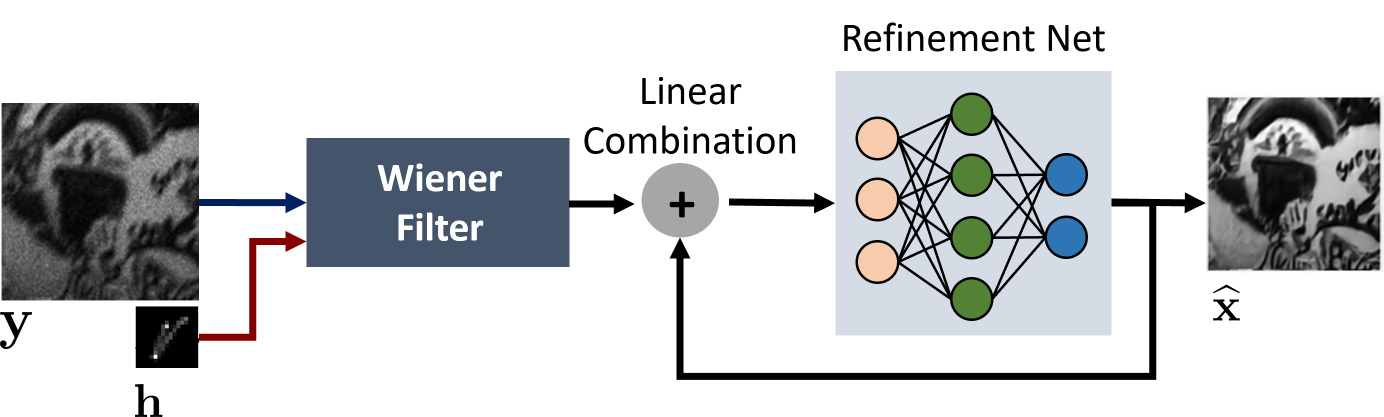}\\
    (b) USRNet modified with VSTP type iterations.
\end{tabular}
\caption{\textbf{What kind of iterations help?} By comparing USRNet and VSTP, we observe two types of iterations. (a) USRNet sends $\widehat{\vx}$ back to the data module iteratively to improve the estimate. (b) VSTP uses $\widehat{\vx}$ to form a linear combination with the Wiener filter outputs. We find that iteration in (a) is more effective.}
\label{fig:nns_cant_deblur_networks_types}
\end{figure}

Another question we ask is the \emph{type} of iterations. Among the methods reported in \tref{tab:SOTA_comparison}, there are two different kinds of iterative schemes as shown in \fref{fig:nns_cant_deblur_networks_types}. The first one is the USRNet where the estimate $\widehat{\vx}$ is fed back to the data module (i.e., the inversion module), and is combined with the raw input $\vy$ and kernel $\vh$ to construct a new intermediate estimate. The second iterative scheme is the one used in VSTP. In this scheme, the estimate $\widehat{\vx}$ is used to form a linear combination with the output of the Wiener filter.

\begin{figure}[!h]
    \centering
    \includegraphics[width=0.8\linewidth]{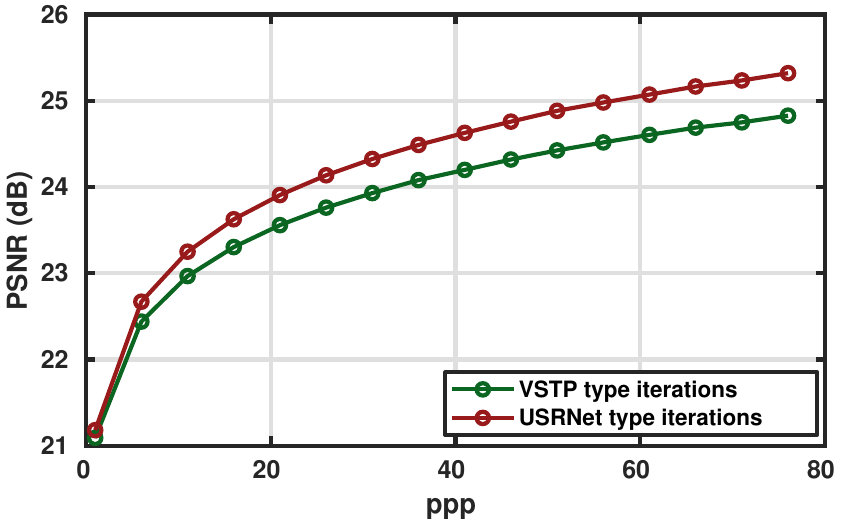}
    \caption{\textbf{What kind of iterations help?} This figure is a follow-up of \fref{fig:nns_cant_deblur_networks_types} where here we plot the PSNR as a function of the photon level.}
    \label{fig:plot_PSNRl_itr3}
\end{figure}

To evaluate the performance of the two schemes, we modify USRNet to incorporate the VSTP mechanism. We argue that this is a fairer comparison than directly using VSTP because VSTP uses a traditional denoiser BM3D. In our modification, we ensure that the two networks are trained with the same type and same amount of data. The results are shown in \fref{fig:plot_PSNRl_itr3}, where we see that the iterative scheme by USRNet has a clear advantage over the VSTP scheme.

Based on the above experiments, our recommendation regarding the iterative scheme is that iterative schemes offer better performance than one-shoot methods. Among the different iterative schemes, we recommend feeding back the estimate $\widehat{\vx}$ directly to the inversion module so that the features of $\widehat{\vx}$ will be utilized better.

\subsection{Secret 3: Feature space is recommended}
The next secret about Poisson deconvolution is that it is better to deconvolve the image in the \emph{feature space} instead of the image space. This observation is based on the difference between PURE-LET and DWDN in \tref{tab:SOTA_comparison}. Both PURE-LET and DWDN use multiple Wiener filters. PURE-LET uses different deblurring strengths (as specified by the hyper-parameter $\lambda$), whereas DWDN uses the same Wiener filter for different feature maps. But the biggest difference is that PURE-LET performs the deconvolution in the image space whereas DWDN performs the deconvolution in the feature space. We show in this subsection that the superior performance of DWDN is partially driven by feature space deconvolution.

\begin{figure}[!h]
\centering
\begin{tabular}{c}
    \includegraphics[width=0.9\linewidth]{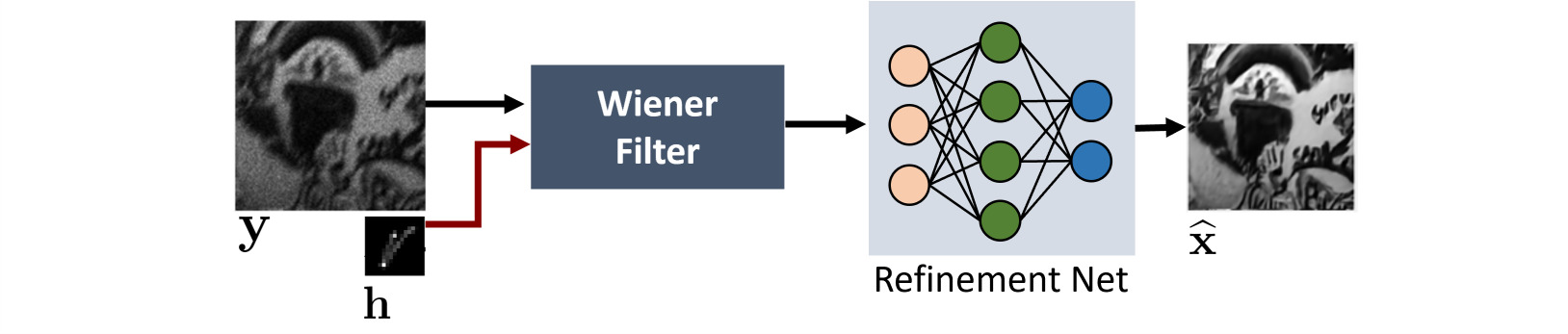}\\
    \scriptsize{(a) Config I. Single Wiener filter}\\
    \includegraphics[width=0.9\linewidth]{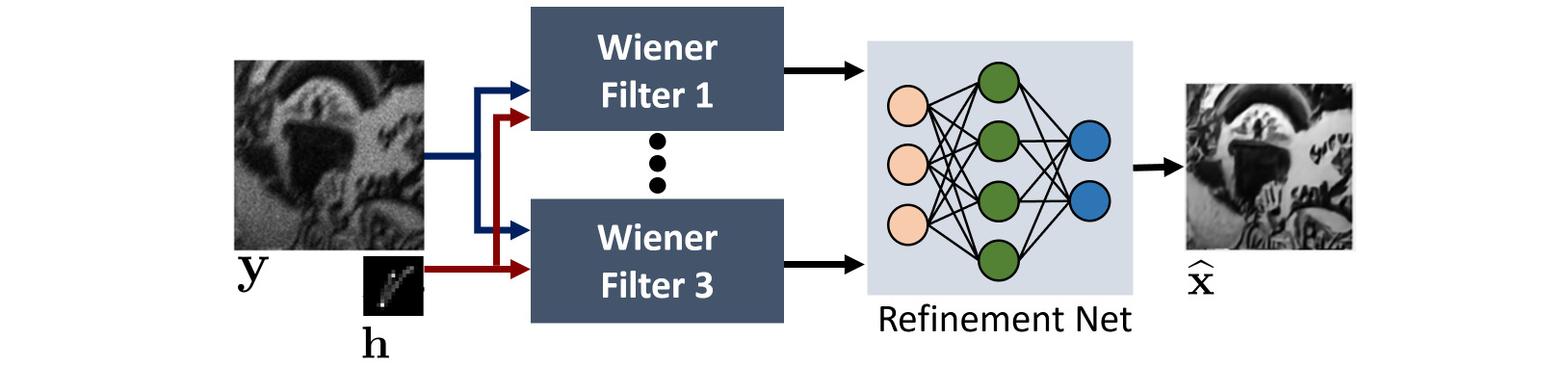}\\
    \scriptsize{(b) Config II. Three Wiener filters}\\
    \includegraphics[width=0.9\linewidth]{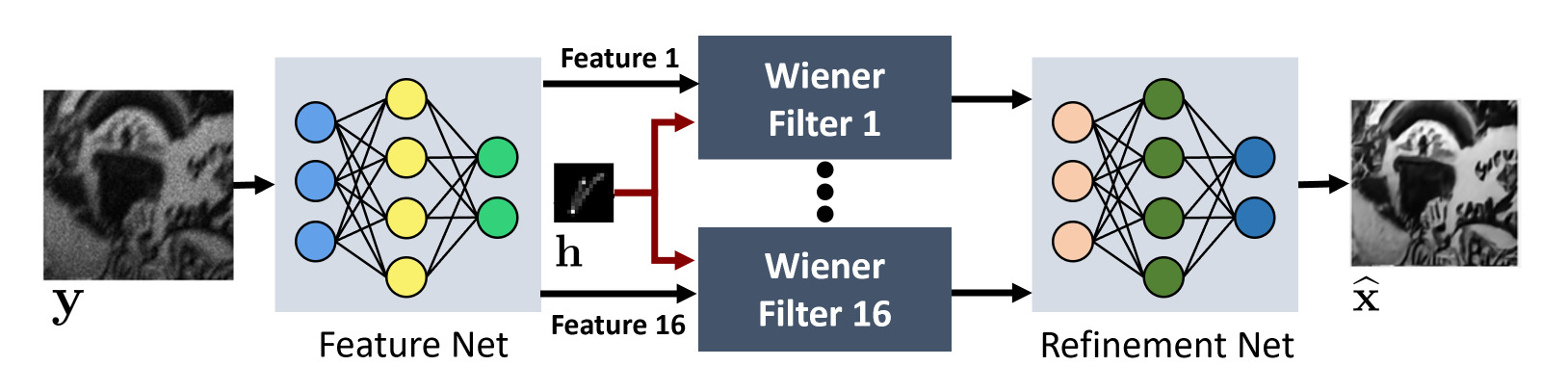}\\
    \scriptsize{(c) Config III. 16 Features w/ 1 Wiener filter each}\\
    \includegraphics[width=0.9\linewidth]{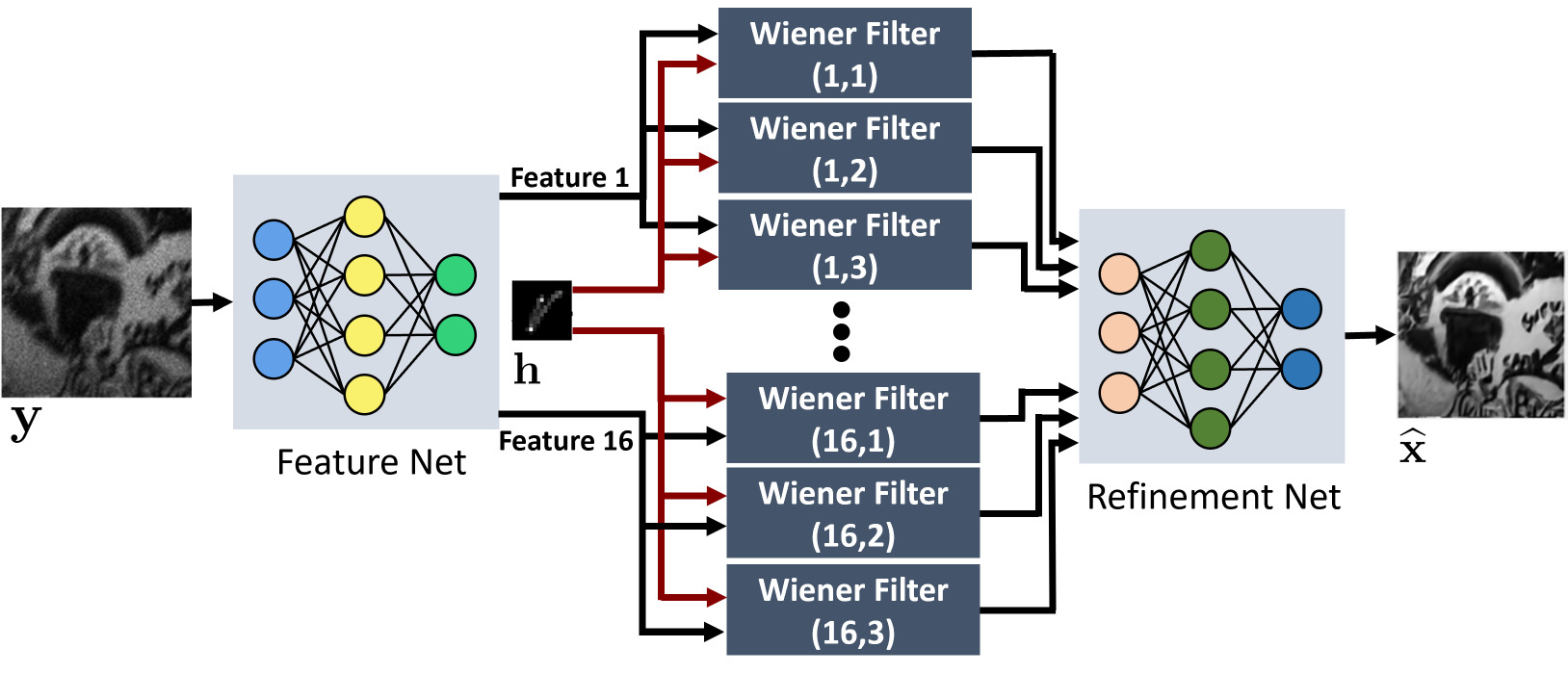}\\
    \scriptsize{(d) Config IV. 16 Features w/ 3 Wiener filters each}\\
\end{tabular}
\caption{\textbf{Secret 3: Feature space is recommended}. The two types of deconvolutions: image space and feature space. (a)-(b) perform deconvolution in the spatial domain, whereas (c)-(d) perform deconvolution in the feature space.}
\label{fig:improve_deconvolution_types}
\end{figure}

To prove the usefulness of feature space deconvolution instead of image space, we consider the following four modifications of DWDN by placing the Wiener filters in different ways.
\begin{enumerate}
    \item \textbf{Configuration I} in \fref{fig:improve_deconvolution_types}(a) uses a single Wiener filter followed by a refinement network. This is the vanilla network for baseline analysis.
    \item \textbf{Configuration II} in \fref{fig:improve_deconvolution_types}(b) uses three Wiener filters as in PURE-LET. Each Wiener filter uses a different regularization parameter $\lambda$. We use a deep neural network as the refinement step so that it is a fair comparison with DWDN.
    \item \textbf{Configuration III} in \fref{fig:improve_deconvolution_types}(c) uses a feature extraction unit to pull the features before sending them to Wiener filters. This is the same as DWDN. In our experiment, there are 16 feature maps. The regularization parameter $\lambda$ is the same across the 16 Wiener filters. 
    \item \textbf{Configuration IV} in \fref{fig:improve_deconvolution_types}(d) uses 16 Wiener filters where each has three sub-configurations. Each sub-configuration uses a different $\lambda$. We regard Configuration IV as the ultimate modification we can make within the context of our analysis.
\end{enumerate}

The comparisons between these configurations are shown in \tref{tab:dwdn_table}. We see that across the different photon levels, the ones that perform deconvolution in the feature space are \emph{significantly} better. Our intuitive argument is that in the feature space, the signals are already \emph{decomposed}. If the feature extraction unit is powerful, signals will be captured in a few leading feature dimensions whereas noise will be concentrated in the other dimensions. Therefore, the strong signal features will be deconvolved well by the Wiener filter with a smaller $\lambda$, whereas the noise features will be attenuated by a large $\lambda$. As a result, the overall deconvolution will be better. As for how much regularization $\lambda$ is needed, Configuration III and IV tell us that the benefit is marginal.

\begin{table}[!htb]
\caption{Comparison of deconvolution methods in the spatial domain or the feature space.}
    \label{tab:dwdn_table}
    \centering
    \scalebox{0.9}{
    \begin{tabular}{ccccc}
    &Config. I & Config II & Config III & Config IV\\
    \hline
         Features& \xmark & \xmark & \cmark & \cmark \Tstrut\\
         3 Wieners& \xmark & \cmark & \xmark & \cmark \Bstrut\\
          \hline \hline
         10 ppp & 22.08 dB & 22.13 dB & 22.43 dB & \textbf{22.47 dB}\Tstrut\\
         30 ppp & 23.09 dB & 23.11 dB & 23.41 dB & \textbf{23.47 dB}\\
         50 ppp & 23.59 dB & 23.64 dB & 23.92 dB & \textbf{23.97 dB}\Bstrut\\
         \hline
    \end{tabular}}
\end{table}

Based on these findings, our recommendation here is that whenever possible, deconvolution should be performed in the feature space. Using different regularization parameter $\lambda$ does not seem to have a significant difference.

\subsection{Secret 4: Poisson likelihood is not needed}
By virtue of Poisson deconvolution, the likelihood function should be Poisson. However, several observations make us believe that the Poisson likelihood is not needed in a neural network based solution.

\begin{figure}[!h]
\centering
\begin{tabular}{c}
    \includegraphics[height=2.4cm]{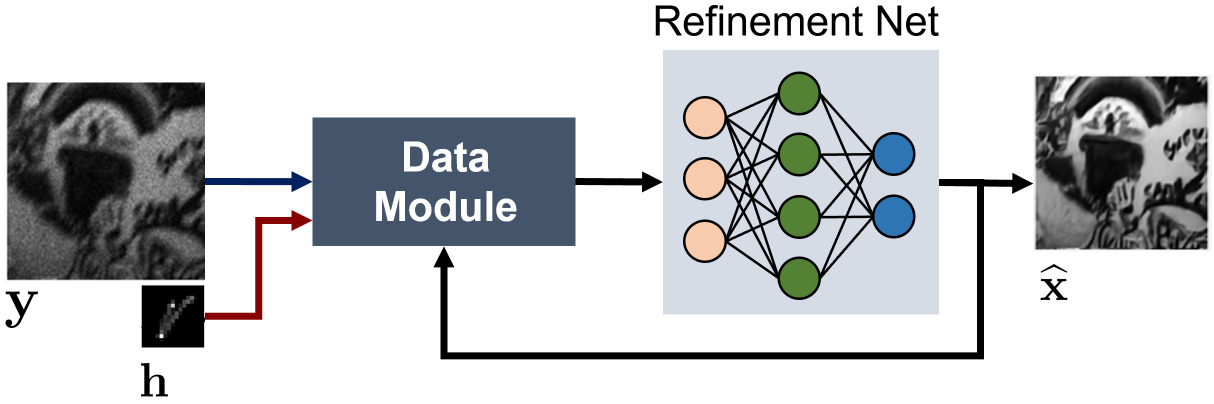}\\
    (a) USRNet \\
    \includegraphics[height=2.4cm]{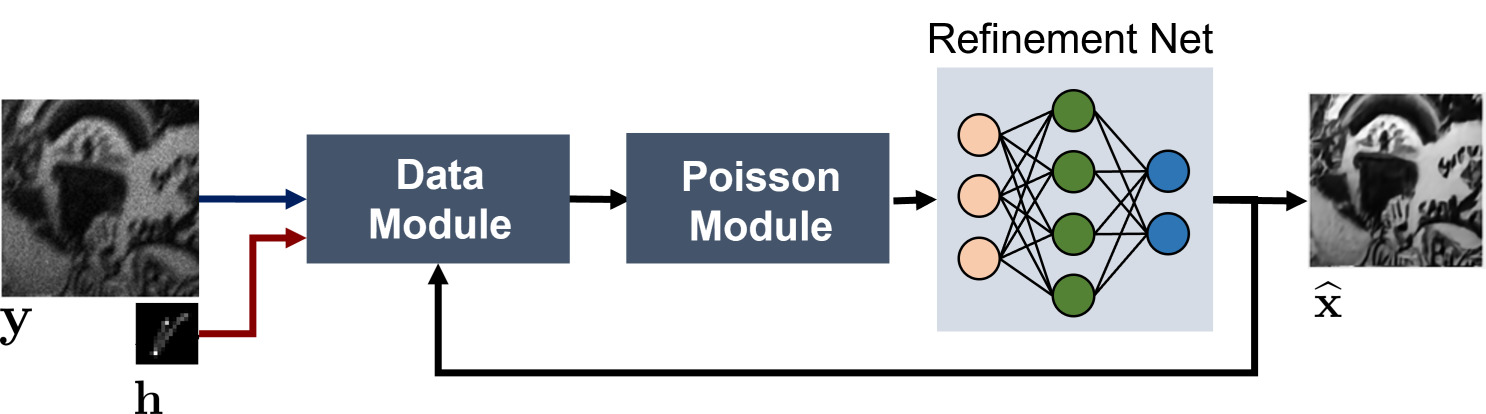}\\
     (b) PhDNet
\end{tabular}
\caption{\textbf{Secret 4: Poisson likelihood is not needed}. USRNet and PhDNet are both iterative unrolled networks. The difference is that in PhDNet, an explicit Poisson module is used to handle the Poisson noise. }
\label{fig:Poisson_design_models}
\end{figure}

Our first observation is the comparison between USRNet and PhDNet in \tref{tab:SOTA_comparison}. USRNet uses a Gaussian likelihood whereas PhDNet uses a Poisson likelihood. Because of the Poisson likelihood, PhDNet needs to introduce a variable splitting technique to specifically handle the Poisson part, see the added Poisson module illustrated in \fref{fig:Poisson_design_models}. However, from Table \ref{tab:Poisson_design_table} we observe that the difference in performance between the two methods is negligible.

Readers may argue that the vanishing performance gap is due to the iterations, i.e., as the number of iterations increases, the network capacity increases and hence they are more capable of handling the Poisson statistics. To prove that this is not the case, we train four versions of USRNet and PhDNet with a fixed number of 1, 2, 4, and 8 iterative blocks in their unrolled networks. We can see from \tref{tab:Poisson_design_table} that irrespective of the number of iterations used by the method, USRNet performs as well as PhDNet. Therefore, whether or not we use an explicit Poisson module does not matter.

\begin{table}[!h]
\caption{Comparison of two unrolled iterative algorithms. USRNet uses a Gaussian likelihood whereas PhDNet uses a Poisson likelihood. The medium photon level of the images in this experiment is set to 30 ppp.}
    \label{tab:Poisson_design_table}
    \centering
    \scalebox{0.9}{
    \begin{tabular}{ccccc}
    \hline
     & {1 Itr.} & {2 Itr.} & {4 Itr.} & {8 Itr.} \Tstrut\\
     \hline \hline
     USRNet \cite{Zhang_2020_USRNet}&\hspace{-1.0ex}24.40 dB&\hspace{-1.0ex}24.79 dB&\hspace{-1.0ex}24.88 dB&\hspace{-1.0ex}24.89 dB\Tstrut \\
     PhDNet \cite{Sanghvi_2022_TCI}&\hspace{-1.0ex}24.40 dB&\hspace{-1.0ex}24.77 dB&\hspace{-1.0ex}24.87 dB&\hspace{-1.0ex}24.87 dB\\
     \hline
    \end{tabular}}
\end{table}

Another ``indirect'' observation is about the design of PURE-LET. In PURE-LET, the Poisson statistics is used to estimate the PURE score which is an unbiased risk estimator of the mean squared error. However, the actual deconvolution step is performed by a bank of Wiener filters - which is derived from Gaussian statistics.

\begin{figure}[!h]
\centering
\begin{tabular}{c}
    \includegraphics[height=1.5cm]{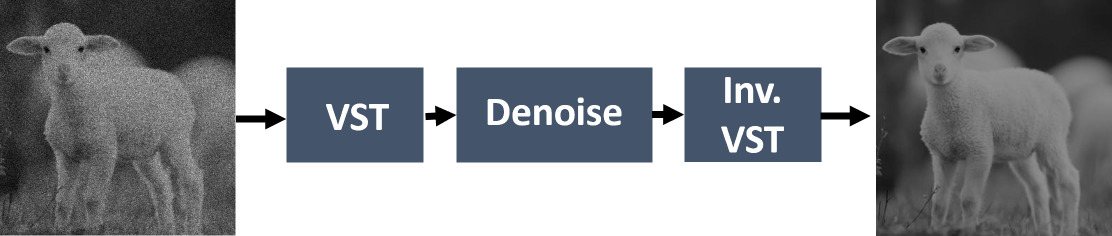}\\
    (a) With VST \\
    \includegraphics[height=1.5cm]{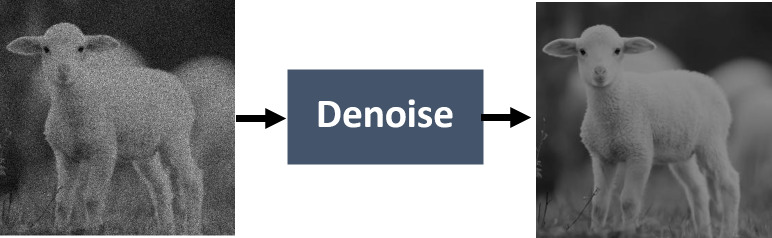}\\
     (b)  Without VST
\end{tabular}
\caption{\textbf{Variance Stabilizing Transform}. VSTs are often used in Poisson noise. (a) A denoising method using VST. (b) A denoising method without VST.}
\label{fig:wVSt_woVST_networks}
\end{figure}

If Poisson modules are not needed, we expect that techniques associated with the Poisson likelihood would not have any significance to the restoration problem. This observation is supported by inspecting methods using the variance stabilizing transform (VST). \fref{fig:wVSt_woVST_networks} shows a typical VST-based image denoising algorithm. In the VST case, we first apply VST to stabilize the Poisson variance. We then denoise the image, and transform back via the inverse VST. In our experiment, we use the ResUNet from \cite{Zhang_2020_USRNet} as the denoiser.

\tref{tab:VST_table} shows the performance between using VST or not. We observe that using VST does not offer the denoiser any advantage. The network without VST even marginally outperforms the denoiser with VST. This finding is consistent with what was reported in \cite{Choi_2018_ICASSP} for binomial noise.

\begin{table}[!h]
\caption{Effect of VST to a low-light denoising task. In this experiment, the denoiser is ResUNet [].}
    \label{tab:VST_table}
    \centering
    \scalebox{0.9}{
    \begin{tabular}{ccc}
    \hline
         ppp & w/ VST & w/o VST\\
          \hline \hline
         10 ppp & 28.28 dB & 28.34 dB\Tstrut\\
         30 ppp & 30.61 dB & 30.67 dB\\
         50 ppp & 31.77 dB & 31.82 dB\Bstrut\\
         \hline
    \end{tabular}}
\end{table}

Based on the above analysis, our conclusion is that when handling the Poisson noise in low-light, network architectures designed for Gaussian likelihood will work just as well. There is no clear advantage of using the more complicated Poisson likelihood and/or variance stabilizing transforms. As long as we can synthesize the Poisson noise for training, the explicit Poisson modules are unimportant.

\subsection{Secret 5: Learning hyperparameters is recommended}
\begin{figure}[!ht]
\centering
    \includegraphics[width=0.9\linewidth]{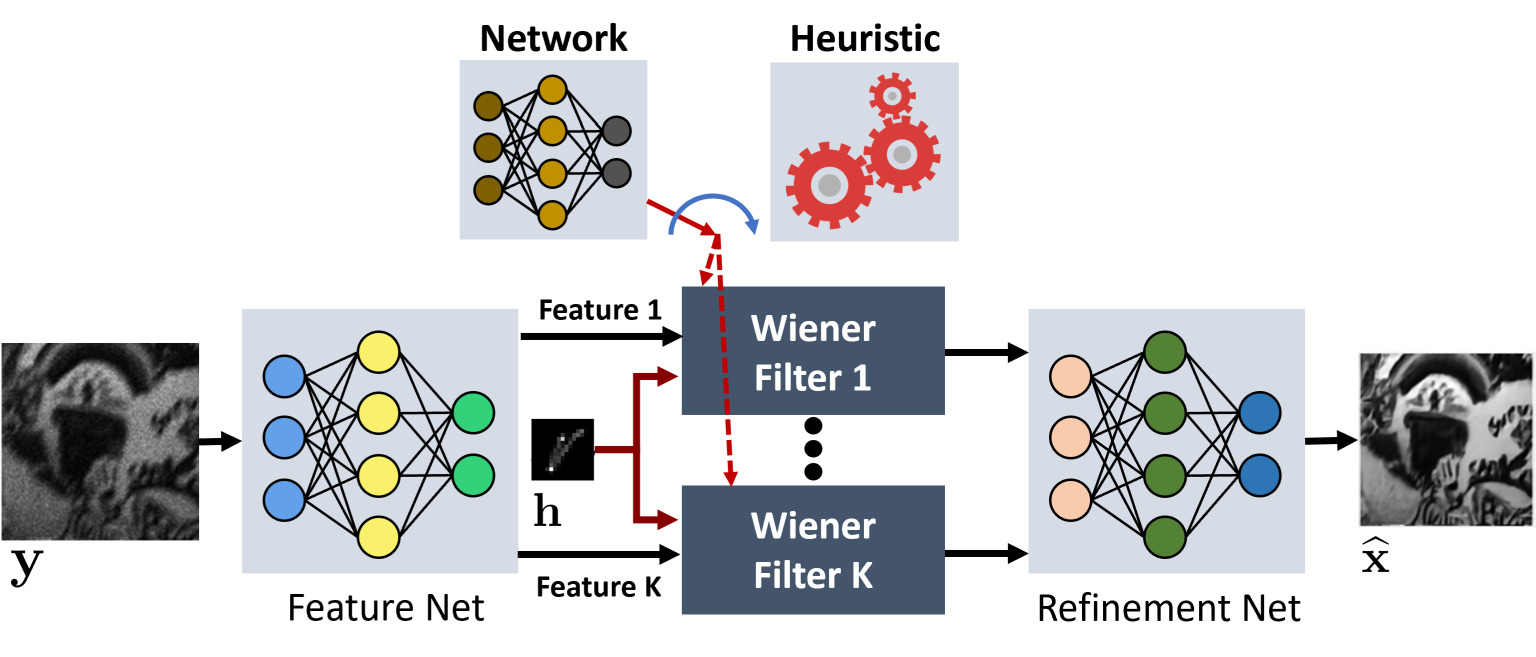}
\caption{\textbf{Secret 5: Hyper-parameter learning is recommended}. We can use heuristics or train a network to select the hyper-parameters. }
\label{fig:secret5}
\end{figure}

Among the learning-based methods, USRNet and PhDNet use networks to learn the hyperparameters that get used in the data module, Poisson module and the refinement net. However, DWDN uses a heuristic method for estimating this parameter. To understand if it is important to learn the hyperparameters, we modify DWDN and learn the hyperparameter that is being input to the Wiener filters using the same network structure used by PhDNet to learn its parameters. \fref{fig:secret5} illustrates the conceptual difference between the two.

The result of this experiment can be found in \tref{tab:parameters_table}. We notice that when DWDN is augmented with a small network for learning the hyperparameters, it performs slightly better than using a heuristic for finding the parameters. The improvement is less than 0.1dB which is not very noticeable. However, since the computational cost of adding a hyper-parameter learning module is so small compared to the whole network, it does not hurt to include it.

\begin{table}[!ht]
\caption{\textbf{Impact of learning hyperparameters}.}
    \label{tab:parameters_table}
    \centering
    \begin{tabular}{ccc}
    \hline
        ppp & DWDN & w/ DWDN + learned para.    \\
          \hline \hline
         10 & 22.43 dB & \textbf{22.50 dB}\Tstrut\\
         30 & 23.41 dB & \textbf{23.49 dB}\\
         50 & 23.92 dB & \textbf{23.99 dB}\Bstrut\\
         \hline
    \end{tabular}
\end{table}

Based on the above experiments, we conclude that hyper-parameter learning is helpful but it is not necessary. We still recommend it because it saves us from hand-tuning the hyper-parameters.

\section{Combining the Secrets} \label{sec:ProposedMethod}
\begin{figure*}
    \centering
    \includegraphics[width=0.6\linewidth]{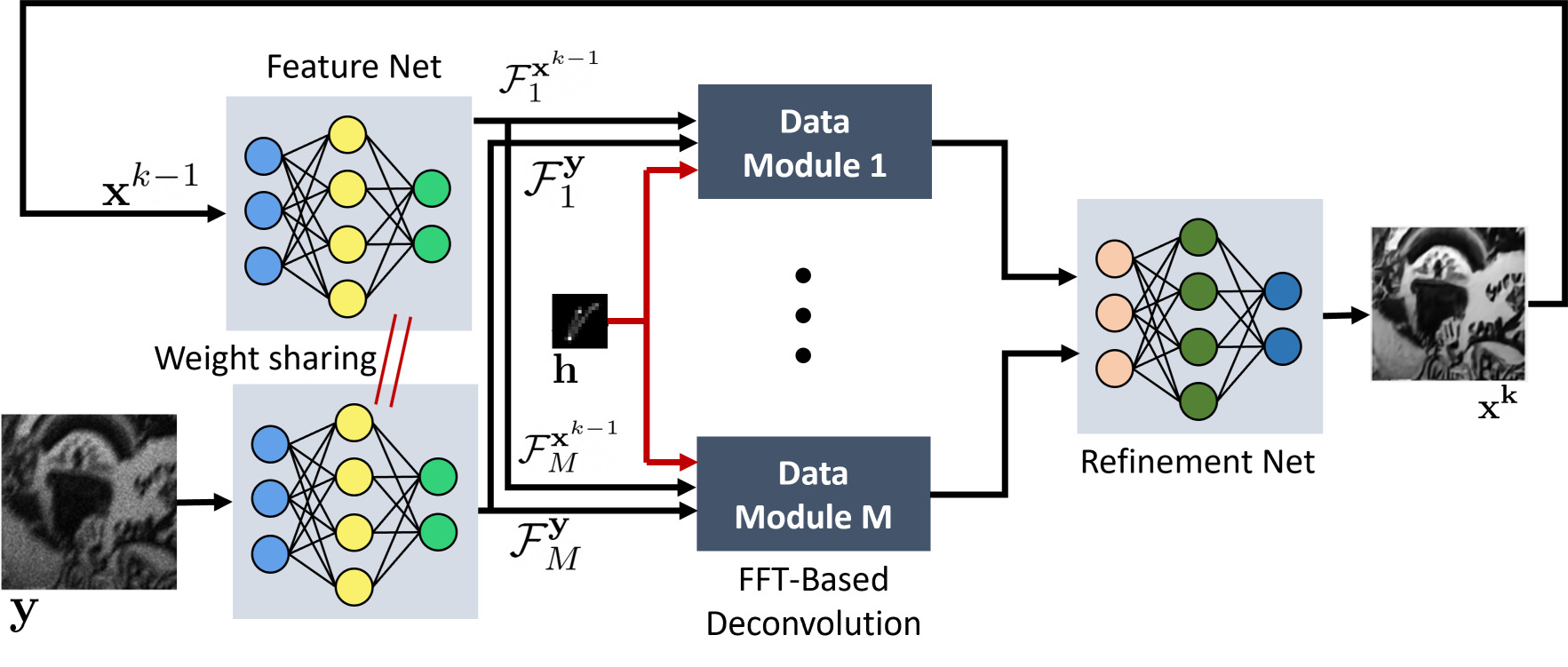}
    \caption{\textbf{Schematic diagram of FIO-Net}. The Five-in-One Network (FIO-Net) utilizes all the five secrets we observed in the previous section. It is an iterative scheme performing deconvolution in the feature space. It uses Wiener filter, but no Poisson likelihood. Hyperparameters are automatically tuned.}
    \label{fig:ProposedMethod}
\end{figure*}

After presenting the five secrets, a natural question is: ``what if we combine these ideas?'' To this end, we create a method called the Five-in-One Network (FIO-Net). We make two remarks before we discuss this network: Firstly, we do not regard FIO-Net as a novel invention or claim it to be a state-of-the-art. We view FIO-Net a check point of the five secrets. We are more interested in checking whether its performance is consistent with the five secrets, rather than expecting it to beat other methods by a big margin. Secondly, although FIO-Net is a combination of the five secrets, it would still require some design because otherwise there is no guarantee it should work. We will present a way to integrate these five ideas.

To elaborate on the design principle of the FIO-Net, we first use Secret 4 to replace Poisson likelihood with the Gaussian likelihood. This implies that as far as the network structure is concerned, we can focus on the Gaussian forward model:
\begin{align}
\vy = \alpha\mH \vx + \vn,
\end{align}
where $\alpha$ defines the photon level. We remark that this is \emph{not} the original Poisson deconvolution problem that we want to solve. However, since Secret 4 tells us that utilizing the Poisson likelihood is not needed, we consider the Gaussian model when designing the neural network. When training the model, we take blurred images and add synthetic Poisson noise instead of Gaussian noise.

Remark: The concept using a sub-optimal forward model in exchange of better reconstruction performance is perhaps counter-intuitive. The general line of argument is known as computational image formation which we refer readers to \cite{Chan_2023_CIF} for detailed elaborations.

Our next step is to use Secret 3, which suggests us to perform the deconvolution in the feature space. To this end, we consider a set of linear filters $\{\mF_i \;|\; i = 1, 2, \hdots, M\}$ and apply them to
\begin{align}
    \mF_i\vy = \alpha\mF_i\mH \vx + \mF_i\vn. \label{eq:noise_model_linearFilter}
\end{align}
Since $\mF_i$ and $\mH$ represent convolutional operations in matrix form, we can switch the order using the commutative property of convolution  to obtain
\begin{align}
    \mF_i\vy = \alpha\mH \mF_i \vx + \mF_i\vn. \label{eq:noise_model_linearFilter_rearranged}
\end{align}
The question now becomes how to recover $\vx$.

Solving \eref{eq:noise_model_linearFilter_rearranged} would require an optimization. In FIO-Net, we consider a generic regularized least squares:
\begin{align}
    \widehat{\vx} = \argmin{\vx}  \sum_{i=1}^M \norm{\mF_i \vy - \alpha  \mH \mF_i \vx}^2  + \lambda g(\vx) \label{eq:optimization_version2}.
\end{align}
where $g(\vx)$ is the prior. Since an unconstrained optimization problem with a sum of two different functions is difficult to optimize, we split the original problem into two simpler subproblems. We introduce a set of new variables $\{\vz_i = \mF_i \vx, i=1,2,\hdots M\}$, and collectively define $\vz = \{\vz_1,\ldots,\vz_M\}$. The new constrained optimization problem now becomes
\begin{align}
\left\{\widehat{\vx},\widehat{\vz}\right\}
= &\argmin{\vx,\vz}  \;\; \sum_{i=1}^M \Big\{\norm{\mF_i \vy - \alpha  \mH \vz_i}^2 + \lambda g(\vx)\Big\} \nonumber \\
  &\text{subject to } \;\; \vz_i = \mF_i \vx, \; i = 1,2,\ldots,M \label{eq:constrained_optimization}.
\end{align}
\eref{eq:constrained_optimization} is a standard optimization that can be solved using the half-quadratic splitting (HQS) \cite{Zhang_2020_USRNet}. HQS formulates an alternative optimization:
\begin{align}
    \left\{\widehat{\vx},\widehat{\vz}\right\} = \argmin{\vx,\vz}  \sum_{i=1}^M \Big\{\norm{\mF_i \vy - \alpha  \mH \vz_i}^2 +& \lambda g(\vx) \nonumber \\+ \mu_i \norm{\mF_i \vx - \vz_i}^2\Big\} \label{eq:HQSmod},
\end{align}
where $\mu_i$ is the penalty strength.

In what follows, we briefly summarize the equations to solve \eref{eq:HQSmod}. During the discussion, we will explain how the secrets are used. The algorithm to solve \eref{eq:HQSmod} involve two steps:
\begin{align}
    \vz_i^k &= \argmin{\vz_i} \norm{\mF_i \vy - \mH \vz_i}^2 + \mu_i^k \norm{\mF_i\vx^{k-1} - \vz_i}^2 \label{eq:data_term}\\
    \vx^k   &= \argmin{\vx}\sum_{i=1}^M \mu_i^k \norm{\mF_i \vx - \vz_i^k}^2 + \lambda g(\vx), \label{eq:prior_term}
\end{align}
where we use the fact that the optimization of $\vz_i$ in \eref{eq:HQSmod} is separable so that we can solve for individual $\vz_i$'s.

Next, we apply Secret 5 which says that we should learn the hyperparameters end-to-end. Thus, we replace the penalty $\mu_i$ with $\mu_i^k$ so that they change over iterations. Similar to \cite{Sanghvi_2022_TCI}, we use a small fully connected neural network for estimating the hyperparamaters $\mu_i^k$ with the kernel $\mH$ and the photon level $\alpha$ used as the input.

Let's solve \eref{eq:data_term} and \eref{eq:prior_term}. \eref{eq:data_term} is a least squares minimization problem, and it has a closed form expression given by
\begin{align}
    \vz_i^k = (\mathbb{I} + \mu_i^k \mH^T \mH)^{-1}(\mF_i\vx^{k-1} +\mu_i^k \mH^T \mF_i \vy). \label{eq:least_squares_soln}
\end{align}
Assuming that the convolution operation represented by $\mH$ is carried out with circular boundary conditions, \eref{eq:least_squares_soln} has a FFT based solution given by
\begin{align}
    \vz_i^k = \mathcal{F}^{-1}\left[\frac{\mathcal{F}(\mF_i \vx^{k-1}) + \mu_i^k \overbar{\mathcal{F}(\mH)}\cdot\mathcal{F}(\mF_i\vy)}{ 1 + \mu_i^k |\mathcal{F}(\mH)|^2}\right]\label{eq:least_squares_soln_FFT},
\end{align}
where $\mathcal{F}(\cdot)$ and $\mathcal{F}^{-1}(\cdot)$ denote the FFT and inverse FFT respectively, and $\overbar{(\cdot)}$ denotes the complex conjugate function. Following the idea of \cite{Dong_2022_DeepWiener_PAMI}, we replace the linear filters with learnable non-linear convolutional neural network $\mathcal{D}_\text{feat}(\cdot)$. Similar to \cite{Dong_2022_DeepWiener_PAMI}, we note that while the solution \eref{eq:least_squares_soln_FFT} was obtained for linear filters, using non-linear neural networks works well and even better than linear filters. Therefore, \eref{eq:least_squares_soln_FFT} is modified as
\begin{align}
    \vz_i^k = \mathcal{F}^{-1}\left[\frac{\mathcal{F}(\mathcal{D}^\text{feat}_i(\vx_{k-1})) + \mu_i^k \overbar{\mathcal{F}(\mH)}\cdot\mathcal{F}(\mathcal{D}^\text{feat}_i(\vy))}{ 1 + \mu_i^k |\mathcal{F}(\mH)|^2}\right],\label{eq:least_squares_soln_FFT_wNN}
\end{align}
where $\{\mathcal{D}^\text{feat}_1(\cdot), \hdots, \mathcal{D}^\text{feat}_M(\cdot)\} = \mathcal{D}^\text{feat}(\cdot)$ represents the features generated by the neural network.

\begin{algorithm}[ht]
\begin{algorithmic}[1]
\State \textbf{Input}: Degraded Image $\vy$, Kernel $\mH$, Photon level $\alpha$
\State $\vx^0 \leftarrow \vy$
\State $\{\calF_i^{\vy}\}_{i=1,\hdots,M} = \mathcal{D}^\text{feat}(\vy)$ \Comment{Feature extraction from $y$}
\State $\{\mu_k\}_{k=1,\hdots,K} = \mathcal{D}^\text{hyp}(\mH, \alpha)$\Comment{Hyperparameters from $\mathcal{D}^\text{hyp}(\cdot)$}
    \For{$k = 1, 2, \cdot\cdot\cdot, K$}
    \State Update $\vz_i^k$ using Equation~\eqref{eq:least_squares_soln_FFT}
    \State Update $\vx^k$ using Equation~\eqref{eq:refine_xk}
    \EndFor
\State return $\vx^{K}$
\end{algorithmic}
\caption{\emph{FIO-Net}: Fixed Iteration Unrolling}
\label{alg:FIONet}
\end{algorithm}

The other subproblem in \eref{eq:prior_term}, in the absence of the filters $\mF_i$, can be thought of as a image denoising problem \cite{chan2016plug}. However, the presence of the filters makes this problem not so straight-forward. However, we can still think of \eref{eq:prior_term} as a restoration task where we want to recover the image $\vx$ from a set of features $\vz_i$. We want to minimize the residue between the input features and the features generated from $\vx$, while enforcing the prior $g(\vx)$. Given the complex nature of restoring the image from a set of features, and the difficulty of defining a good prior term $g(\vx)$, we propose to solve this problem using a convolutional neural network as
\begin{align}
    \vx^k  = \mathcal{D}_{\text{refine}}\left(\vz_1^{k}, \hdots, \vz_M^{k}, \frac{\lambda}{\mu^k}\right), \label{eq:refine_xk}
\end{align}
where we have assumed that the penalties $\mu_i^k = \mu^k$ do not vary over the features. The entire algorithm is summarized in algorithm \ref{alg:FIONet}.

The method is iterative, based on unrolled optimization that uses the convolutional neural networks only for image refinement and a traditional FFT based method is used for deconvolution. The iterative scheme described in Algorithm \ref{alg:FIONet} is unrolled for K = 8 iterations and then trained end-to-end using the same training process as that described in Section III-A. The method incorporates the idea of deconvolving in the feature space, and does not have any specific Poisson design.

\section{Experiments}
After elaborating on the proposed method, we present the quantitative results on BSD300 dataset in  Table~\ref{tab:SOTA_comparison}. We use the same testing process as described in Section~\ref{subsec:training} so that the testing conditions are fair to all methods. We make three comments:
\begin{itemize}
\item Compared to classical methods such as PURE-LET and VSTP, FIO-Net outperforms by a big margin. This should not be a surprise, because all deep learning methods outperform these two classical methods.
\item Compared to a single-pass deep learning method DWDN, the performance of FIO-Net is substantially better, especially for bigger blur kernels. This stresses the importance of iterative methods.
\item Compared to PhD-Net and USR-Net, the performance of FIO-Net is marginal. This is caused by the fact that some of the attributes have overlapping influences, e.g., feature space and iteration. While Secret 3 says that feature space deconvolution could help single-iteration methods, its impact may be diminished when more iterations are used.
\end{itemize}

\begin{figure*}[htp]
\centering
\begin{tabular}{ccccccccc}
\multirow{2}{*}[1.3cm]{\hspace{-2.0ex}\includegraphics[width=0.18\linewidth]{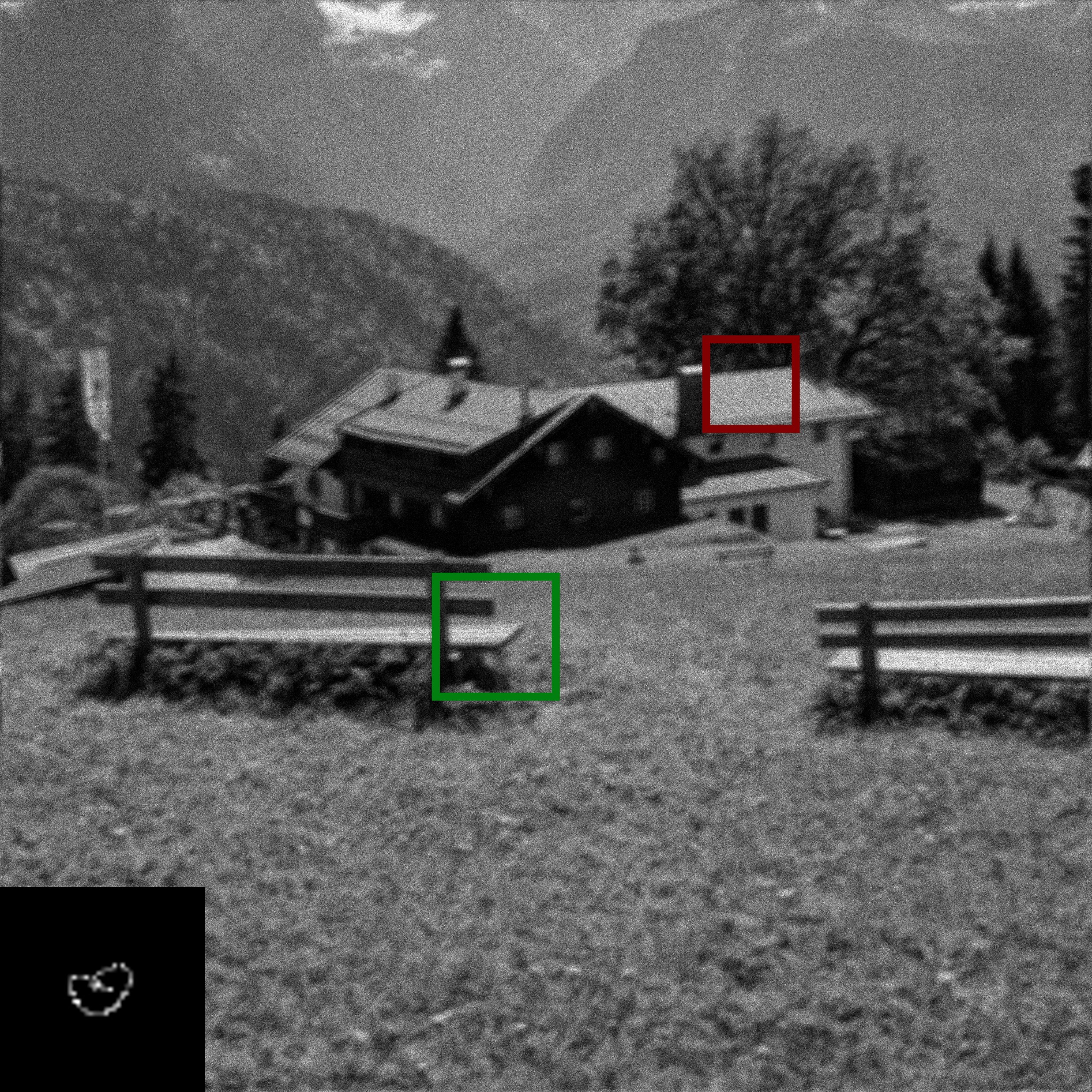}}&
\hspace{-2.0ex}\includegraphics[width=0.09\linewidth]{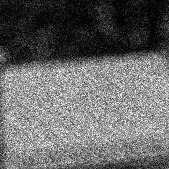}&
\hspace{-2.0ex}\includegraphics[width=0.09\linewidth]{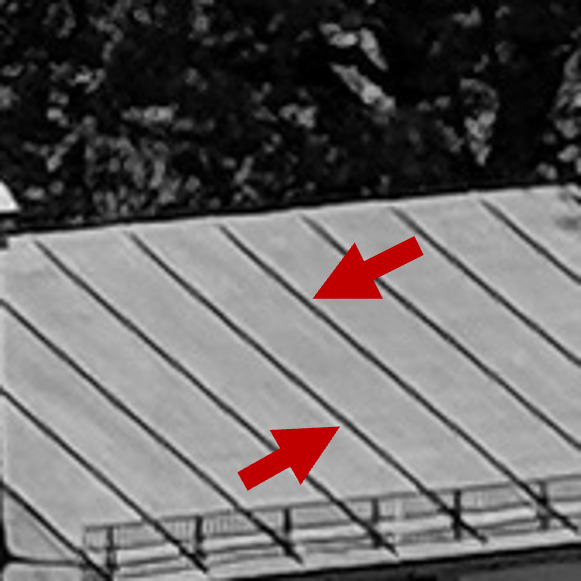}&
\hspace{-2.0ex}\includegraphics[width=0.09\linewidth]{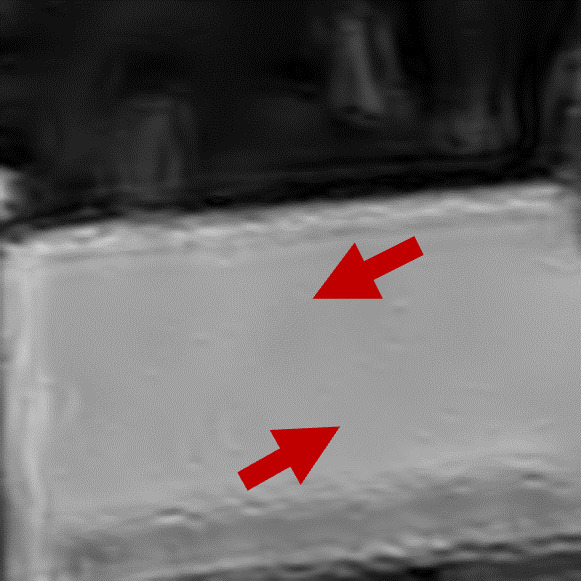}&
\hspace{-2.0ex}\includegraphics[width=0.09\linewidth]{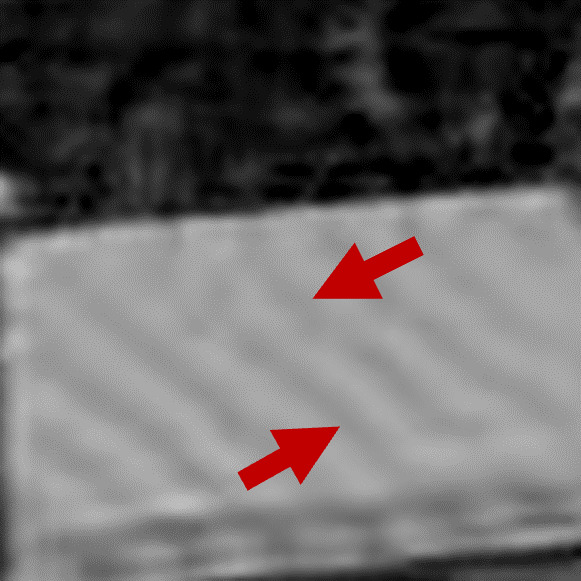}&
\hspace{-2.0ex}\includegraphics[width=0.09\linewidth]{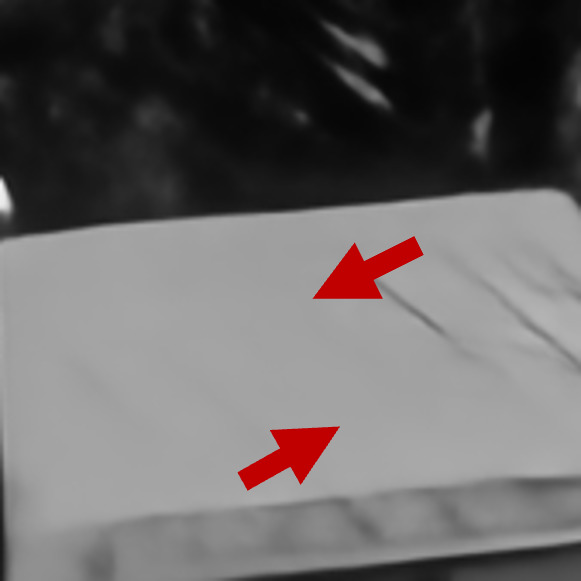}&
\hspace{-2.0ex}\includegraphics[width=0.09\linewidth]{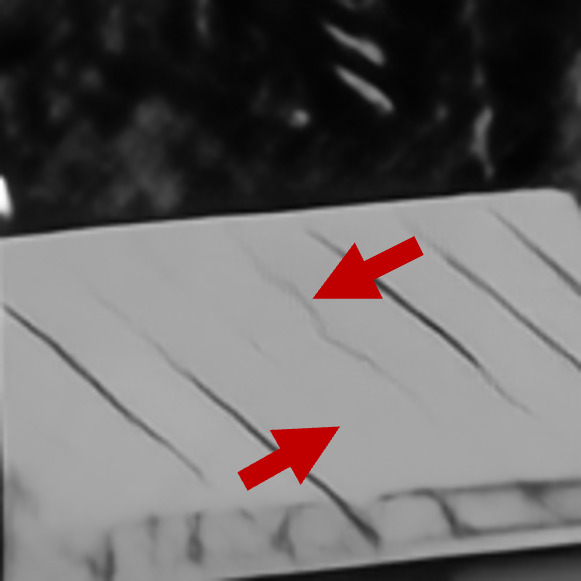}&
\hspace{-2.0ex}\includegraphics[width=0.09\linewidth]{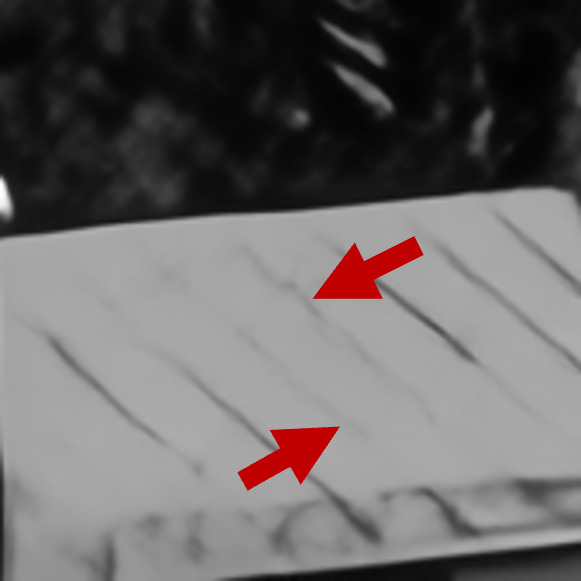}&
\hspace{-2.0ex}\includegraphics[width=0.09\linewidth]{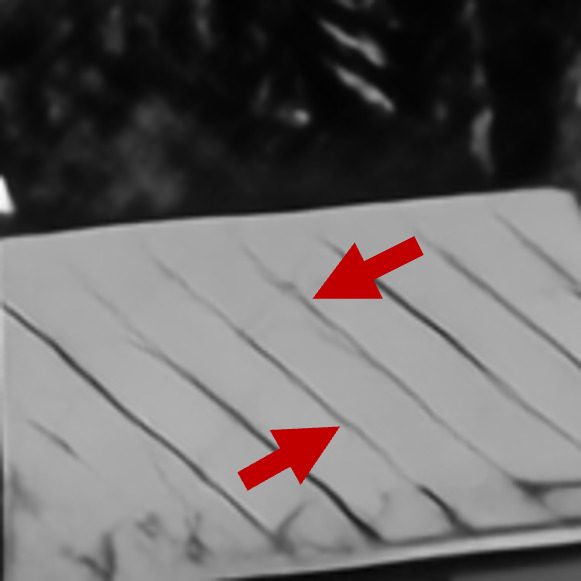}\\
&
\hspace{-2.0ex}\includegraphics[width=0.09\linewidth]{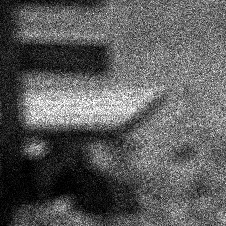}&
\hspace{-2.0ex}\includegraphics[width=0.09\linewidth]{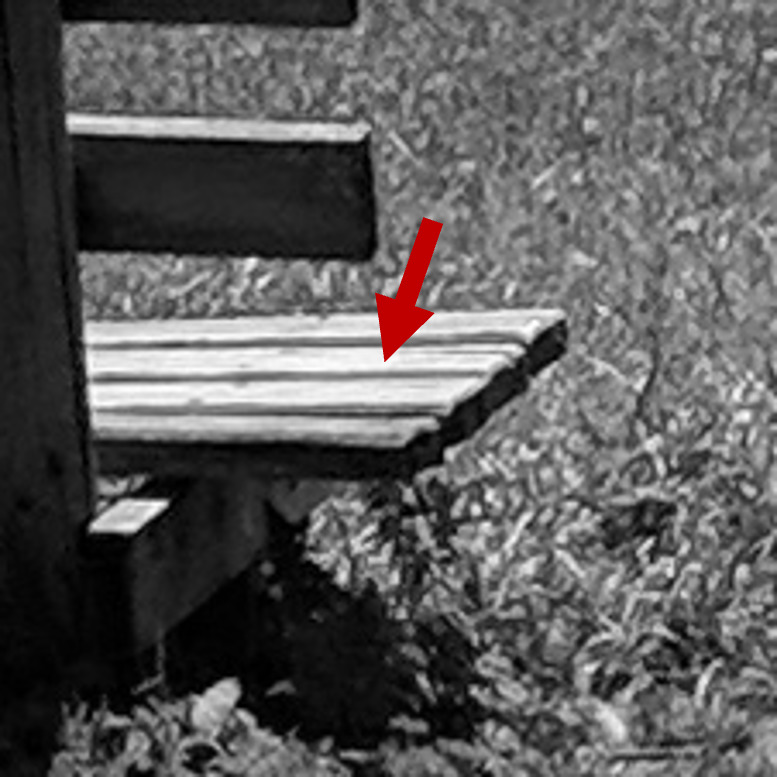}&
\hspace{-2.0ex}\includegraphics[width=0.09\linewidth]{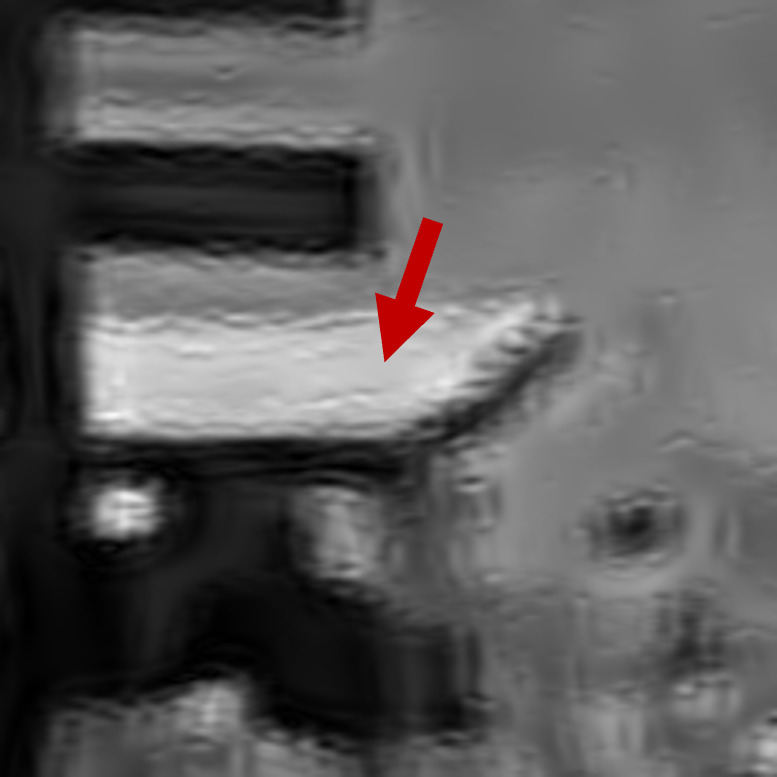}&
\hspace{-2.0ex}\includegraphics[width=0.09\linewidth]{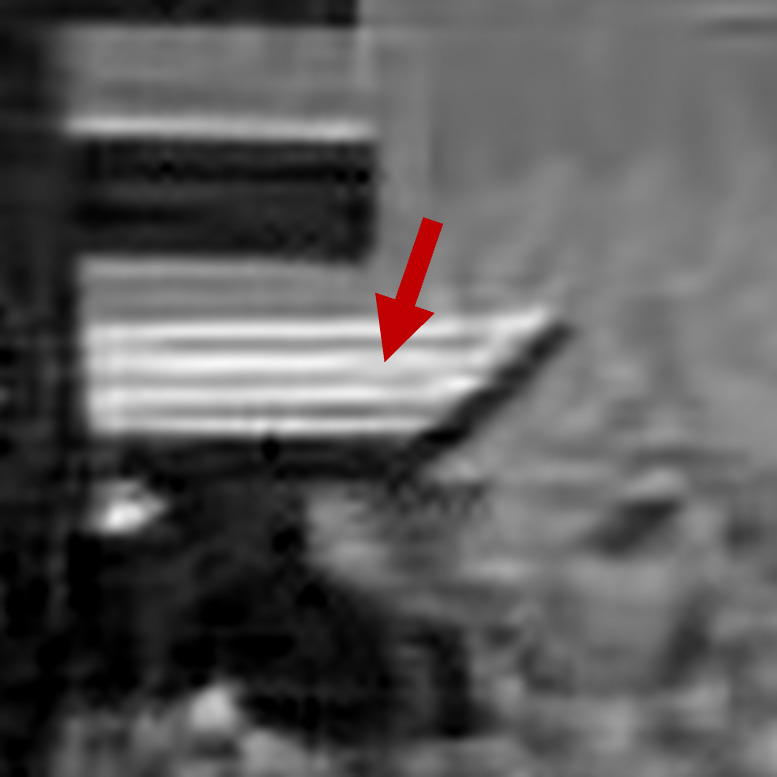}&
\hspace{-2.0ex}\includegraphics[width=0.09\linewidth]{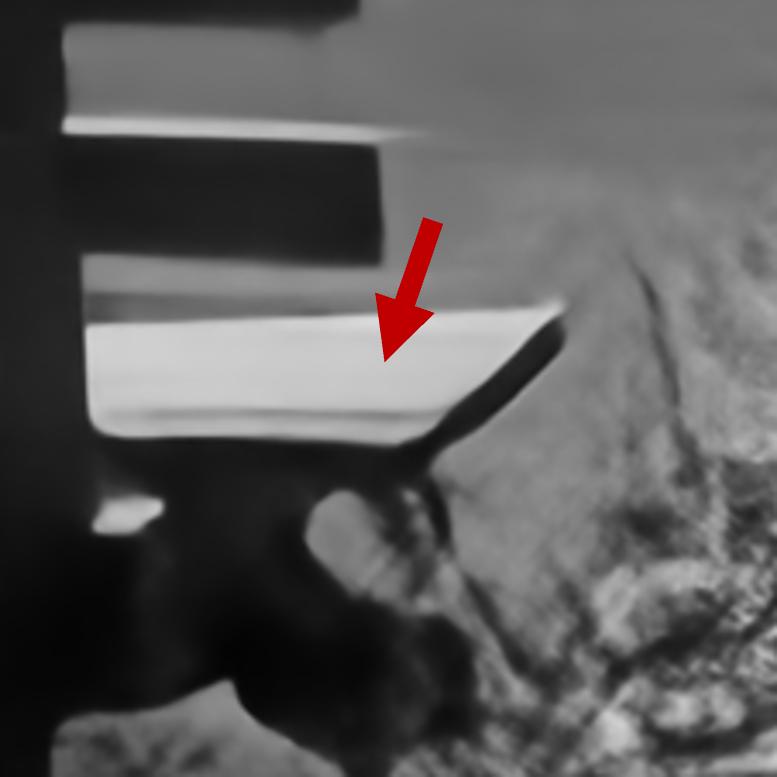}&
\hspace{-2.0ex}\includegraphics[width=0.09\linewidth]{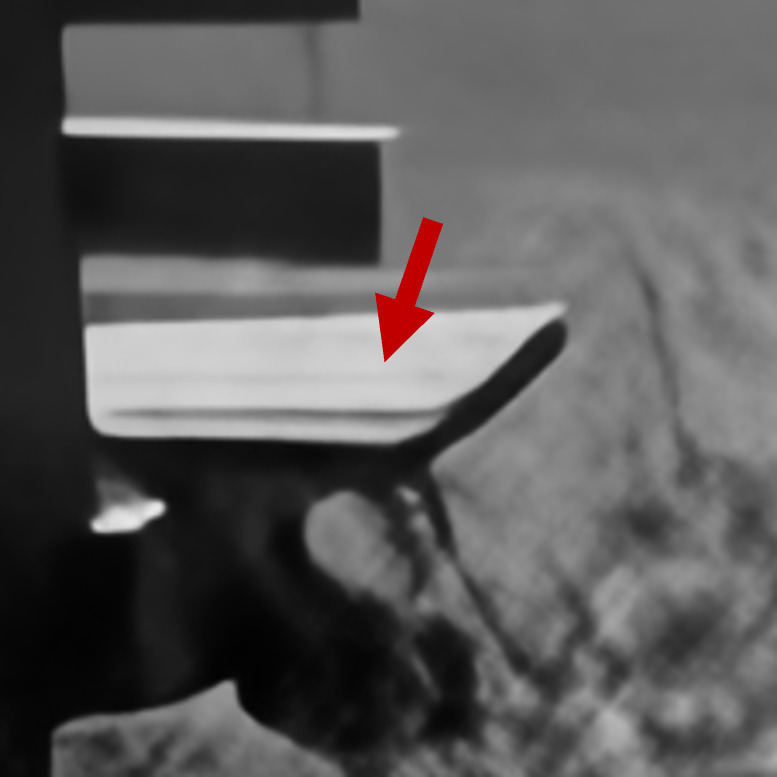}&
\hspace{-2.0ex}\includegraphics[width=0.09\linewidth]{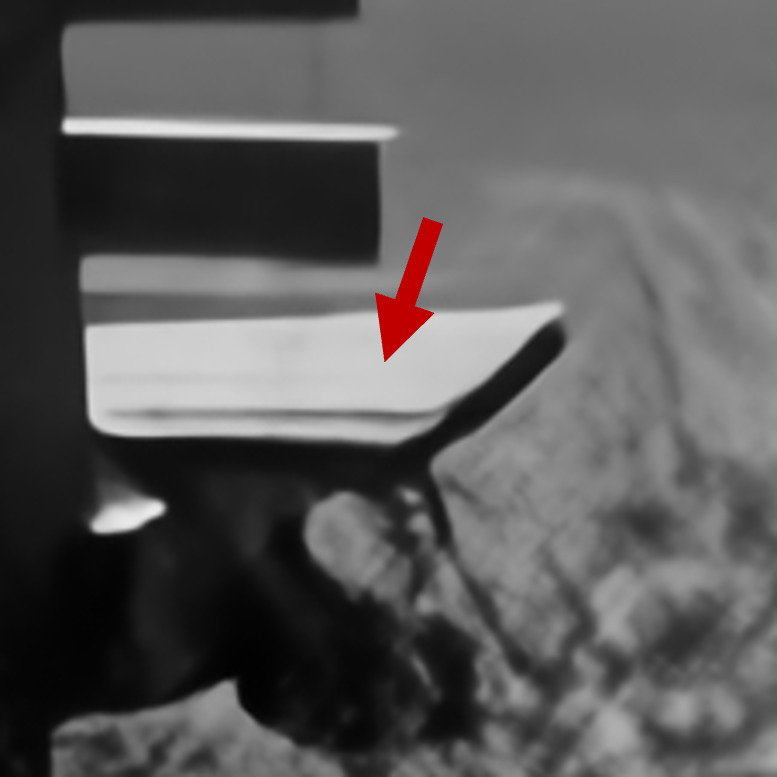}&
\hspace{-2.0ex}\includegraphics[width=0.09\linewidth]{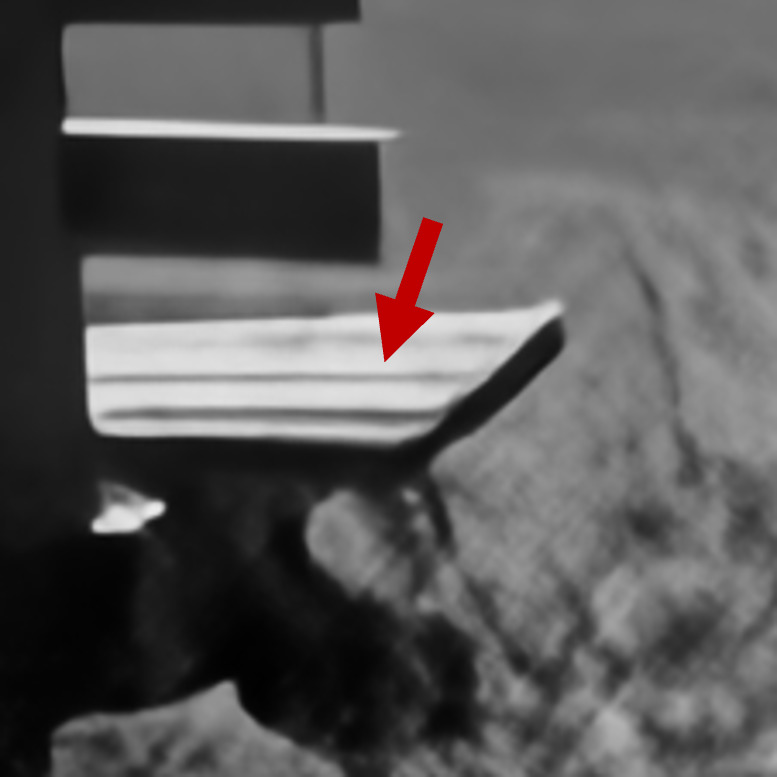}\\
\multirow{2}{*}[1.3cm]{\hspace{-2.0ex}\includegraphics[width=0.18\linewidth]{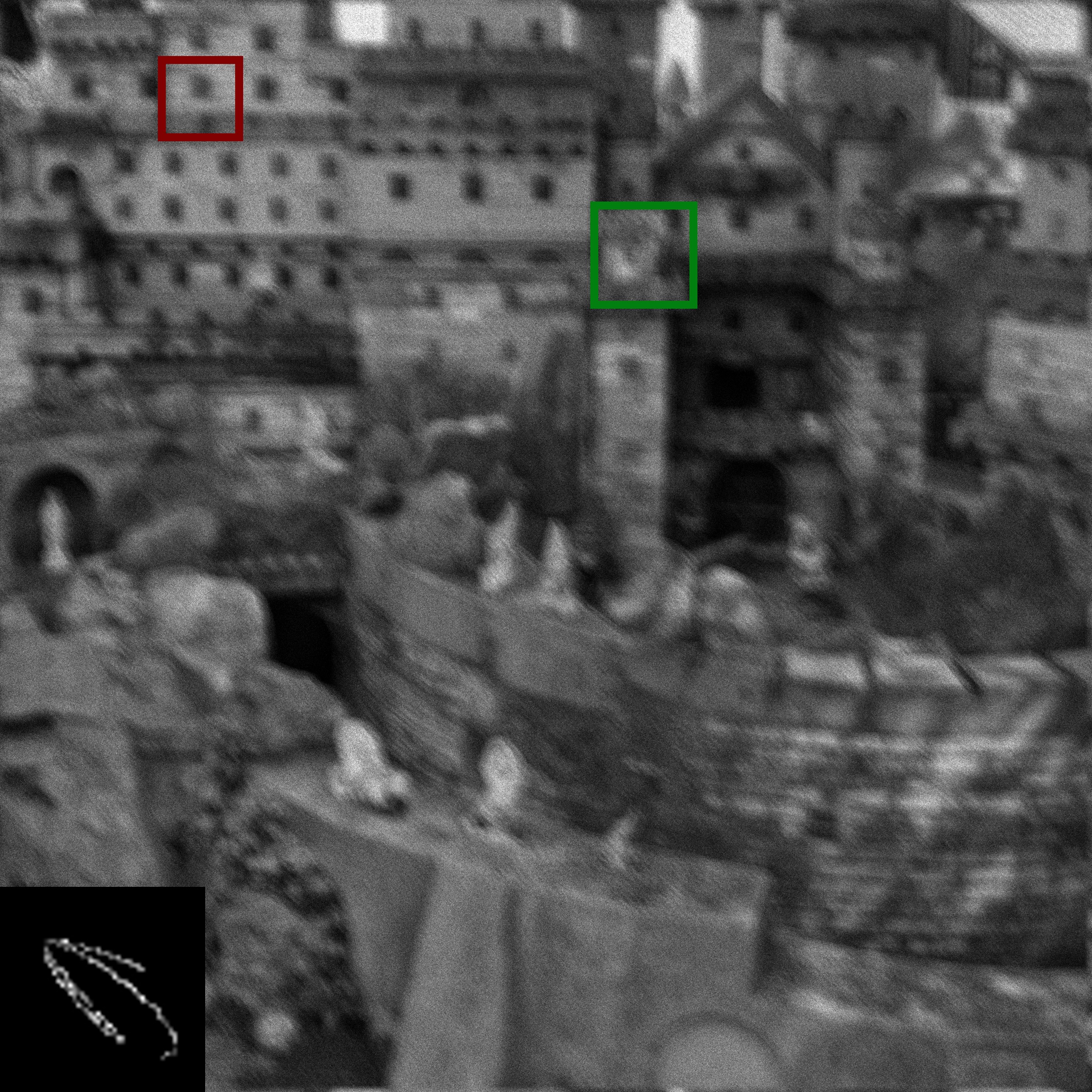}}&
\hspace{-2.0ex}\includegraphics[width=0.09\linewidth]{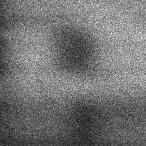}&
\hspace{-2.0ex}\includegraphics[width=0.09\linewidth]{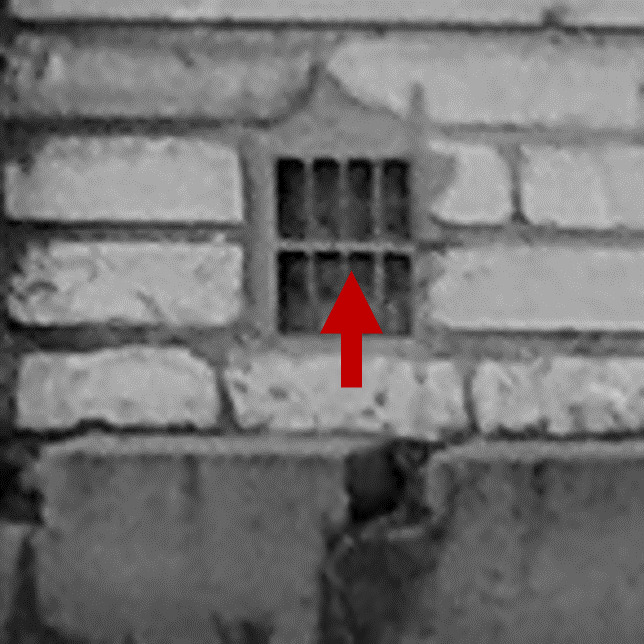}&
\hspace{-2.0ex}\includegraphics[width=0.09\linewidth]{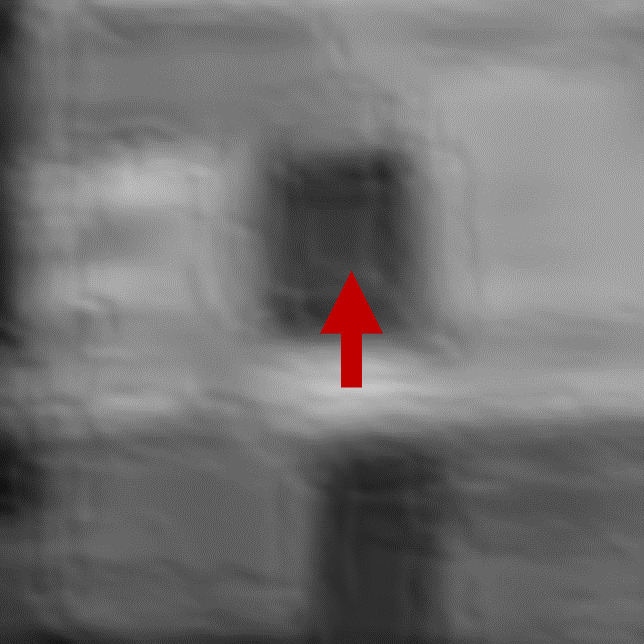}&
\hspace{-2.0ex}\includegraphics[width=0.09\linewidth]{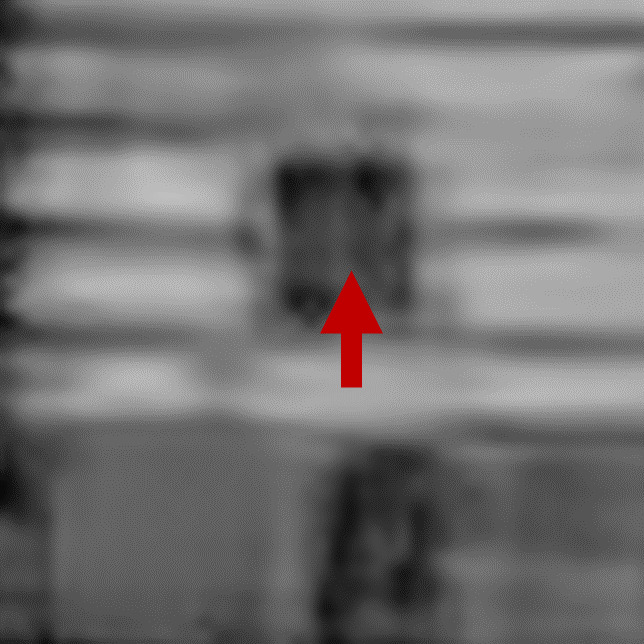}&
\hspace{-2.0ex}\includegraphics[width=0.09\linewidth]{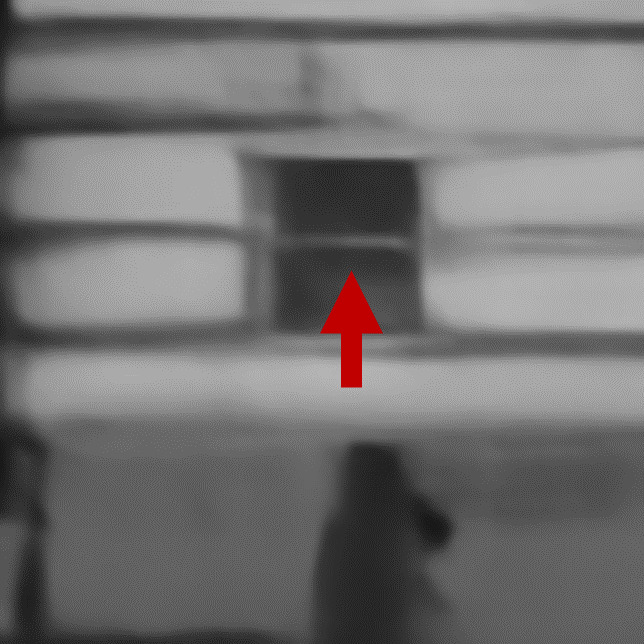}&
\hspace{-2.0ex}\includegraphics[width=0.09\linewidth]{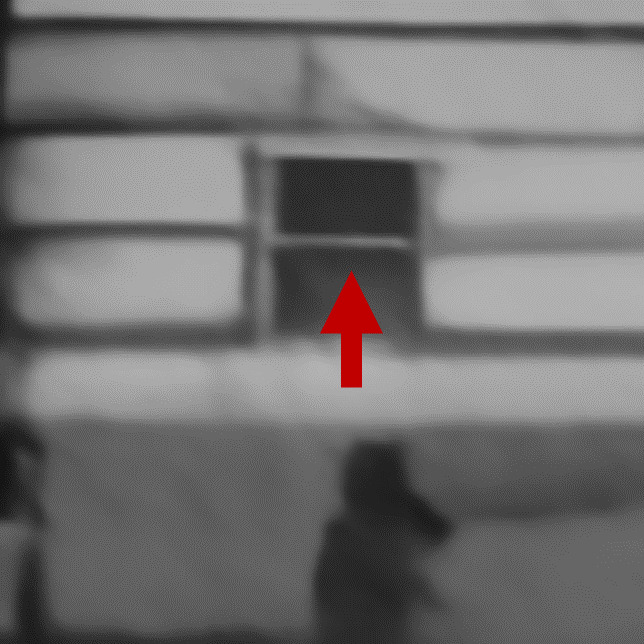}&
\hspace{-2.0ex}\includegraphics[width=0.09\linewidth]{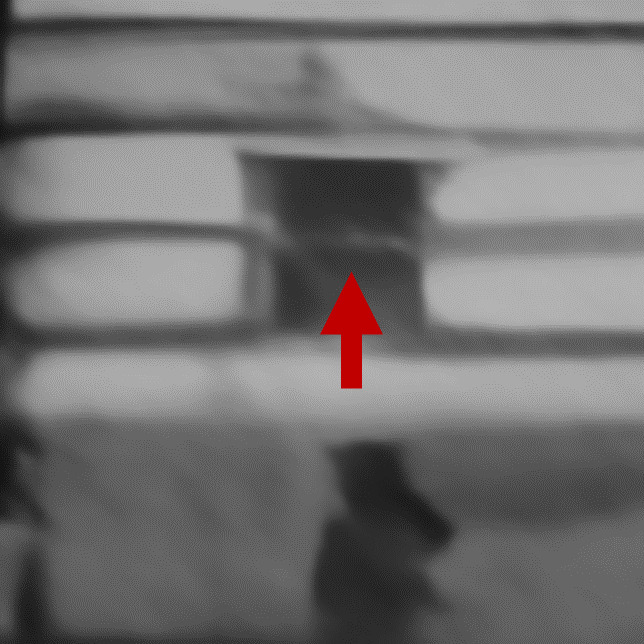}&
\hspace{-2.0ex}\includegraphics[width=0.09\linewidth]{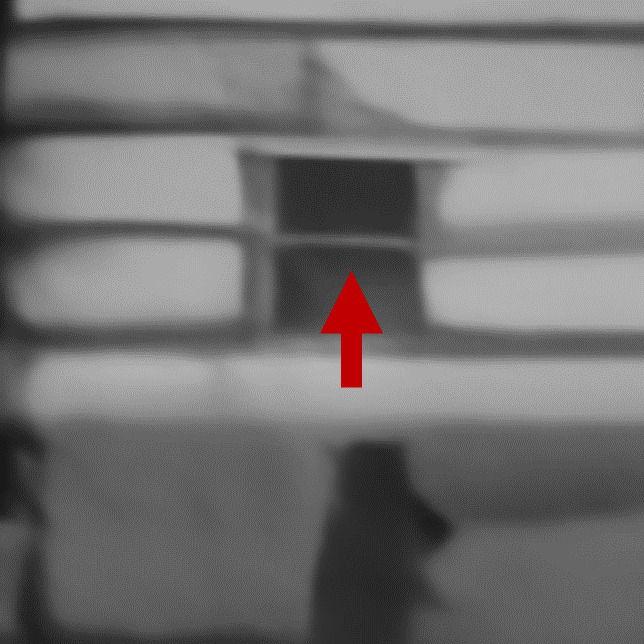}\\
&
\hspace{-2.0ex}\includegraphics[width=0.09\linewidth]{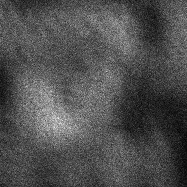}&
\hspace{-2.0ex}\includegraphics[width=0.09\linewidth]{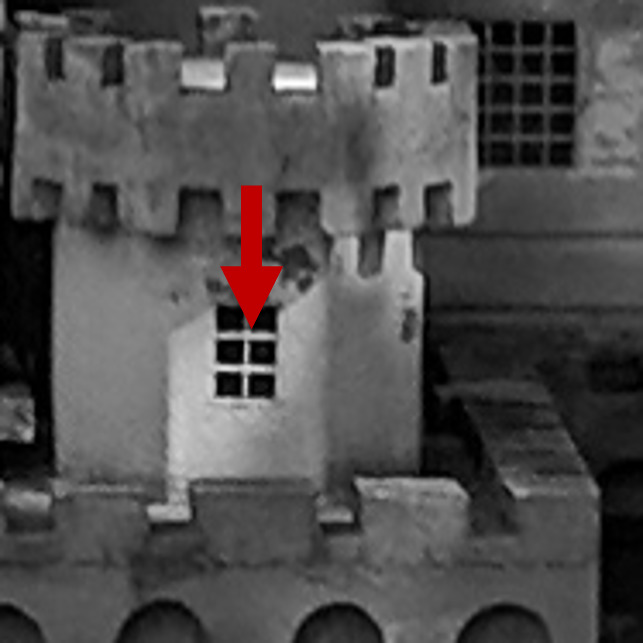}&
\hspace{-2.0ex}\includegraphics[width=0.09\linewidth]{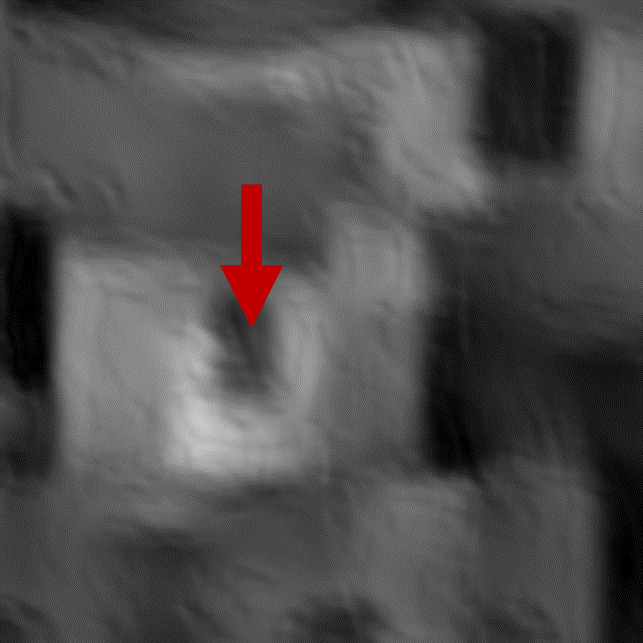}&
\hspace{-2.0ex}\includegraphics[width=0.09\linewidth]{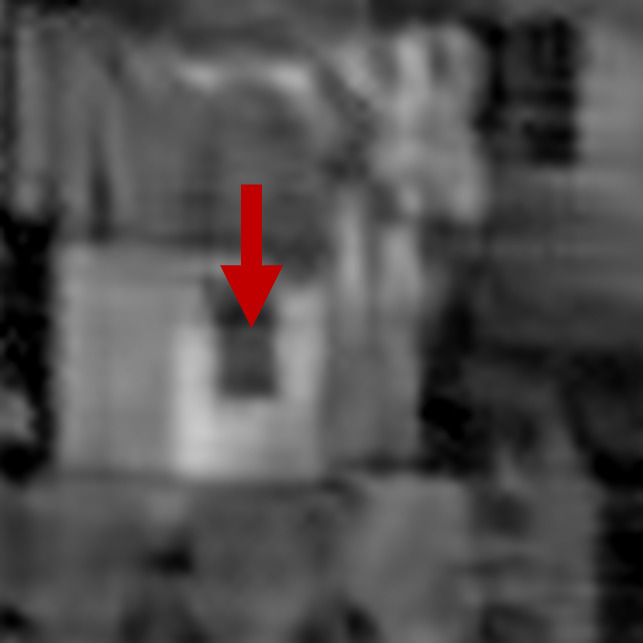}&
\hspace{-2.0ex}\includegraphics[width=0.09\linewidth]{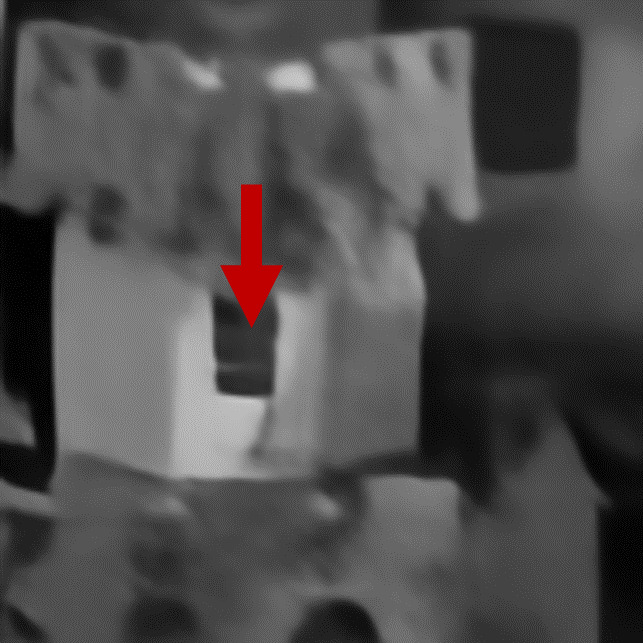}&
\hspace{-2.0ex}\includegraphics[width=0.09\linewidth]{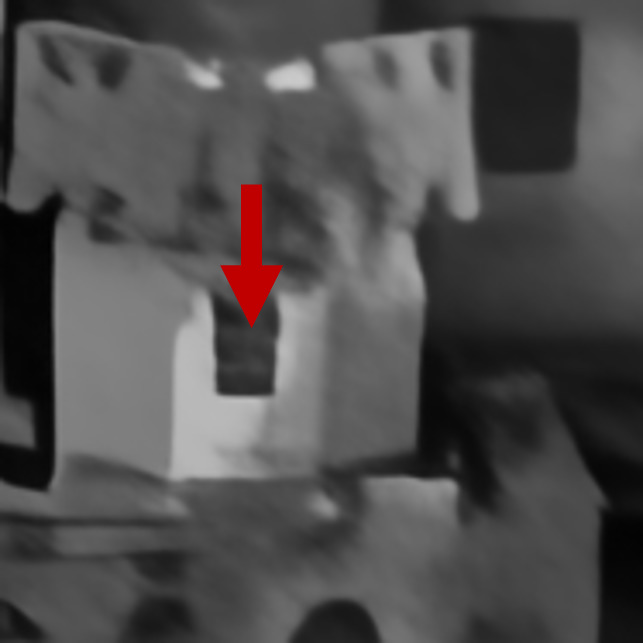}&
\hspace{-2.0ex}\includegraphics[width=0.09\linewidth]{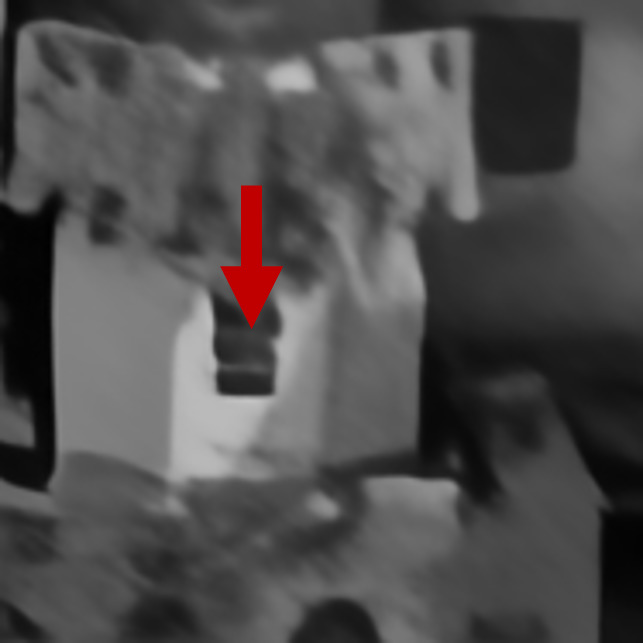}&
\hspace{-2.0ex}\includegraphics[width=0.09\linewidth]{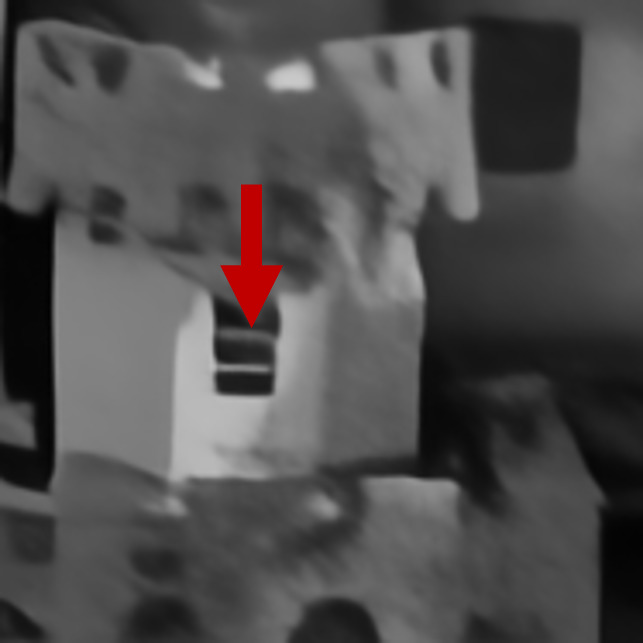}\\
\multirow{2}{*}[1.3cm]{\hspace{-2.0ex}\includegraphics[width=0.18\linewidth]{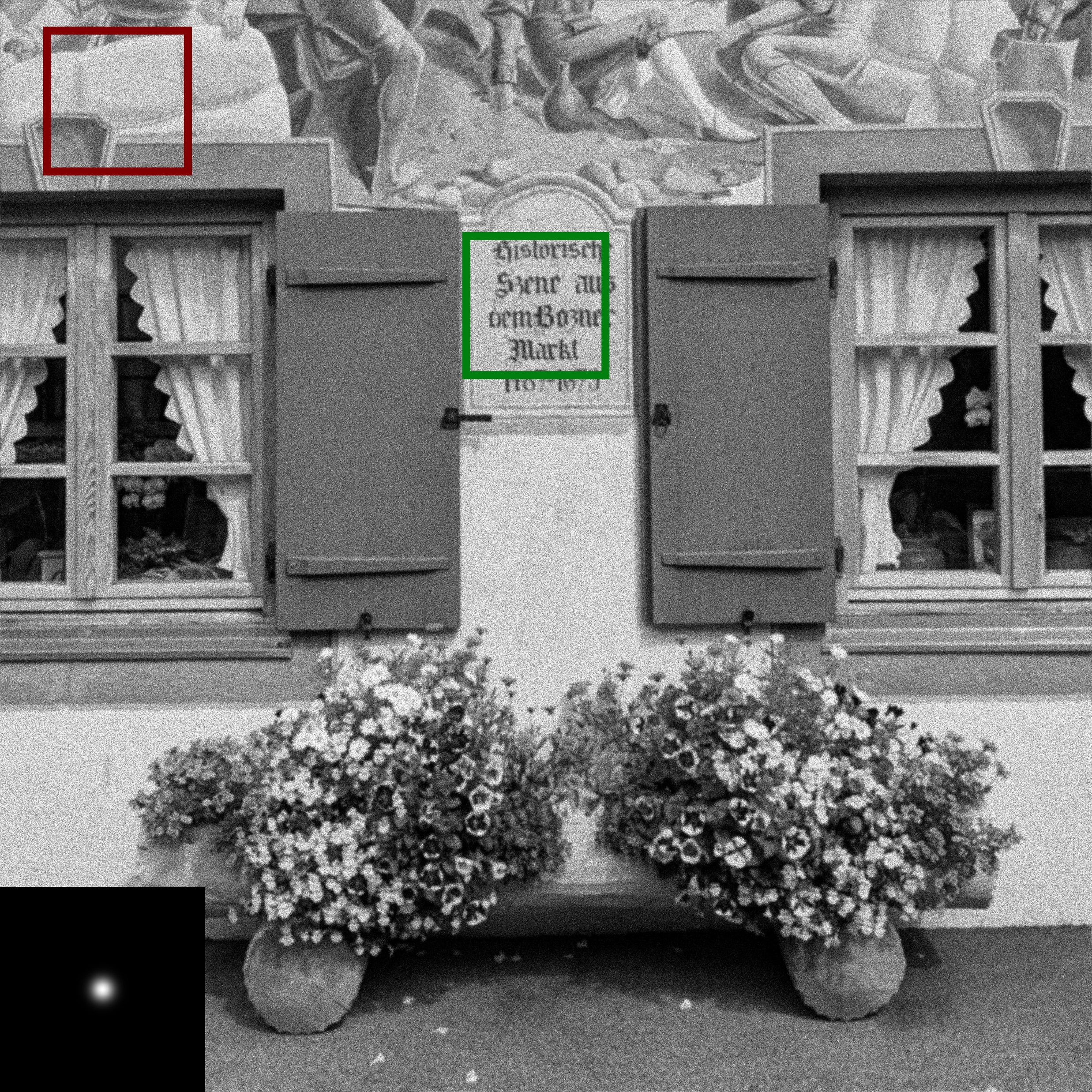}}&
\hspace{-2.0ex}\includegraphics[width=0.09\linewidth]{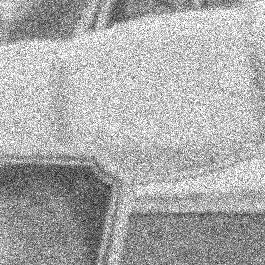}&
\hspace{-2.0ex}\includegraphics[width=0.09\linewidth]{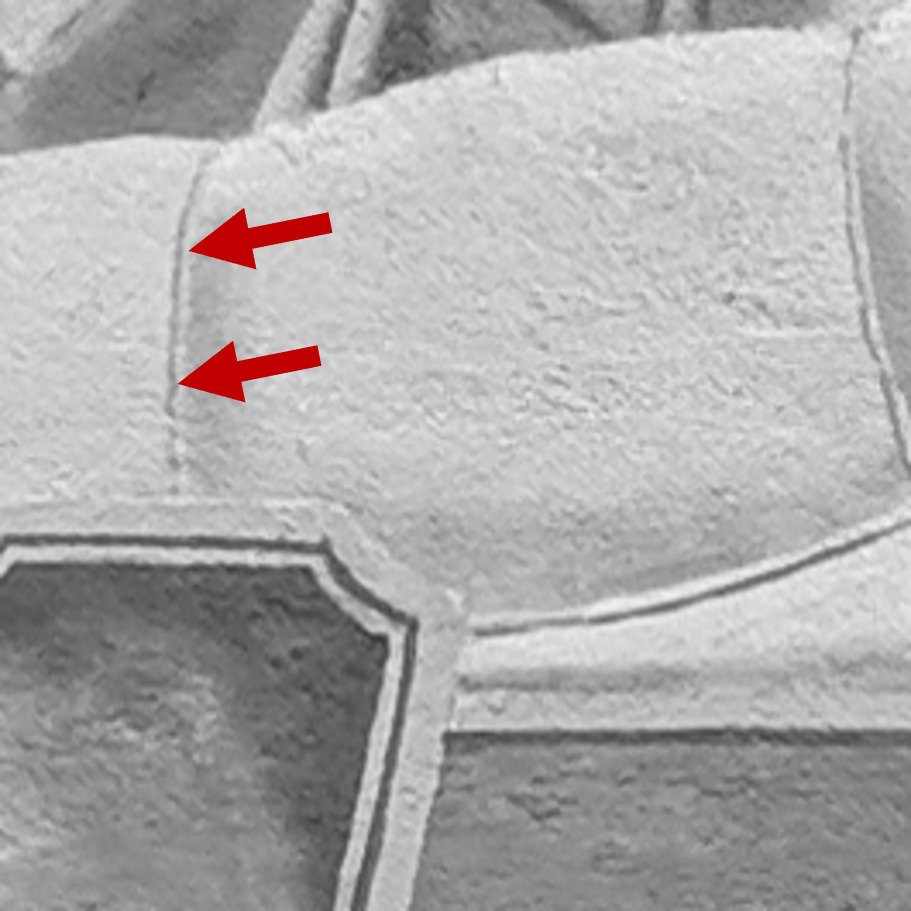}&
\hspace{-2.0ex}\includegraphics[width=0.09\linewidth]{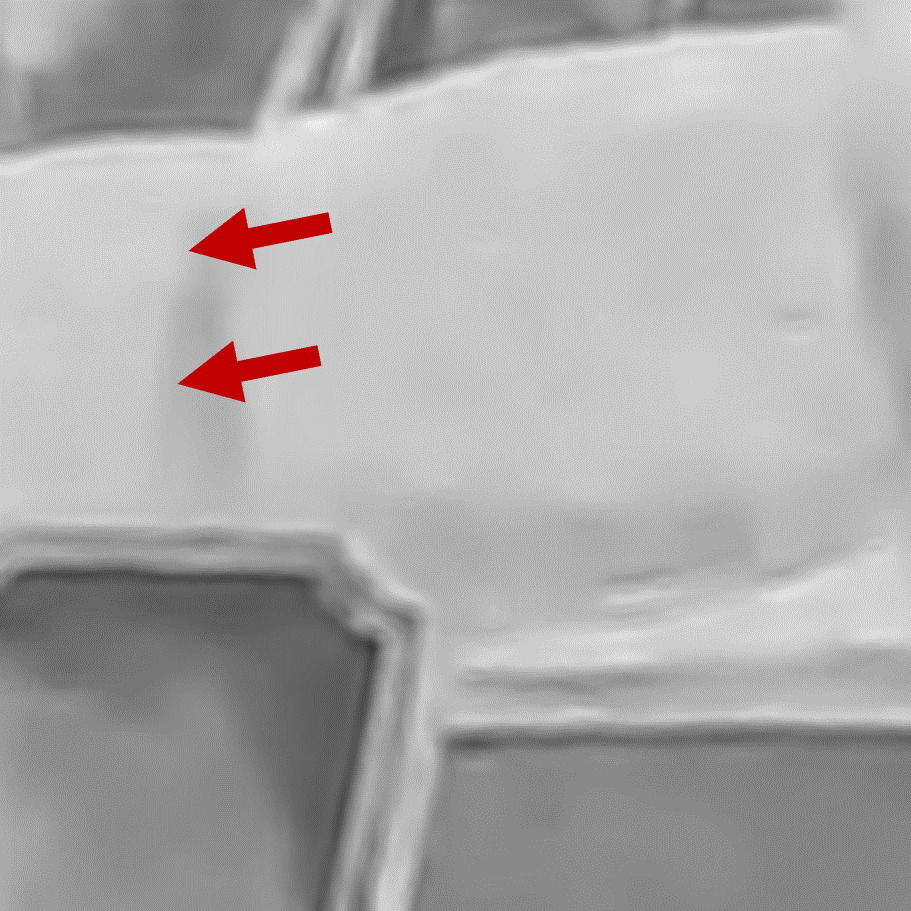}&
\hspace{-2.0ex}\includegraphics[width=0.09\linewidth]{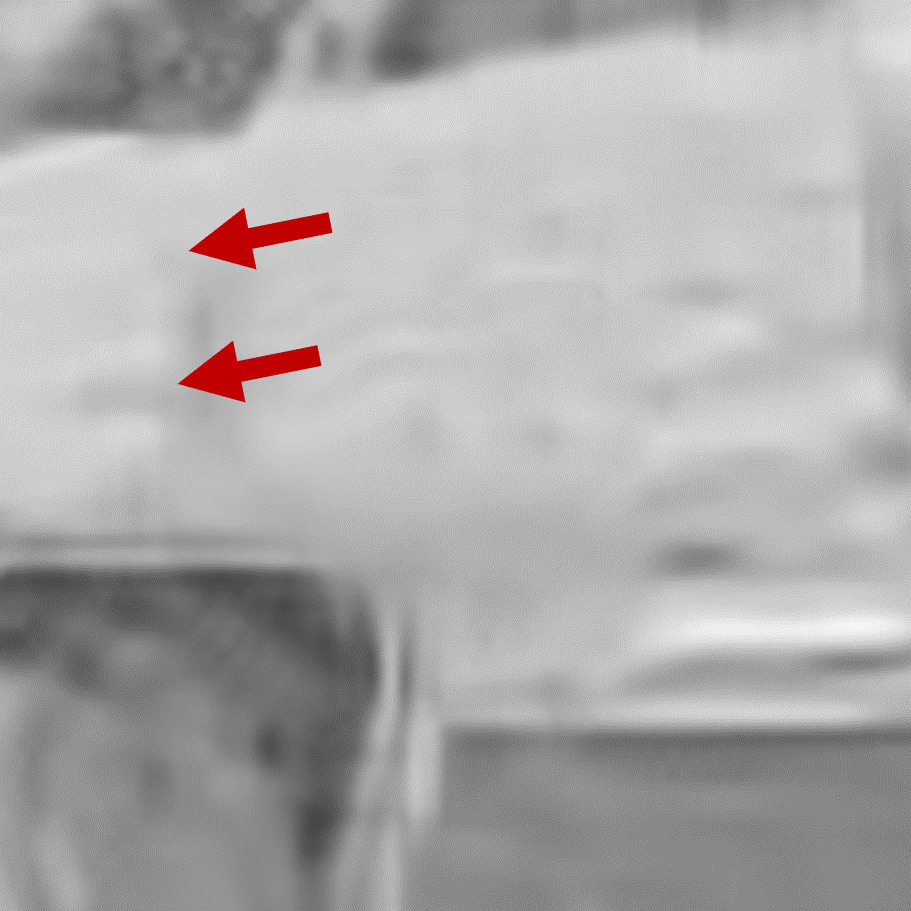}&
\hspace{-2.0ex}\includegraphics[width=0.09\linewidth]{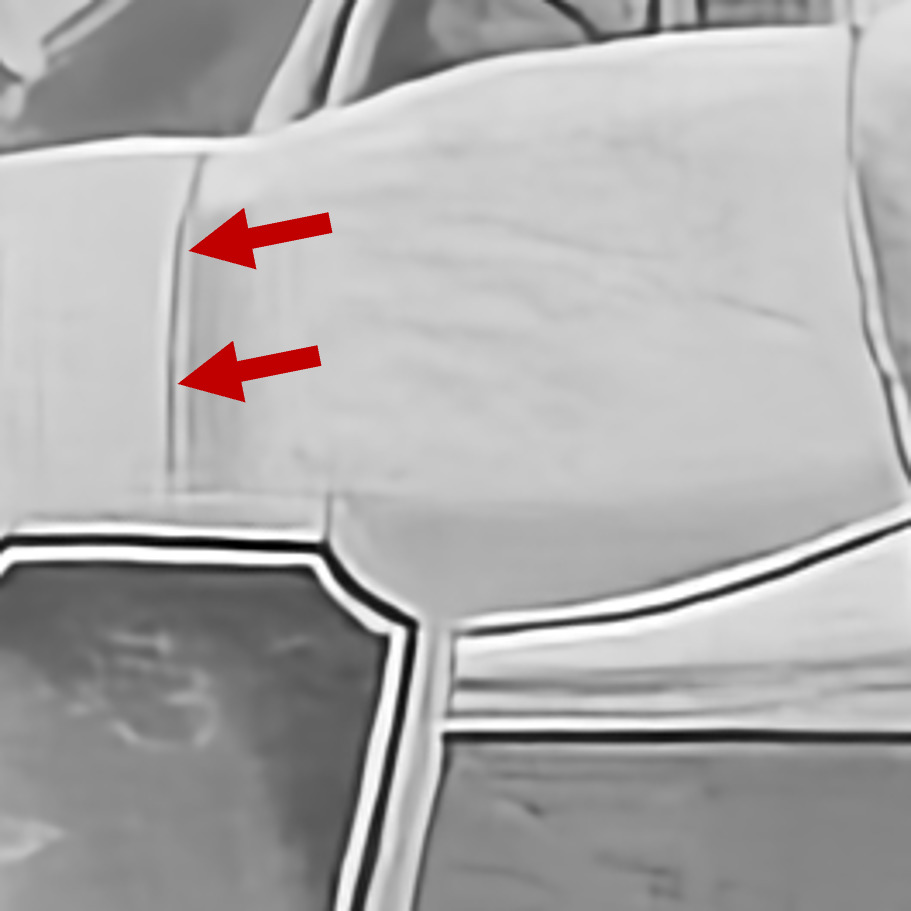}&
\hspace{-2.0ex}\includegraphics[width=0.09\linewidth]{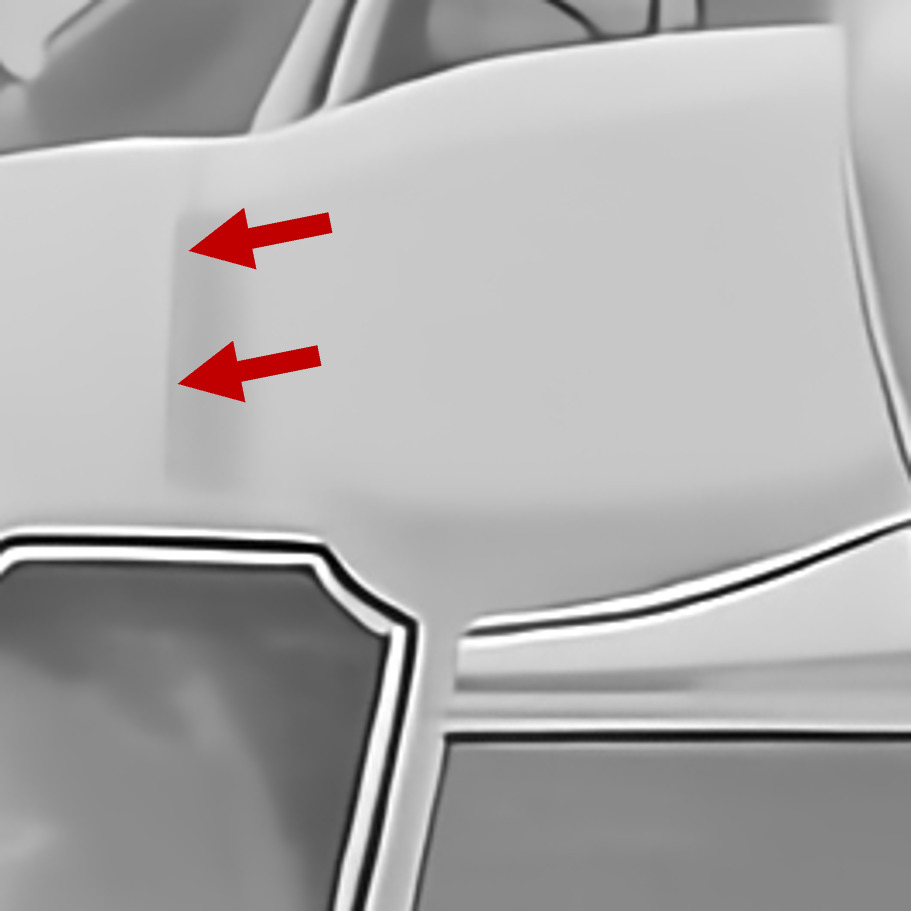}&
\hspace{-2.0ex}\includegraphics[width=0.09\linewidth]{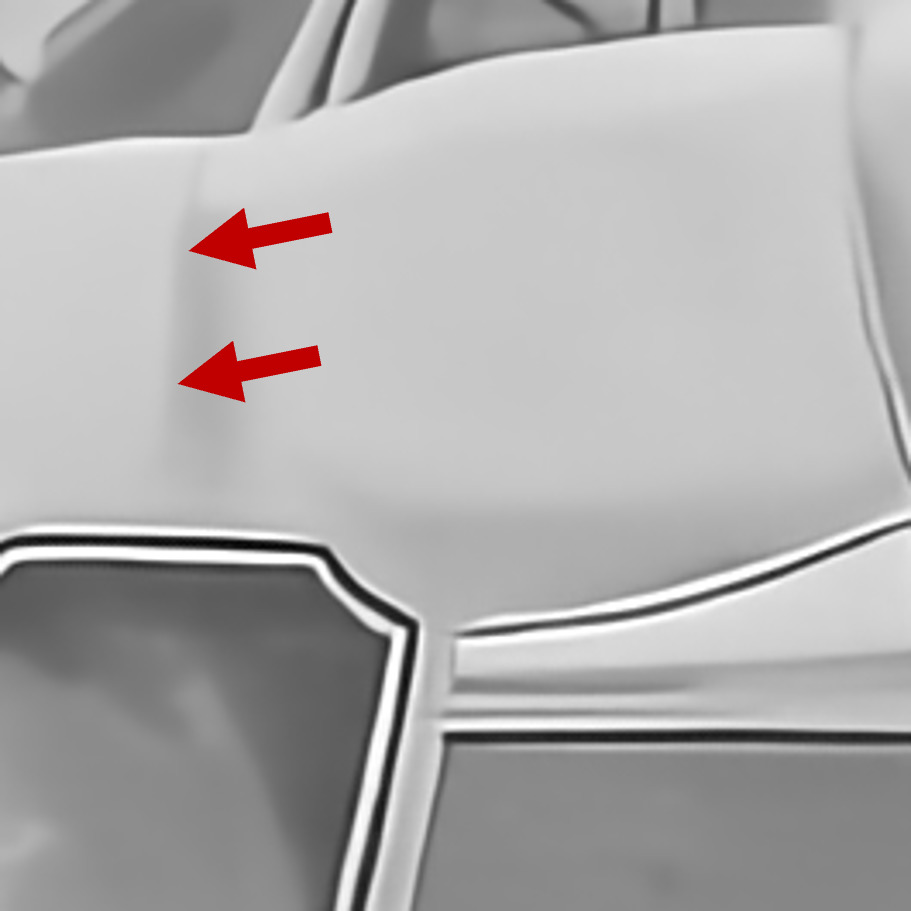}&
\hspace{-2.0ex}\includegraphics[width=0.09\linewidth]{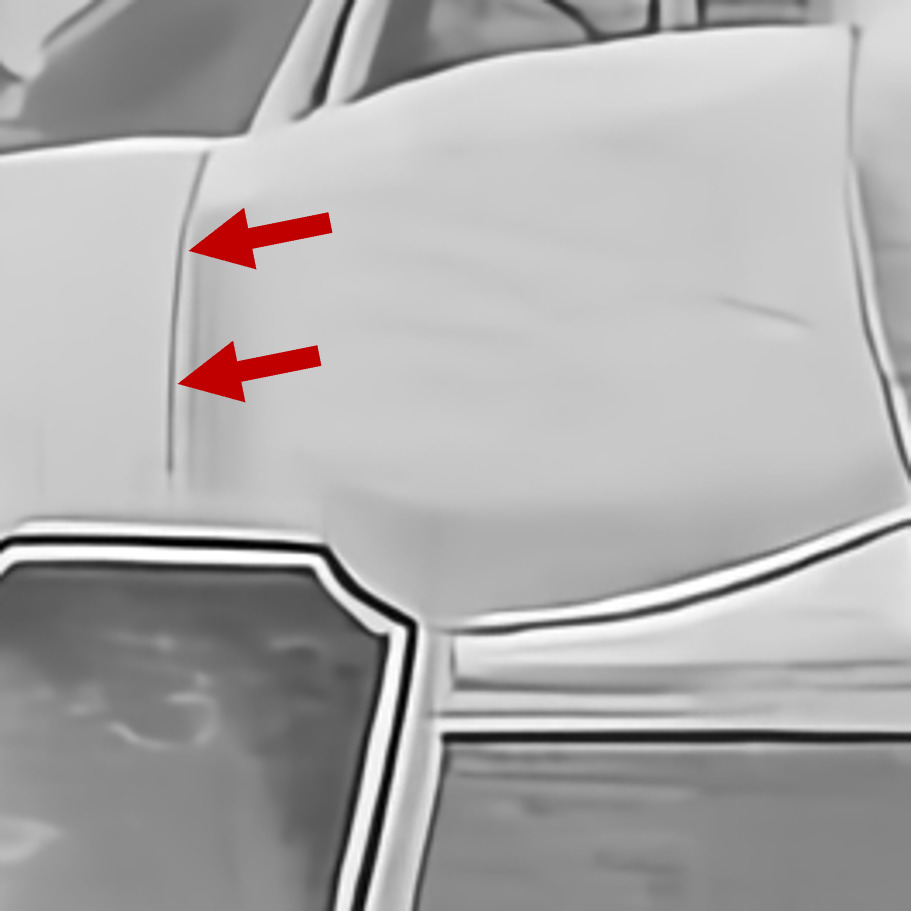}\\
&
\hspace{-2.0ex}\includegraphics[width=0.09\linewidth]{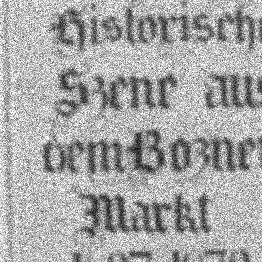}&
\hspace{-2.0ex}\includegraphics[width=0.09\linewidth]{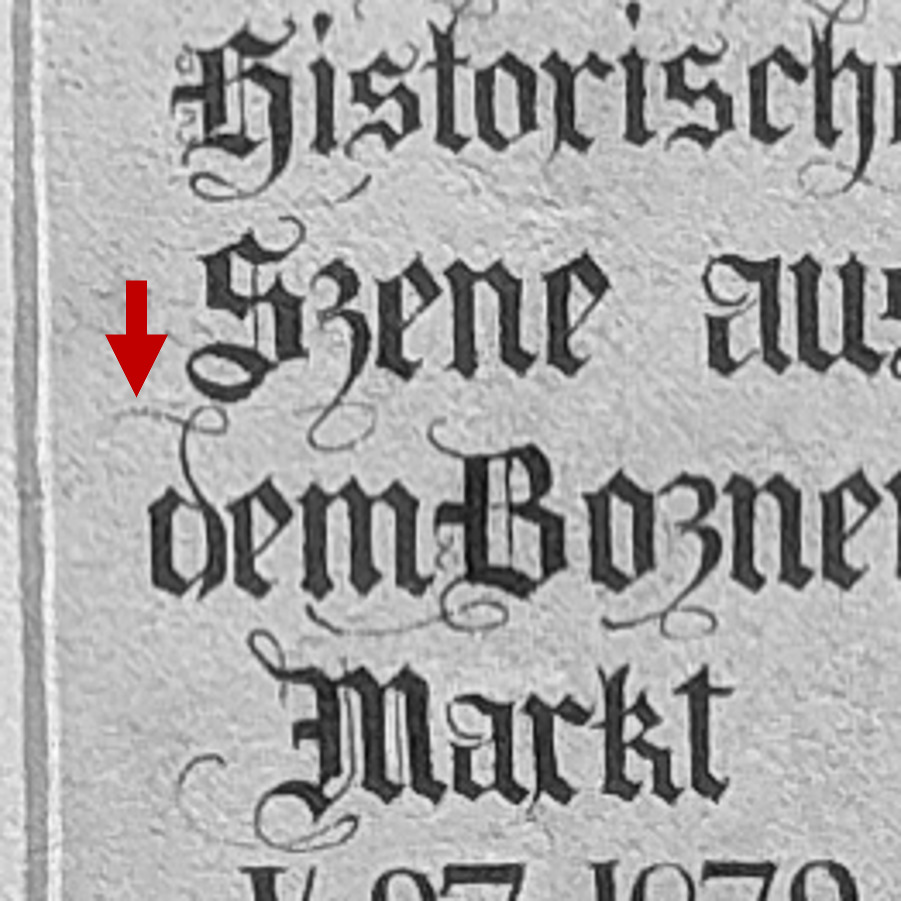}&
\hspace{-2.0ex}\includegraphics[width=0.09\linewidth]{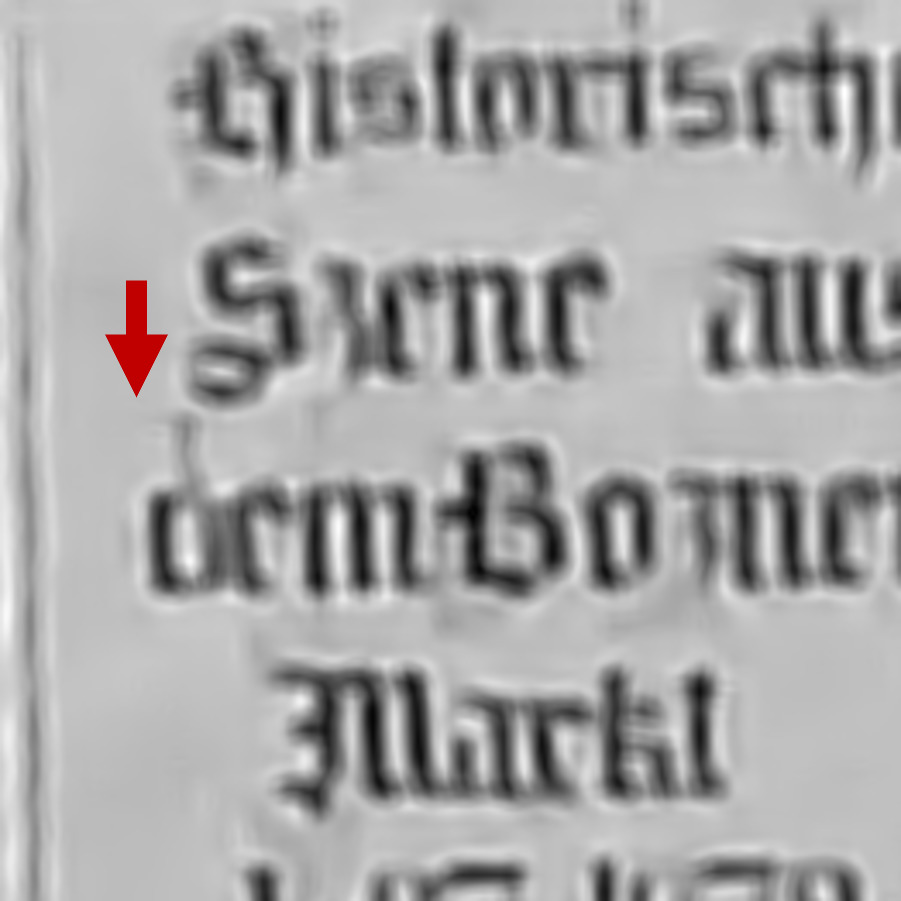}&
\hspace{-2.0ex}\includegraphics[width=0.09\linewidth]{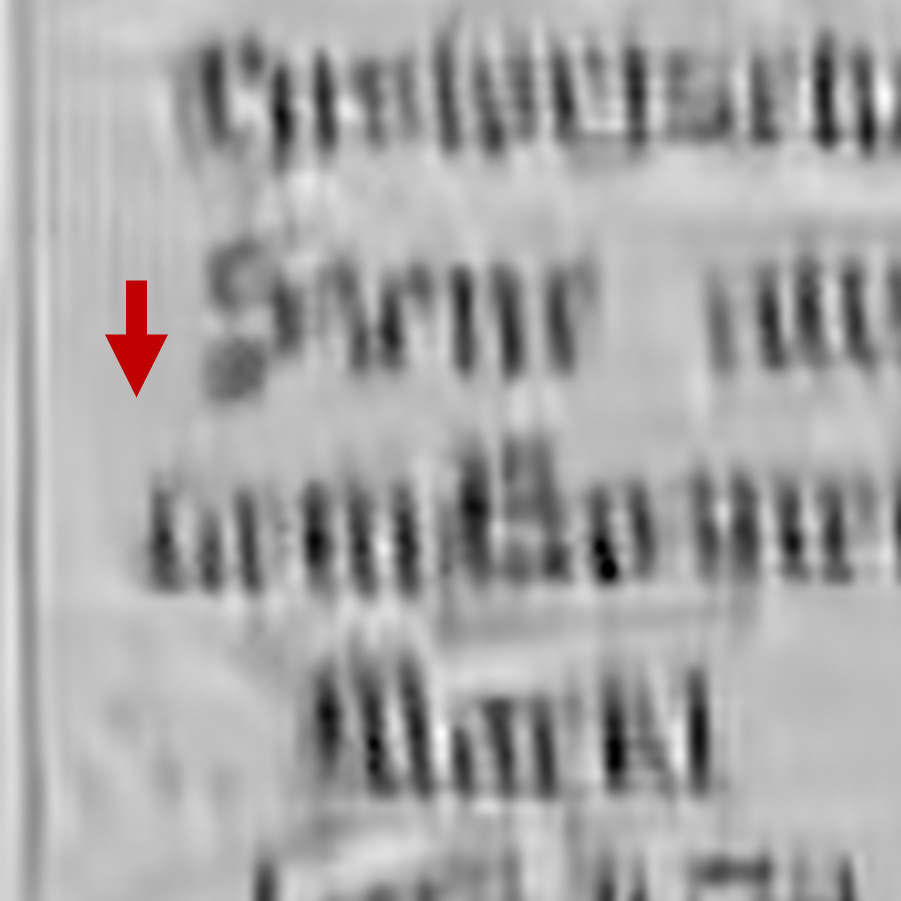}&
\hspace{-2.0ex}\includegraphics[width=0.09\linewidth]{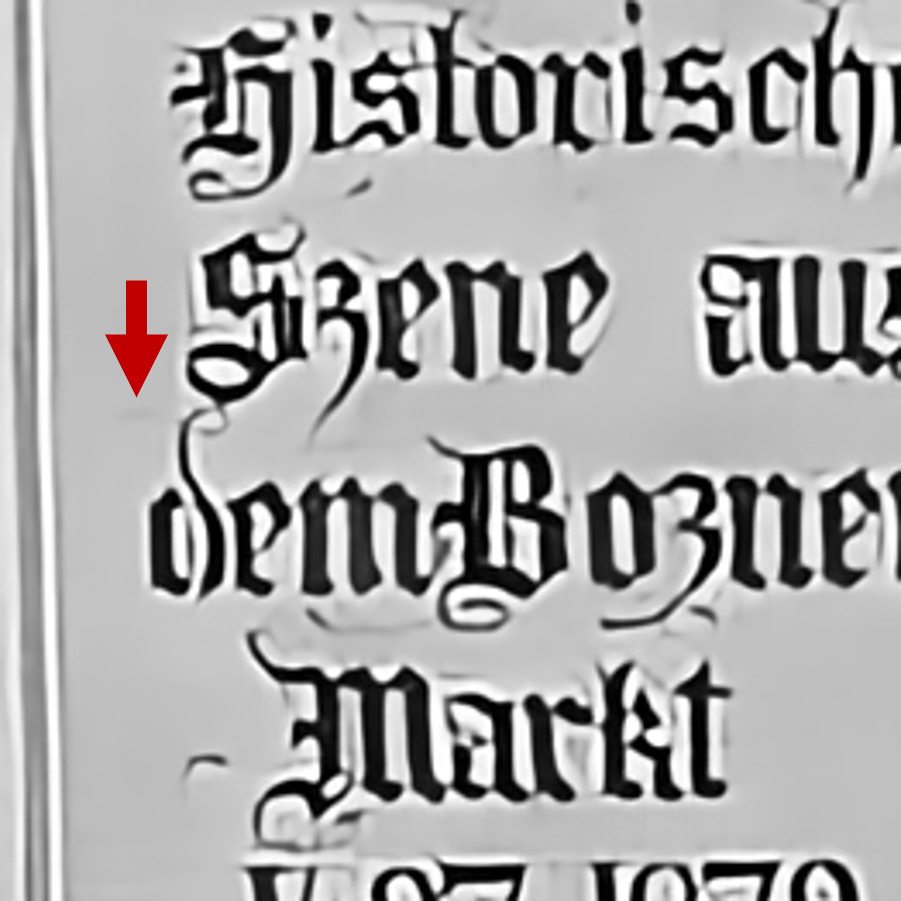}&
\hspace{-2.0ex}\includegraphics[width=0.09\linewidth]{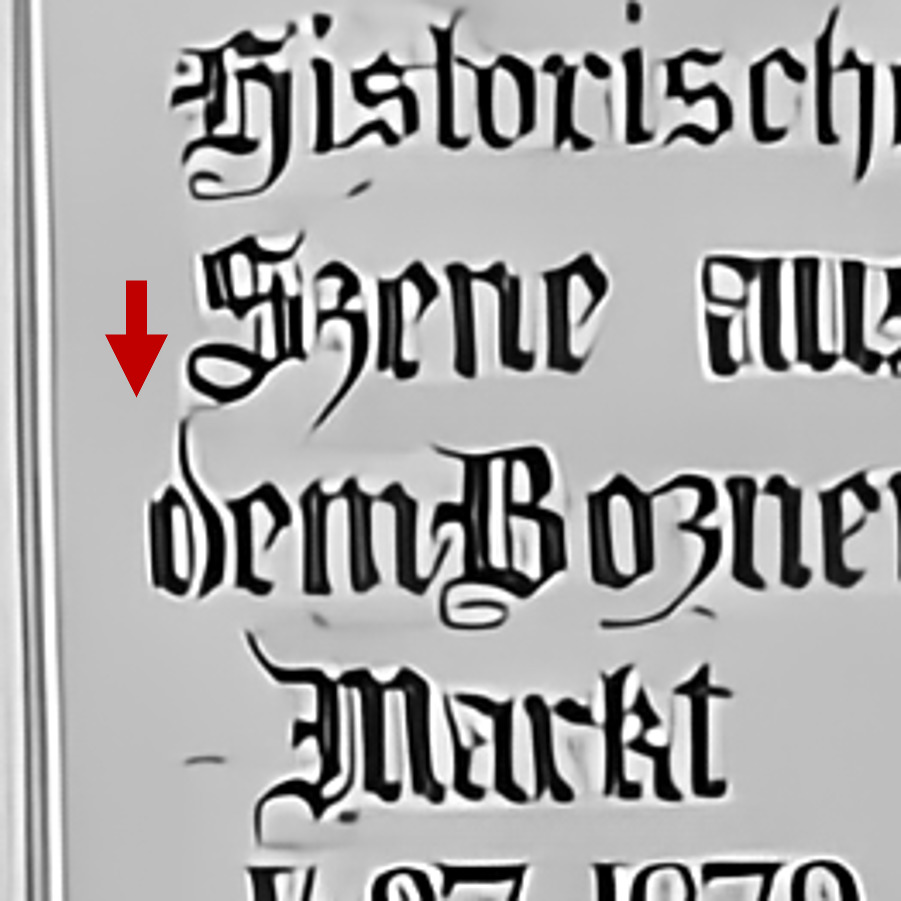}&
\hspace{-2.0ex}\includegraphics[width=0.09\linewidth]{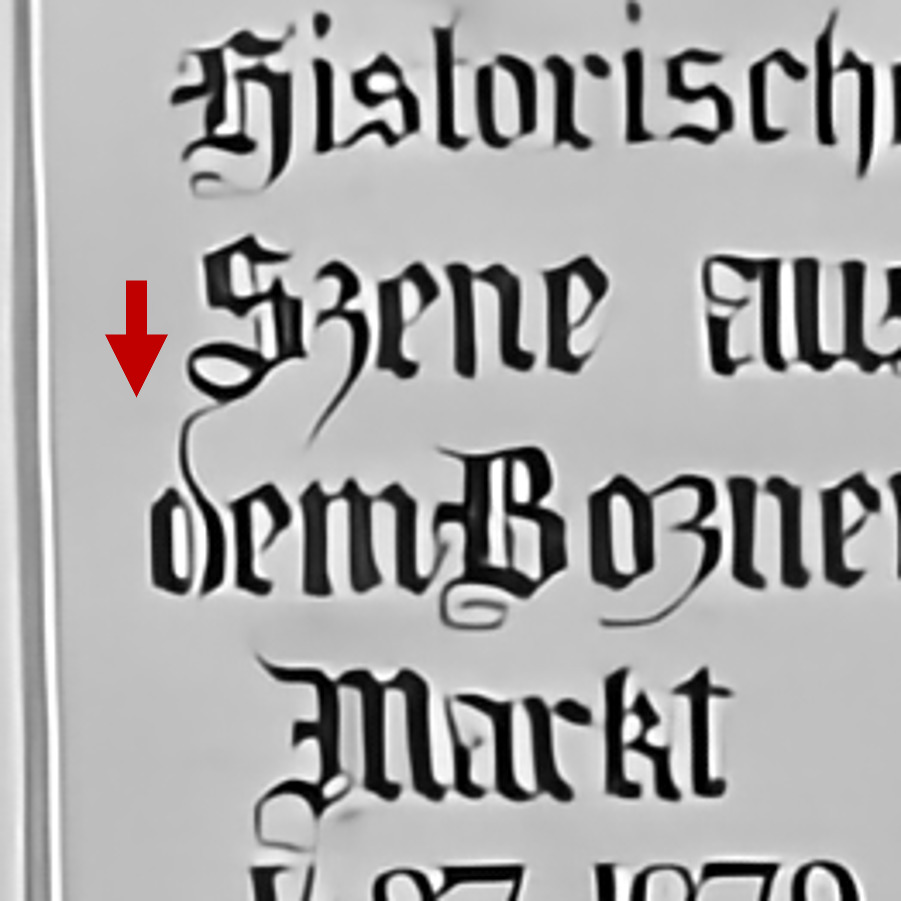}&
\hspace{-2.0ex}\includegraphics[width=0.09\linewidth]{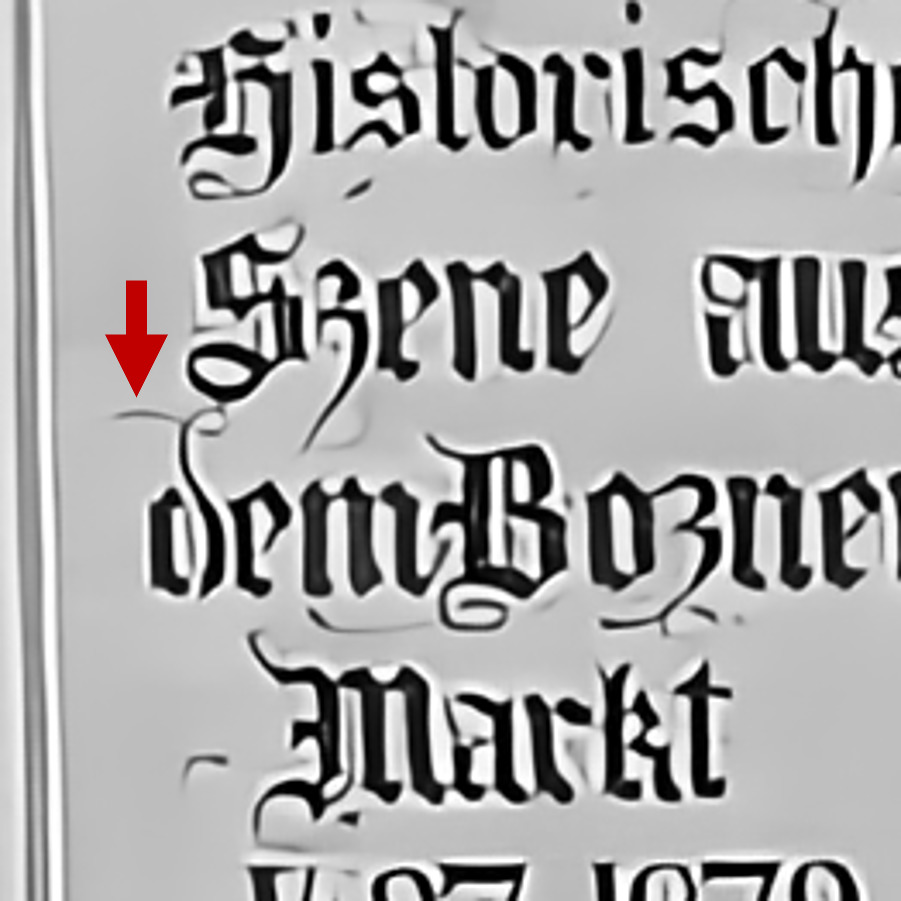}\\
\multirow{2}{*}[1.3cm]{\hspace{-2.0ex}\includegraphics[width=0.18\linewidth]{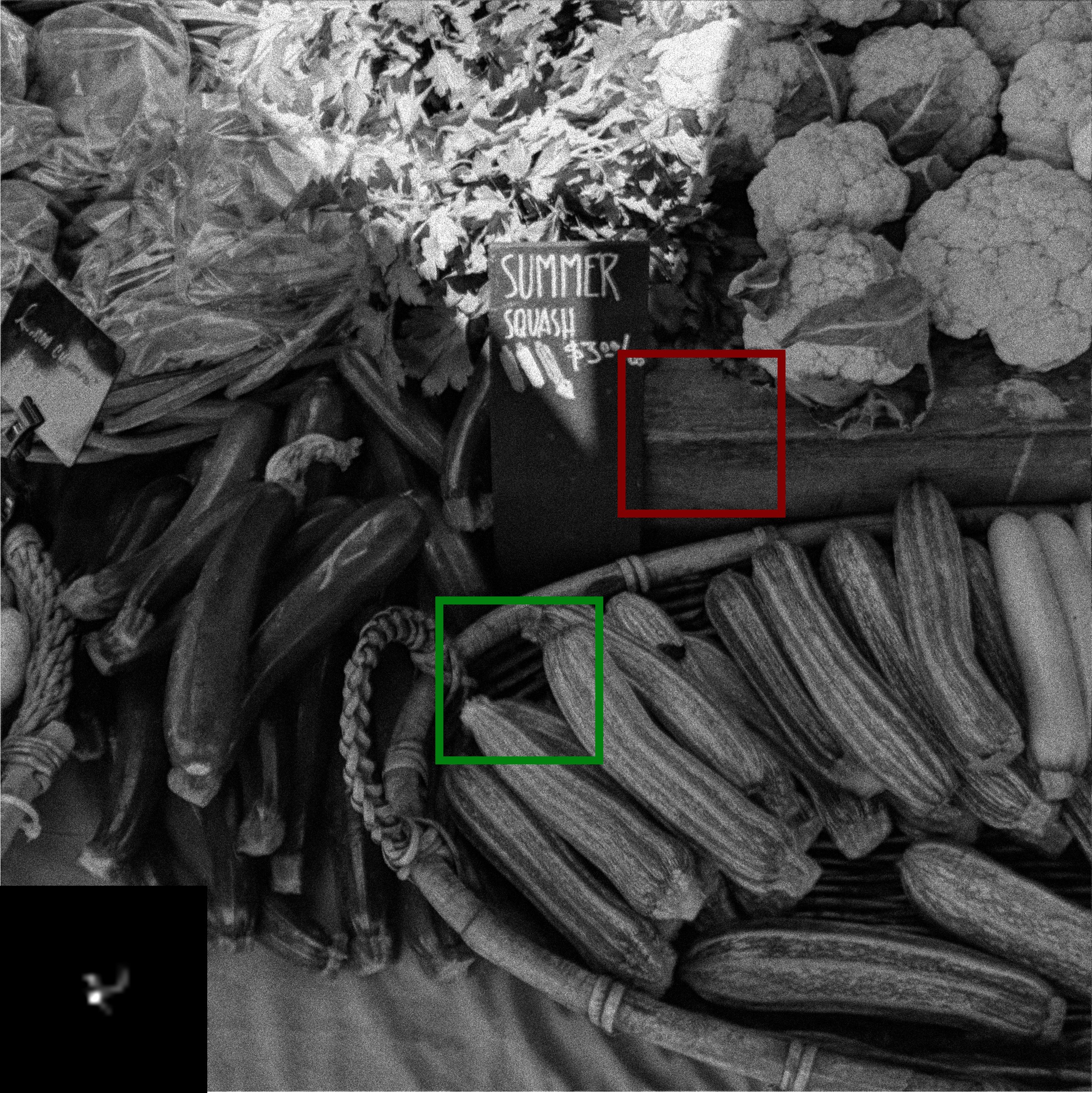}}&
\hspace{-2.0ex}\includegraphics[width=0.09\linewidth]{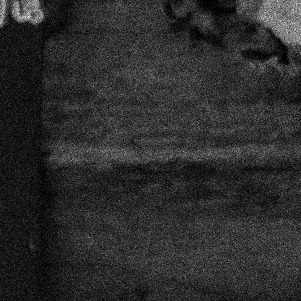}&
\hspace{-2.0ex}\includegraphics[width=0.09\linewidth]{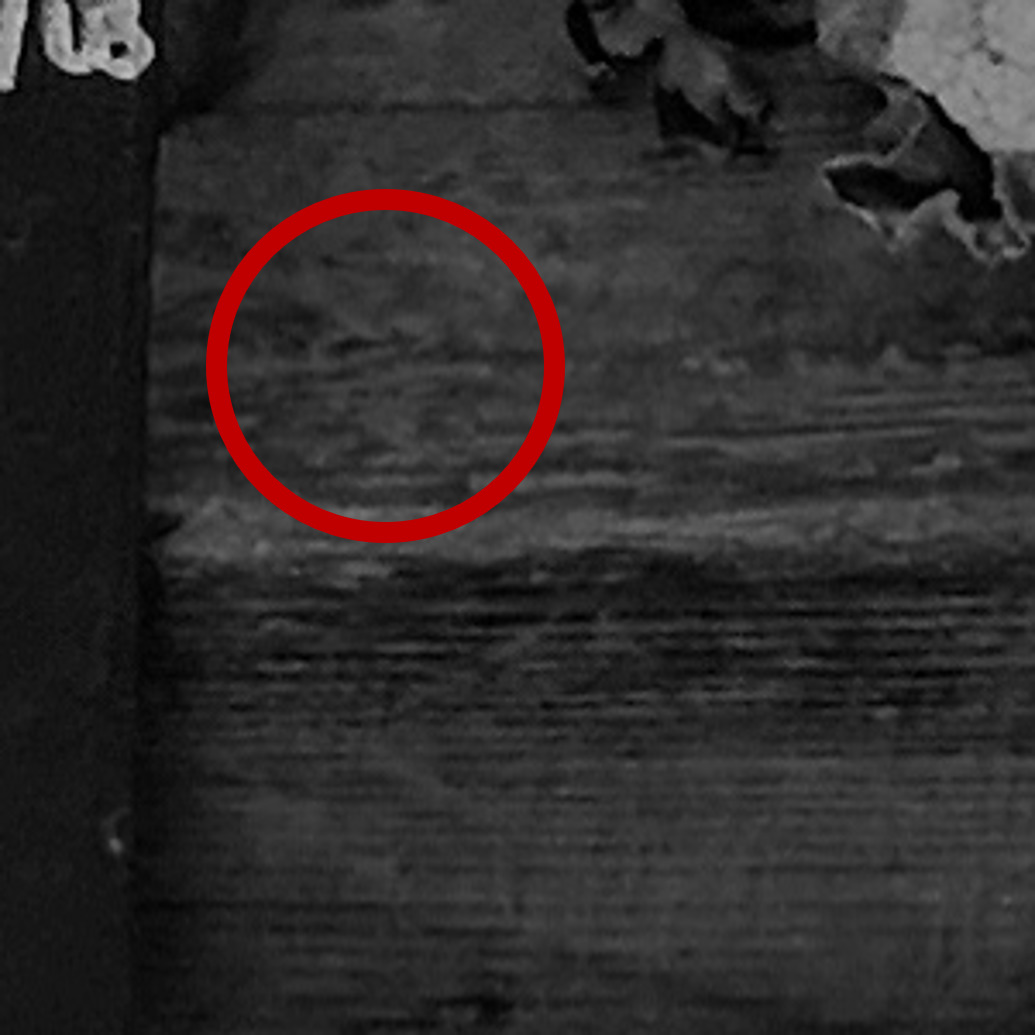}&
\hspace{-2.0ex}\includegraphics[width=0.09\linewidth]{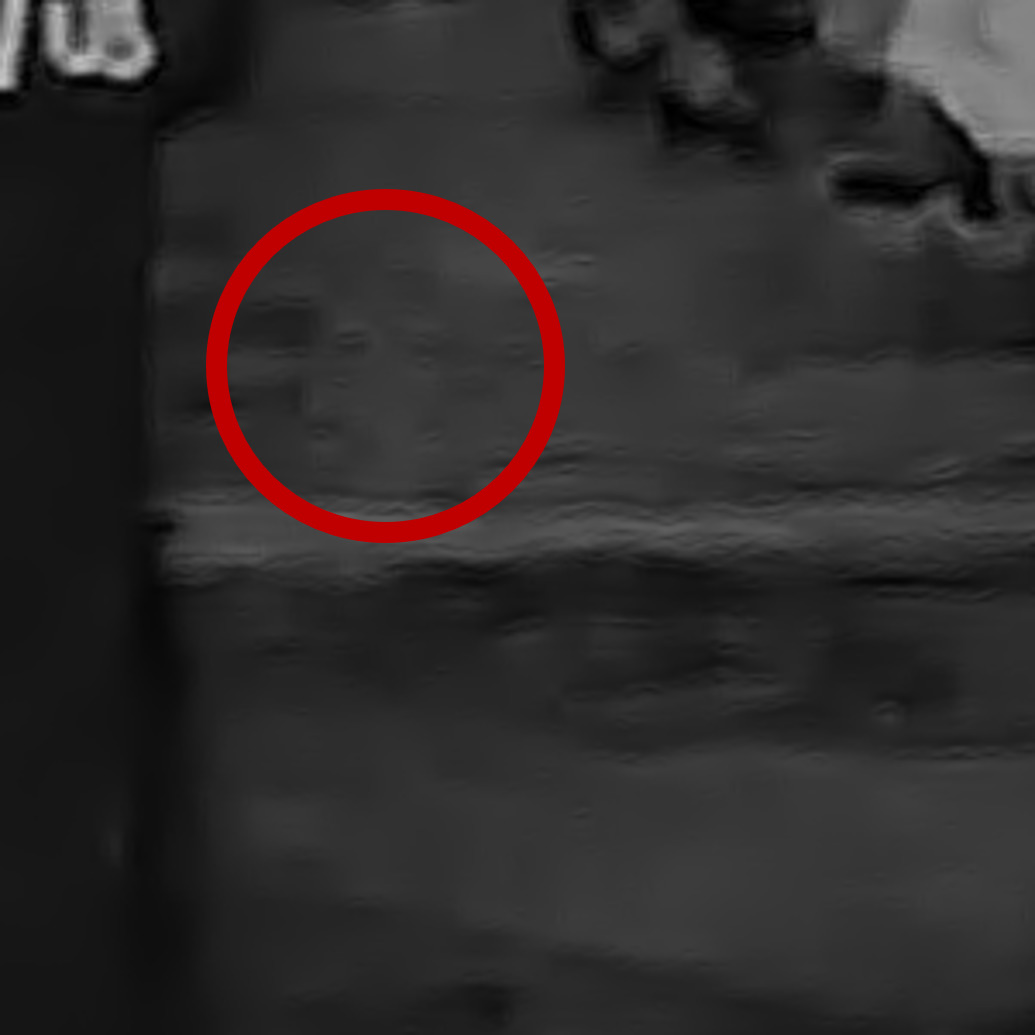}&
\hspace{-2.0ex}\includegraphics[width=0.09\linewidth]{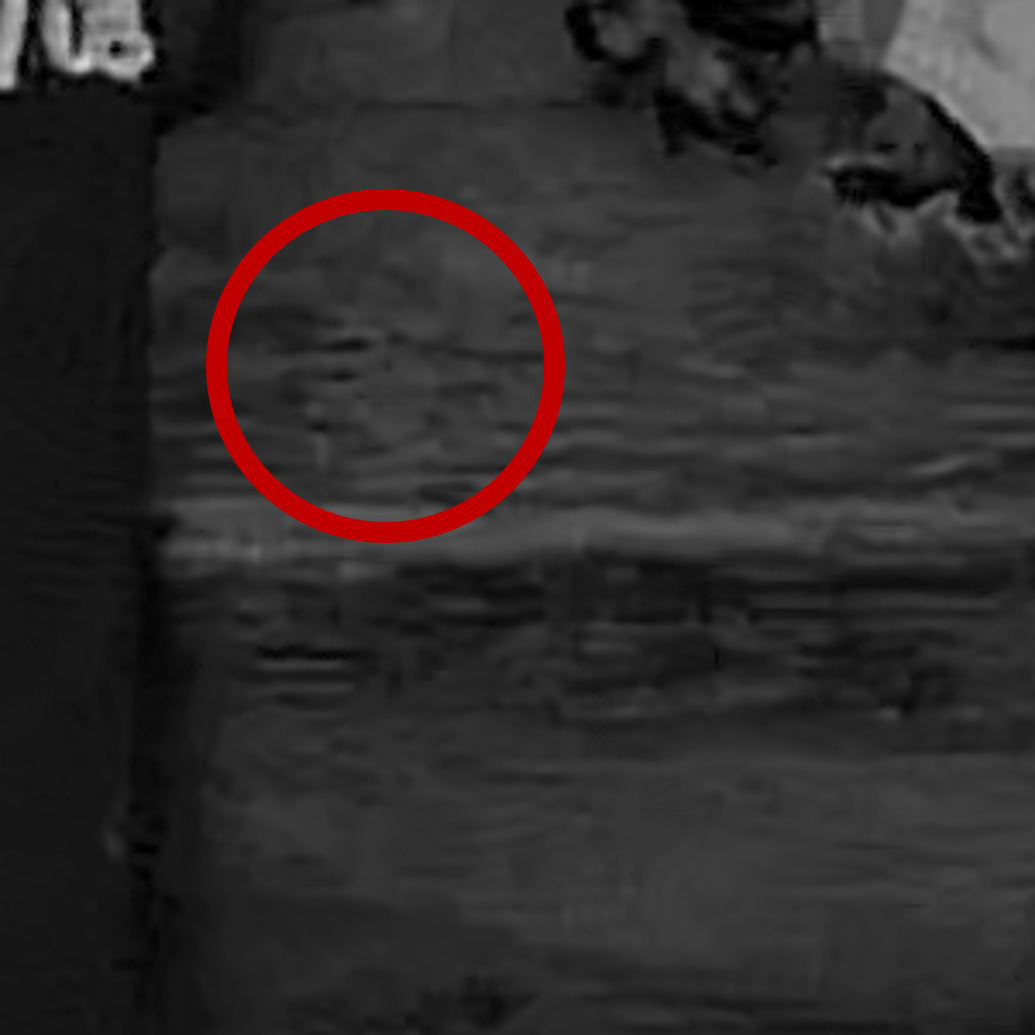}&
\hspace{-2.0ex}\includegraphics[width=0.09\linewidth]{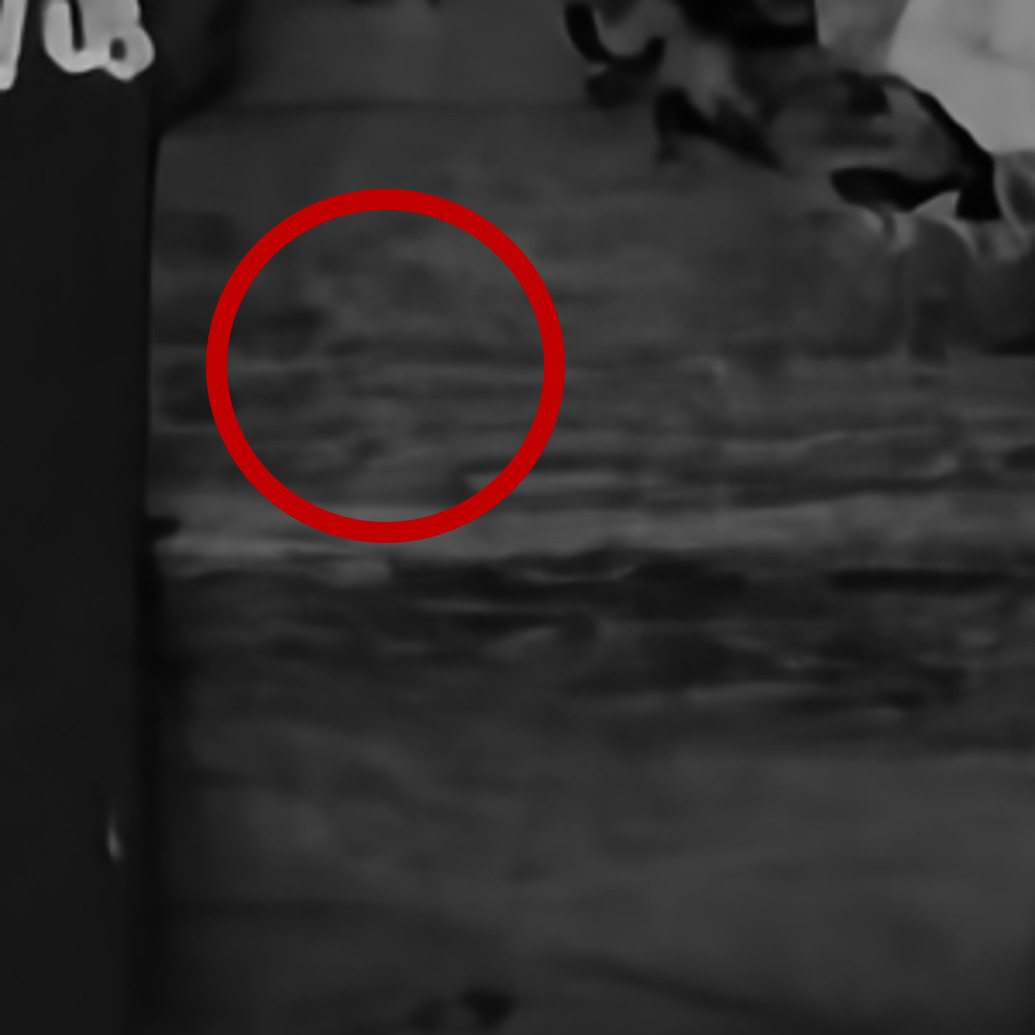}&
\hspace{-2.0ex}\includegraphics[width=0.09\linewidth]{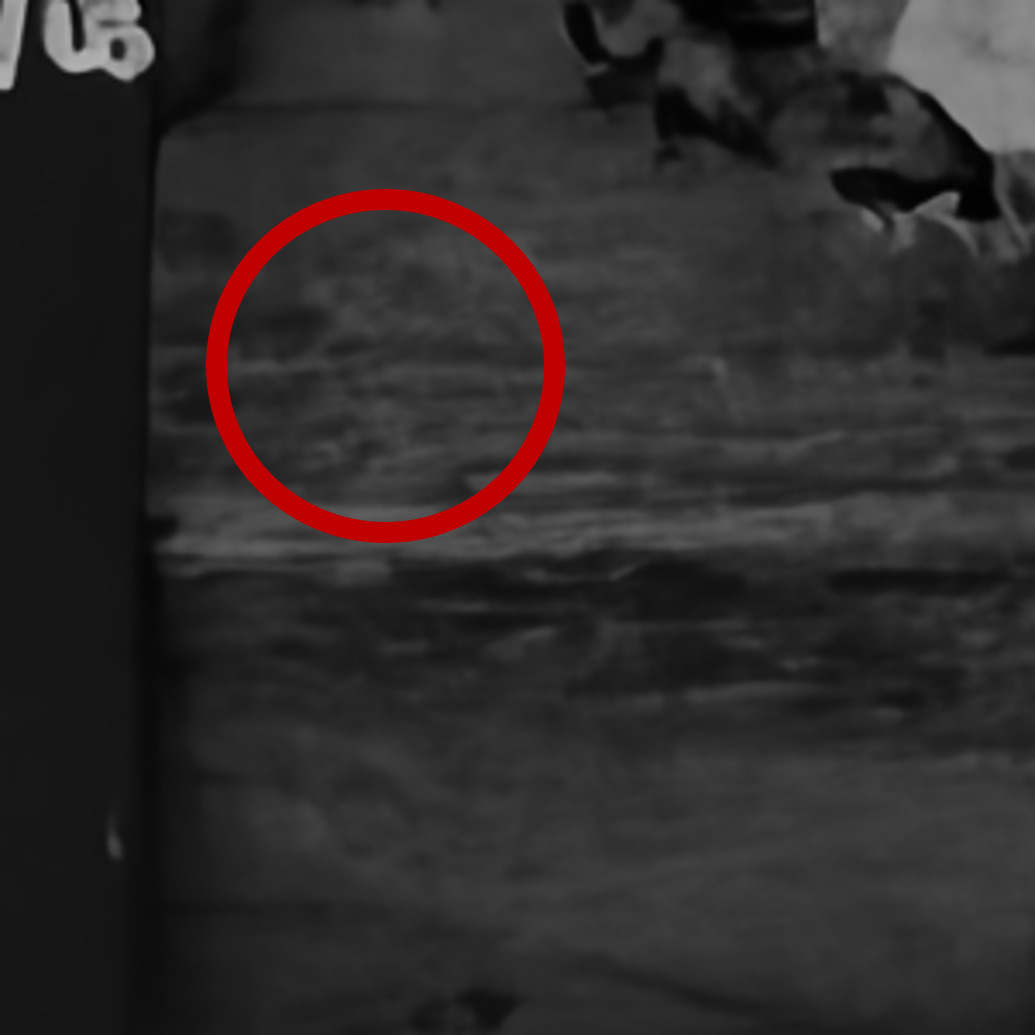}&
\hspace{-2.0ex}\includegraphics[width=0.09\linewidth]{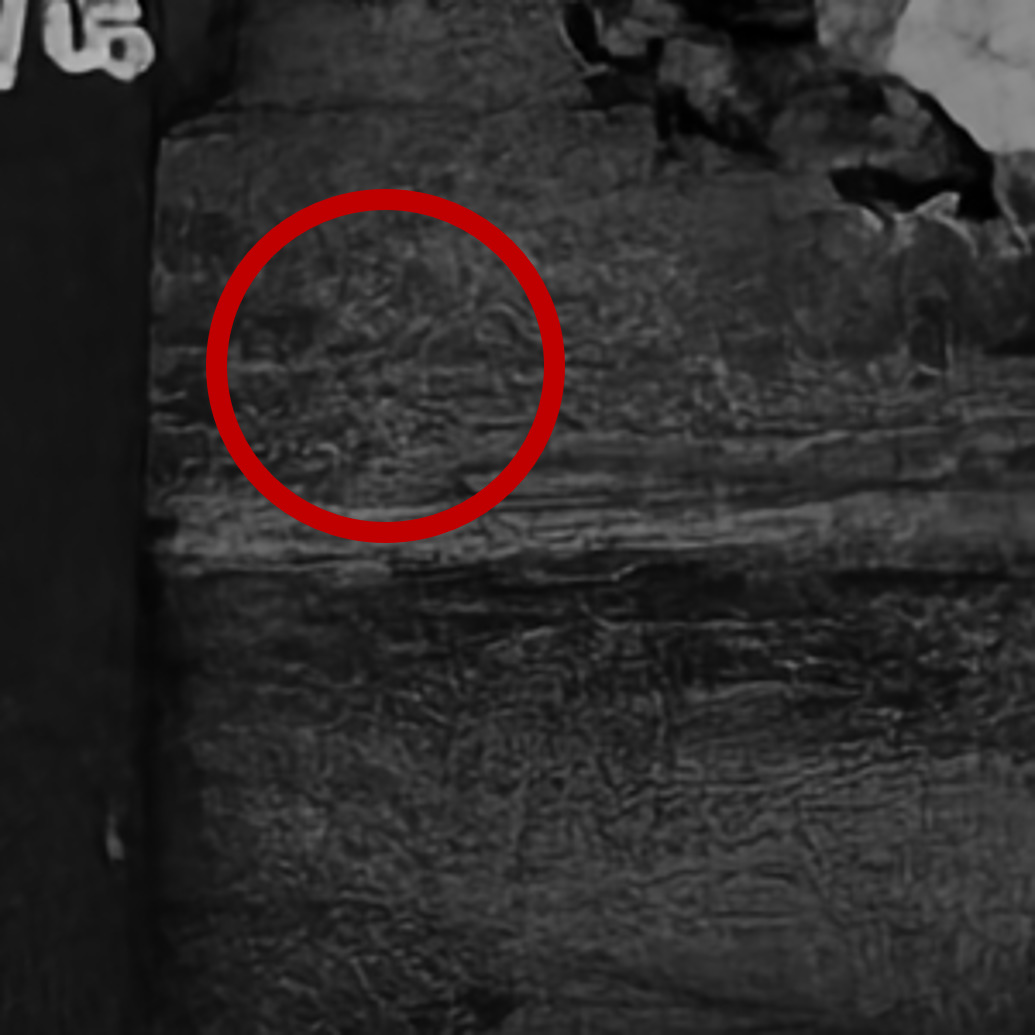}&
\hspace{-2.0ex}\includegraphics[width=0.09\linewidth]{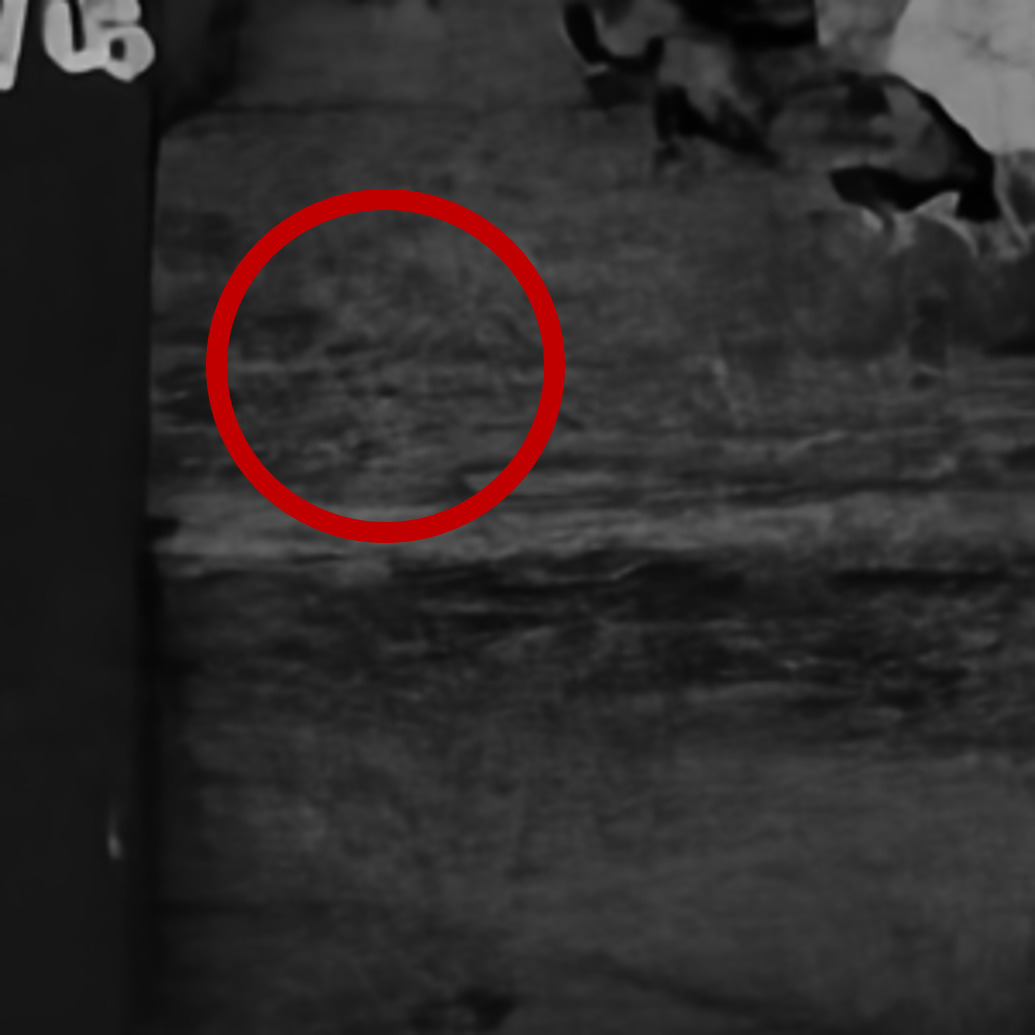}\\
&
\hspace{-2.0ex}\includegraphics[width=0.09\linewidth]{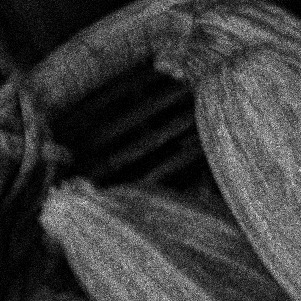}&
\hspace{-2.0ex}\includegraphics[width=0.09\linewidth]{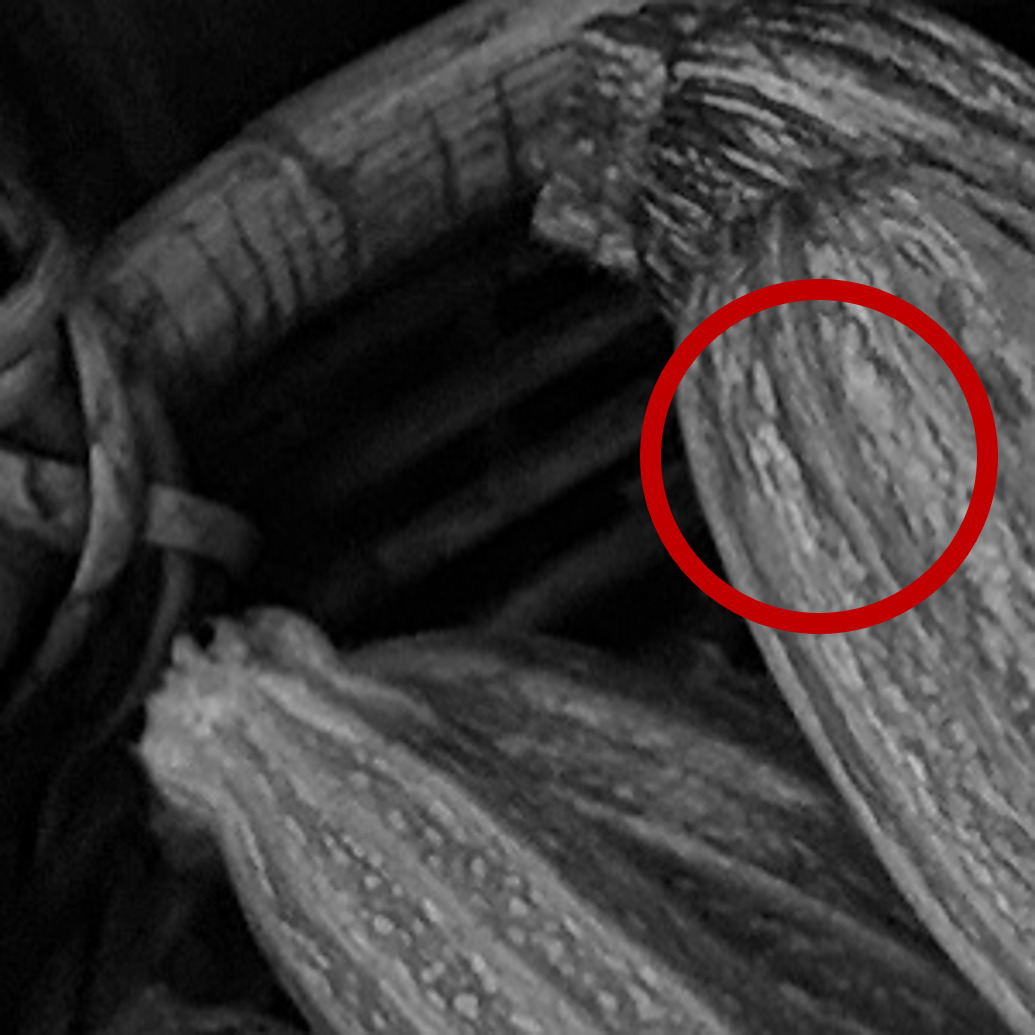}&
\hspace{-2.0ex}\includegraphics[width=0.09\linewidth]{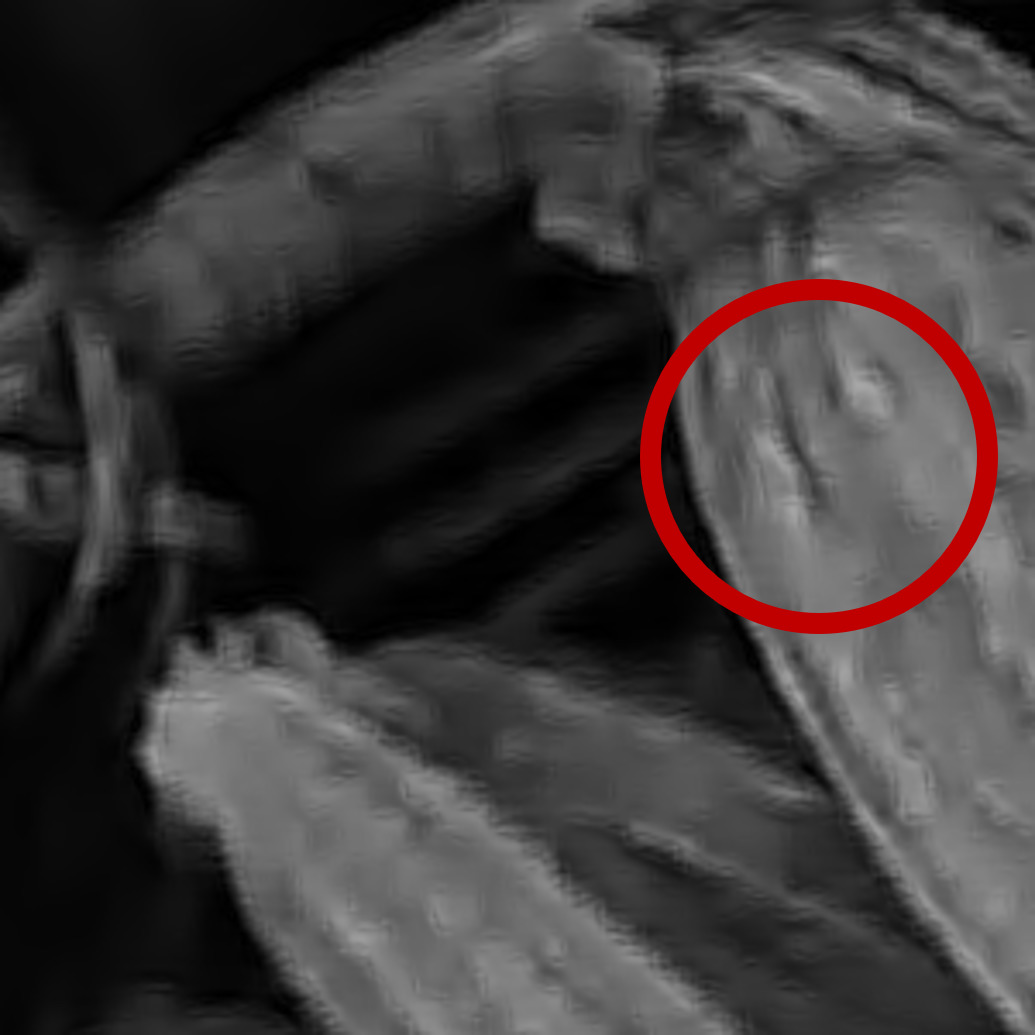}&
\hspace{-2.0ex}\includegraphics[width=0.09\linewidth]{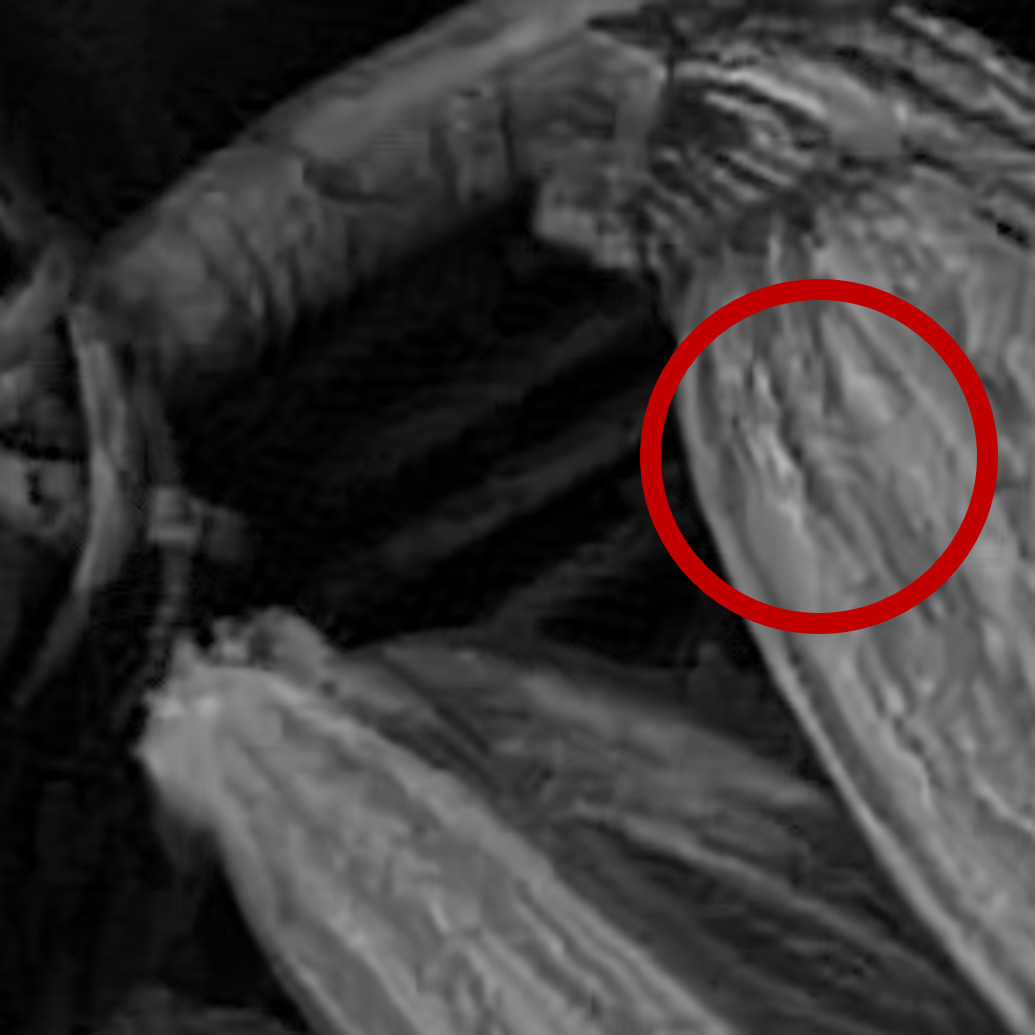}&
\hspace{-2.0ex}\includegraphics[width=0.09\linewidth]{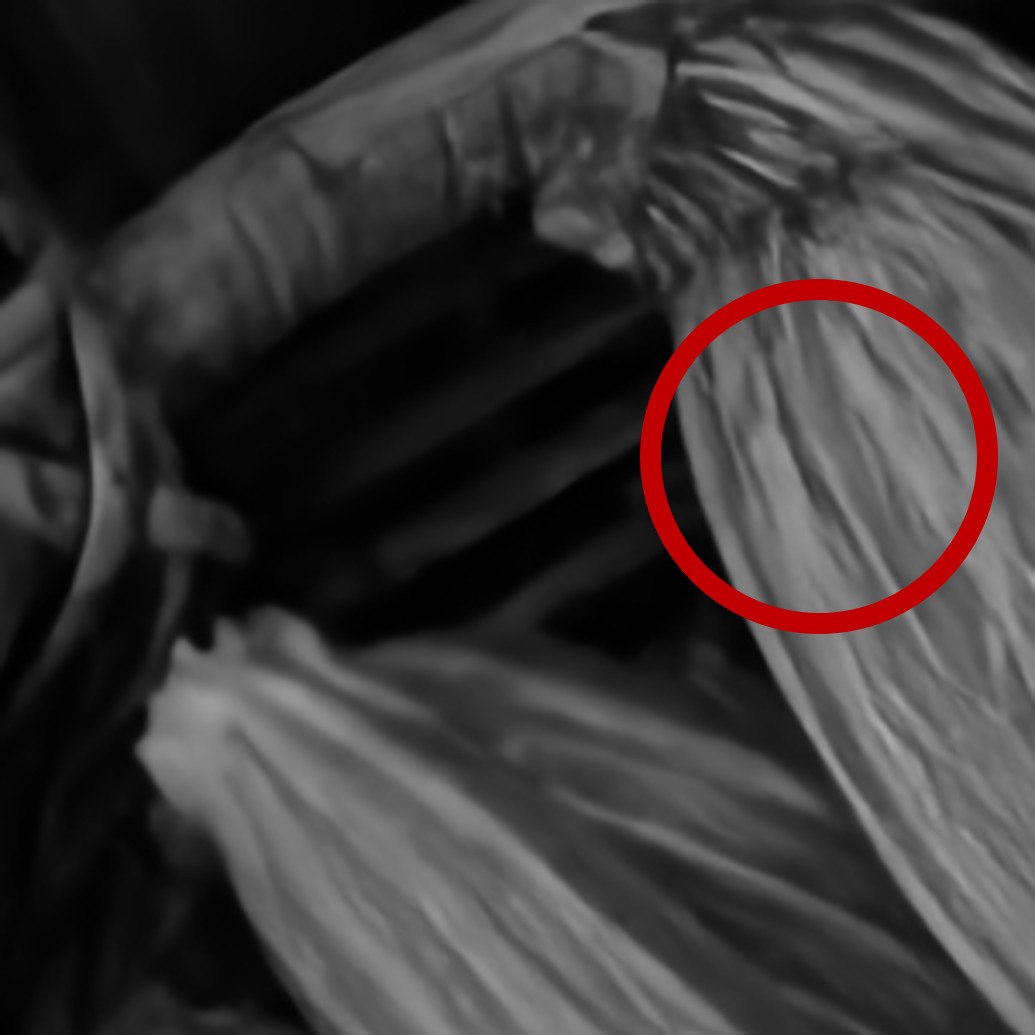}&
\hspace{-2.0ex}\includegraphics[width=0.09\linewidth]{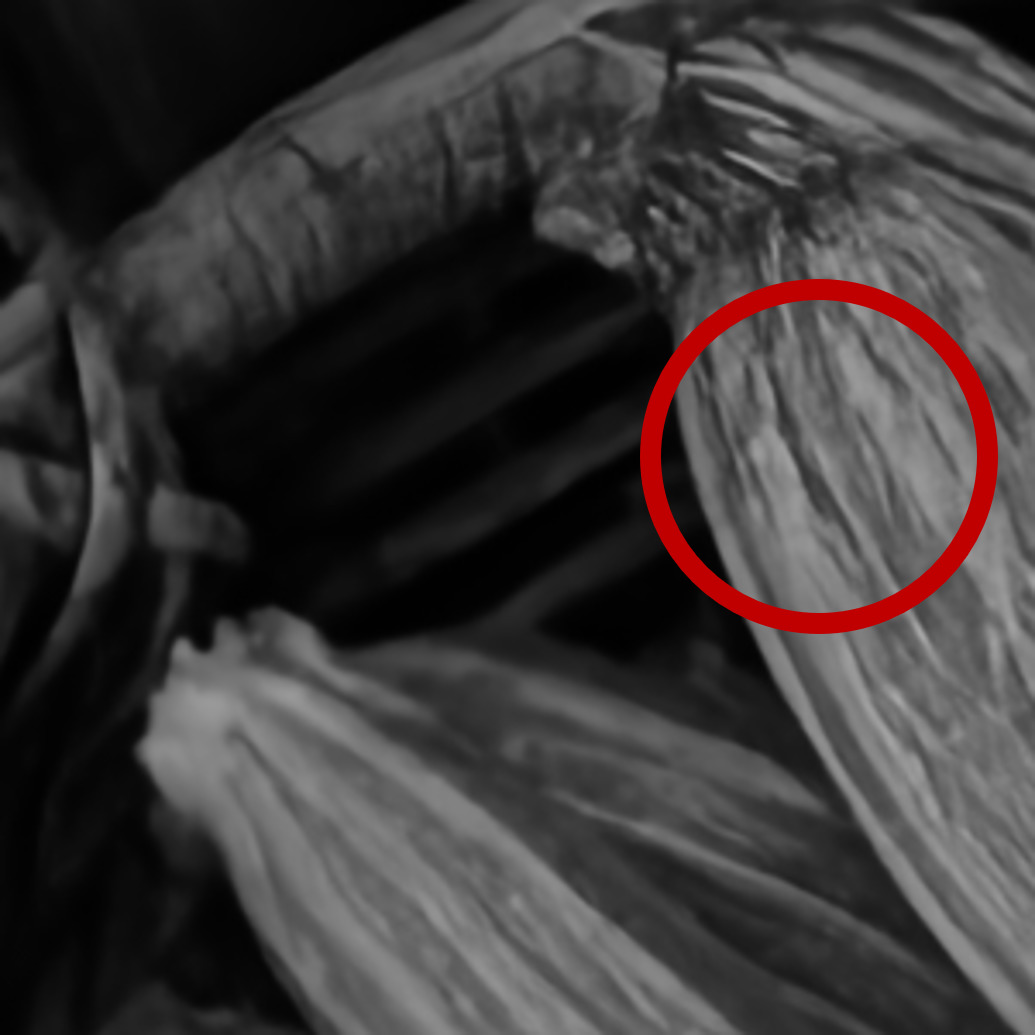}&
\hspace{-2.0ex}\includegraphics[width=0.09\linewidth]{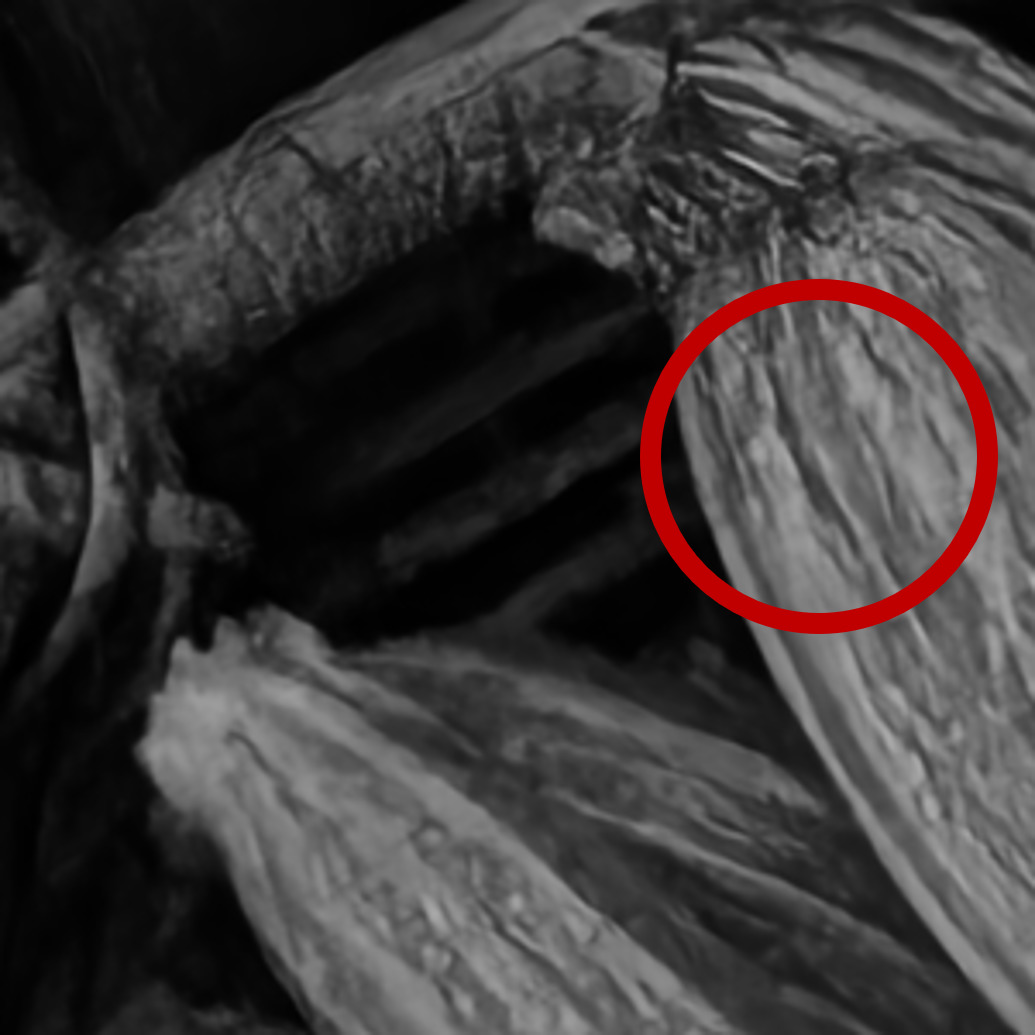}&
\hspace{-2.0ex}\includegraphics[width=0.09\linewidth]{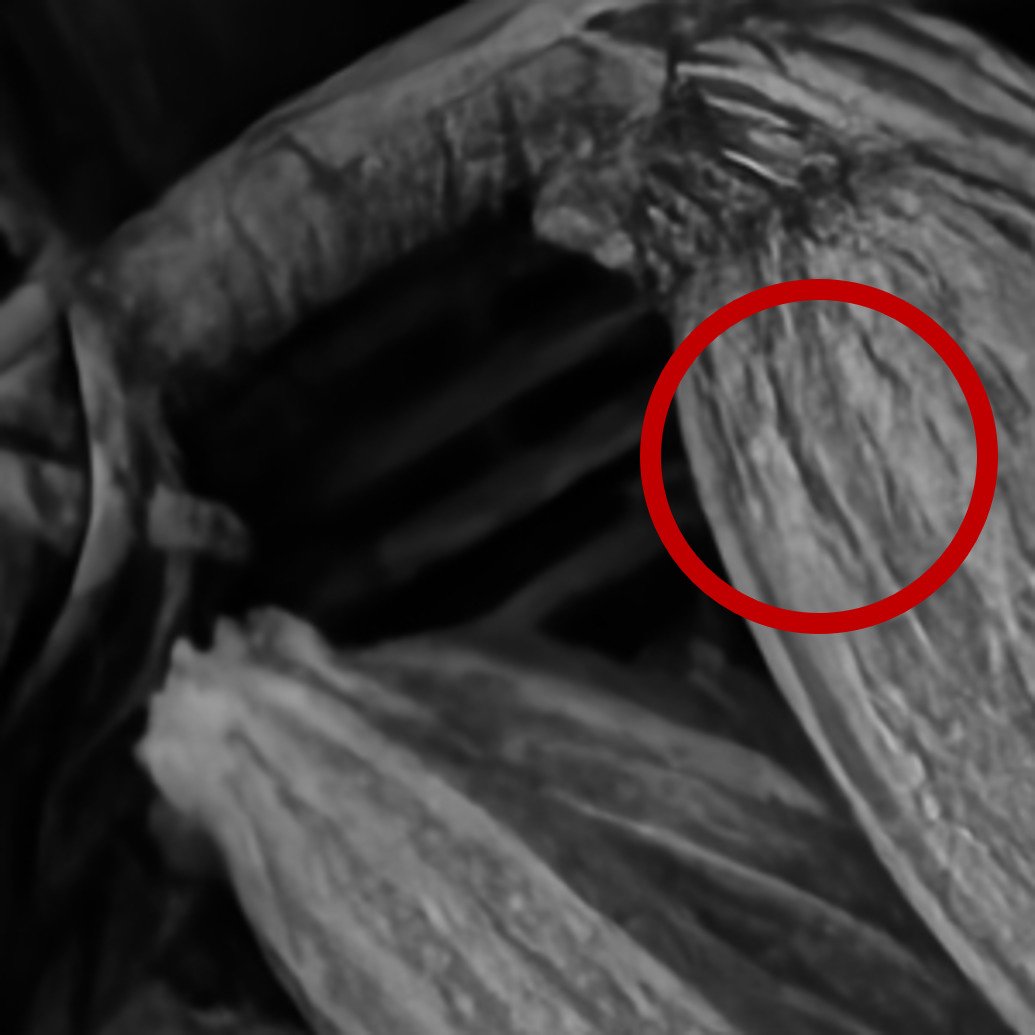}\\
\multirow{2}{*}[1.3cm]{\hspace{-2.0ex}\includegraphics[width=0.18\linewidth]{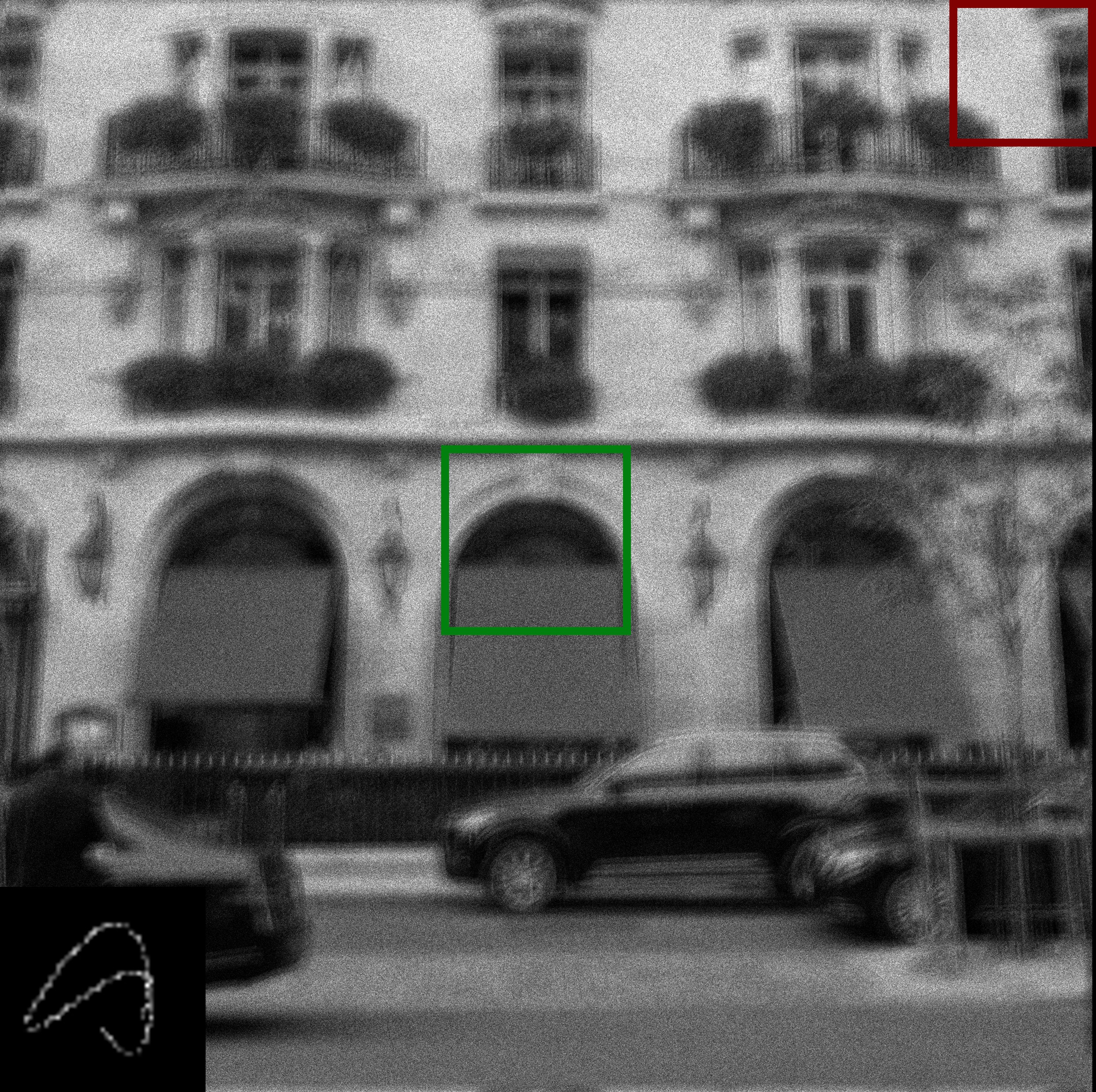}}&
\hspace{-2.0ex}\includegraphics[width=0.09\linewidth]{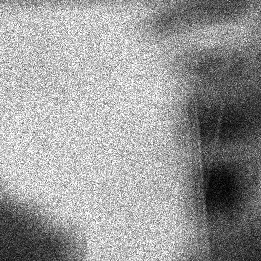}&
\hspace{-2.0ex}\includegraphics[width=0.09\linewidth]{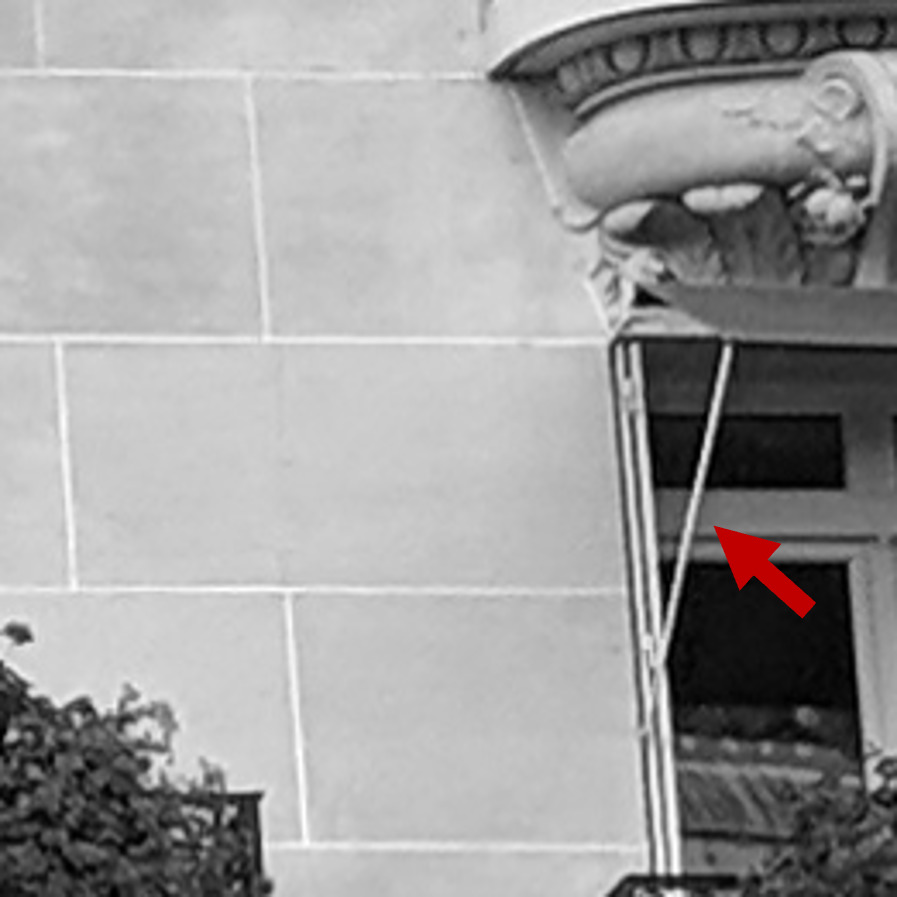}&
\hspace{-2.0ex}\includegraphics[width=0.09\linewidth]{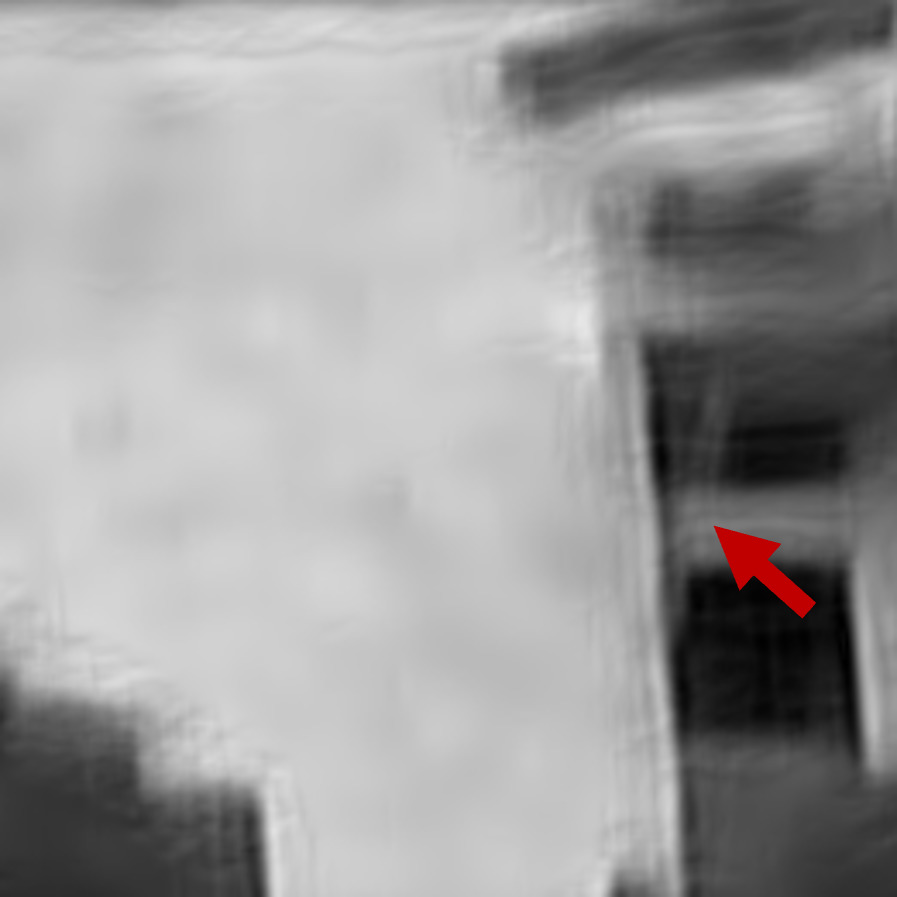}&
\hspace{-2.0ex}\includegraphics[width=0.09\linewidth]{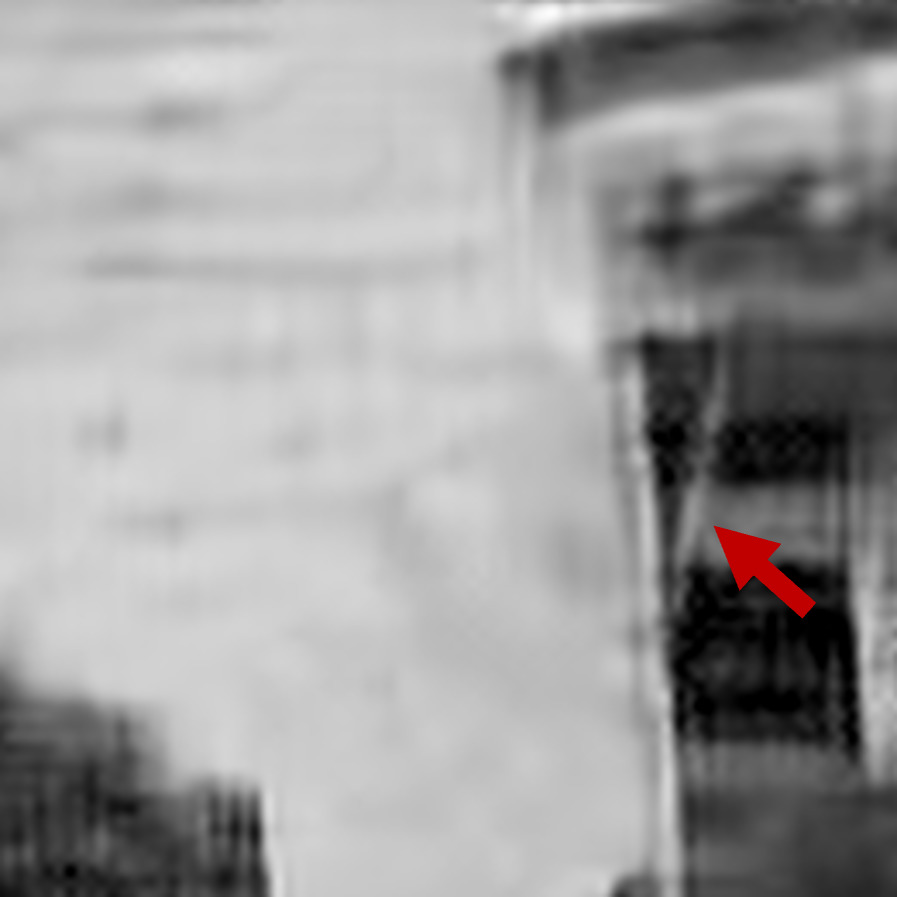}&
\hspace{-2.0ex}\includegraphics[width=0.09\linewidth]{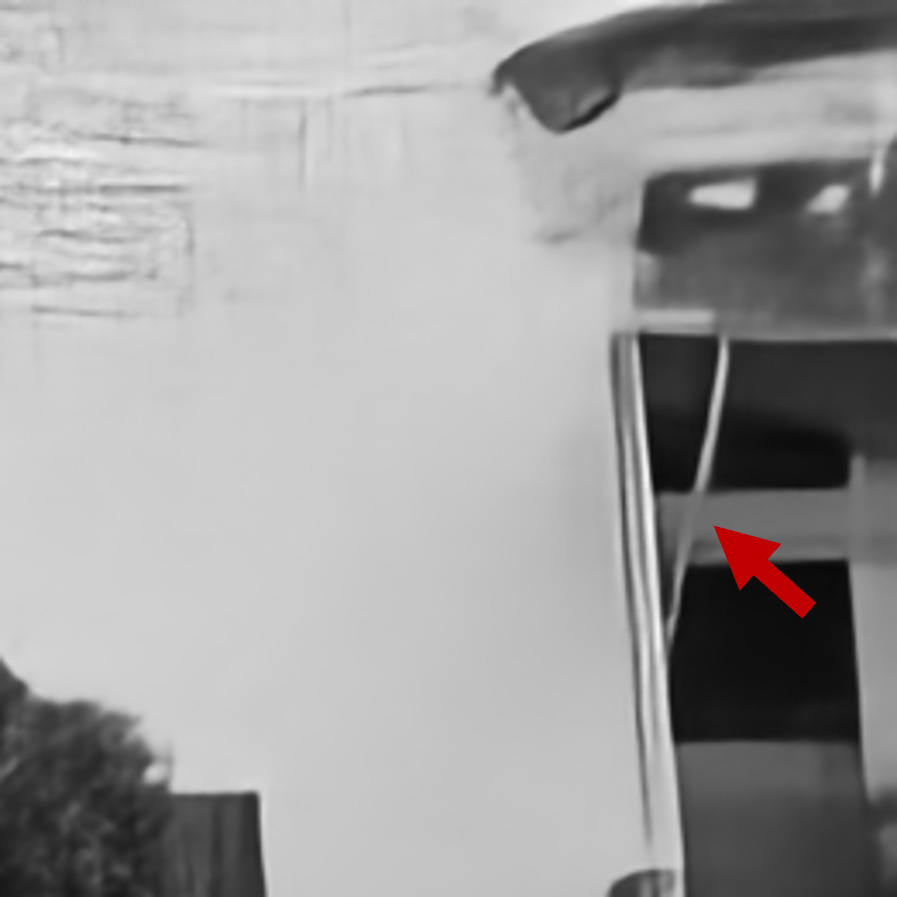}&
\hspace{-2.0ex}\includegraphics[width=0.09\linewidth]{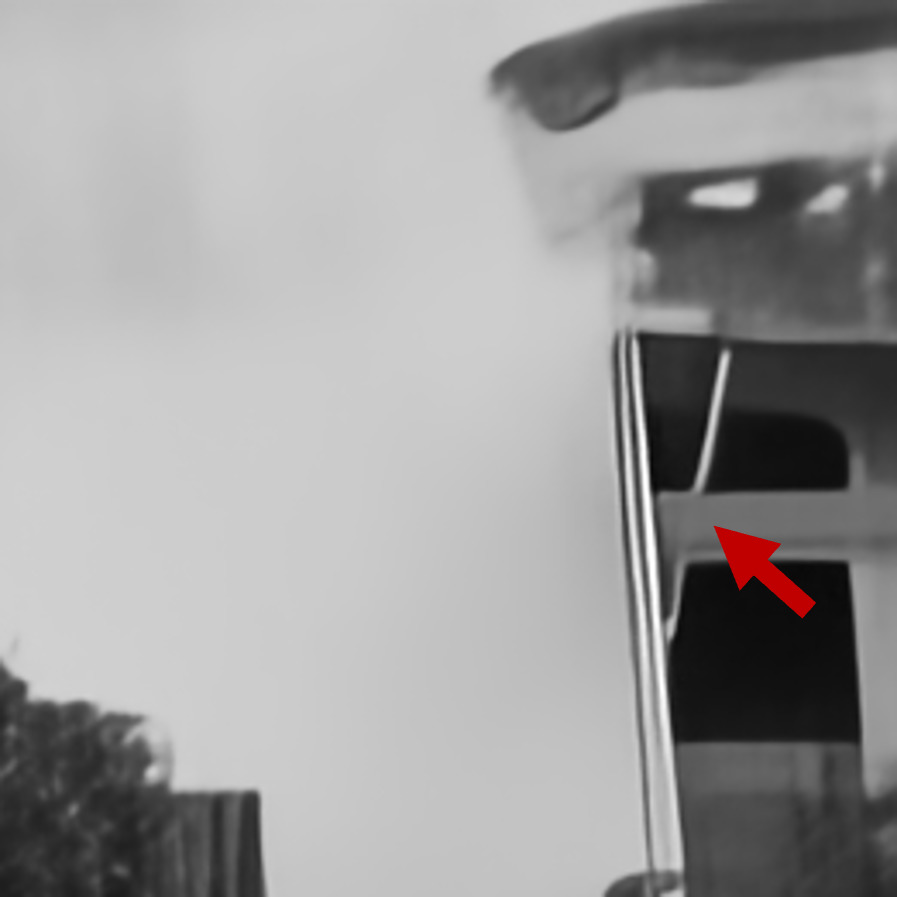}&
\hspace{-2.0ex}\includegraphics[width=0.09\linewidth]{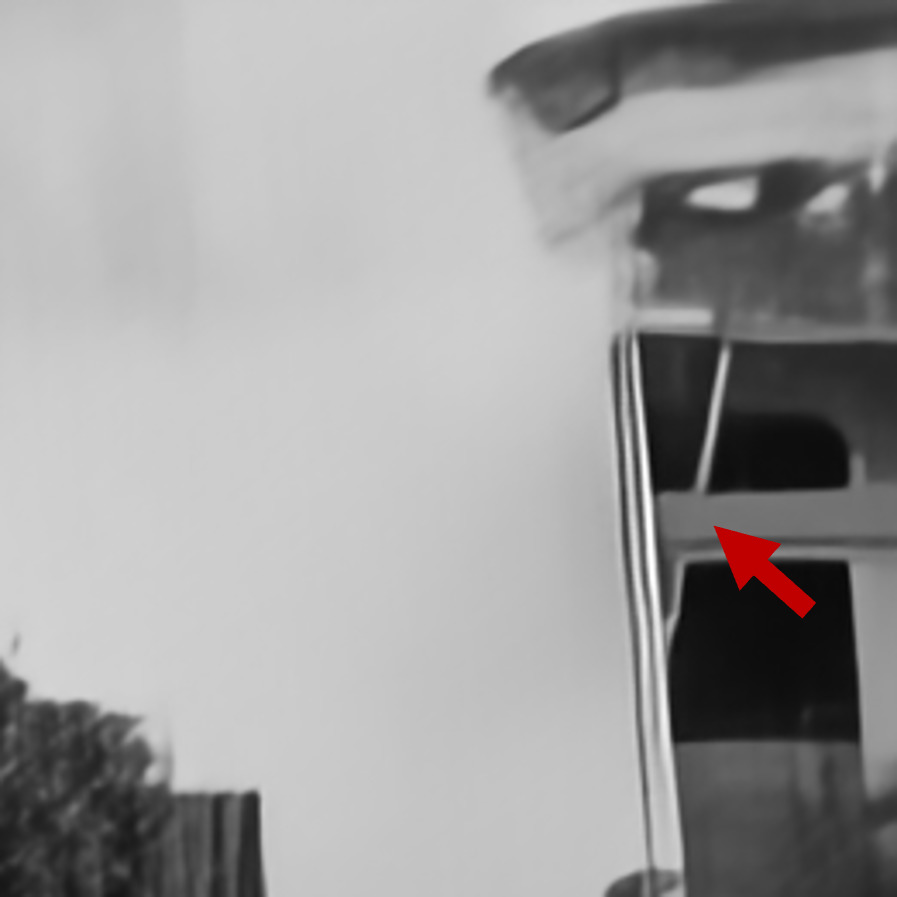}&
\hspace{-2.0ex}\includegraphics[width=0.09\linewidth]{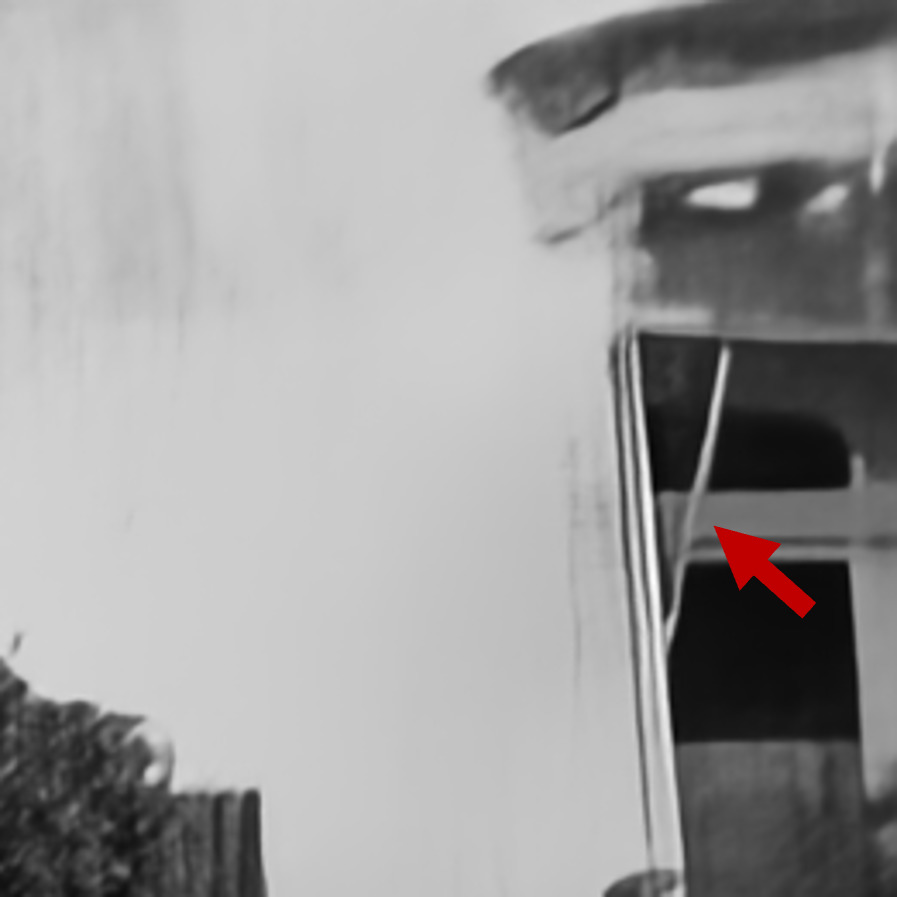}\\
&
\hspace{-2.0ex}\includegraphics[width=0.09\linewidth]{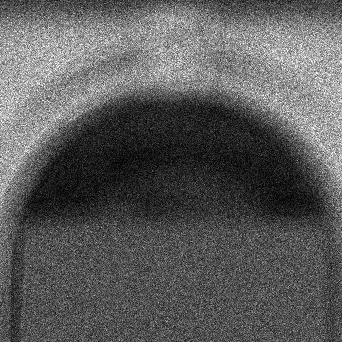}&
\hspace{-2.0ex}\includegraphics[width=0.09\linewidth]{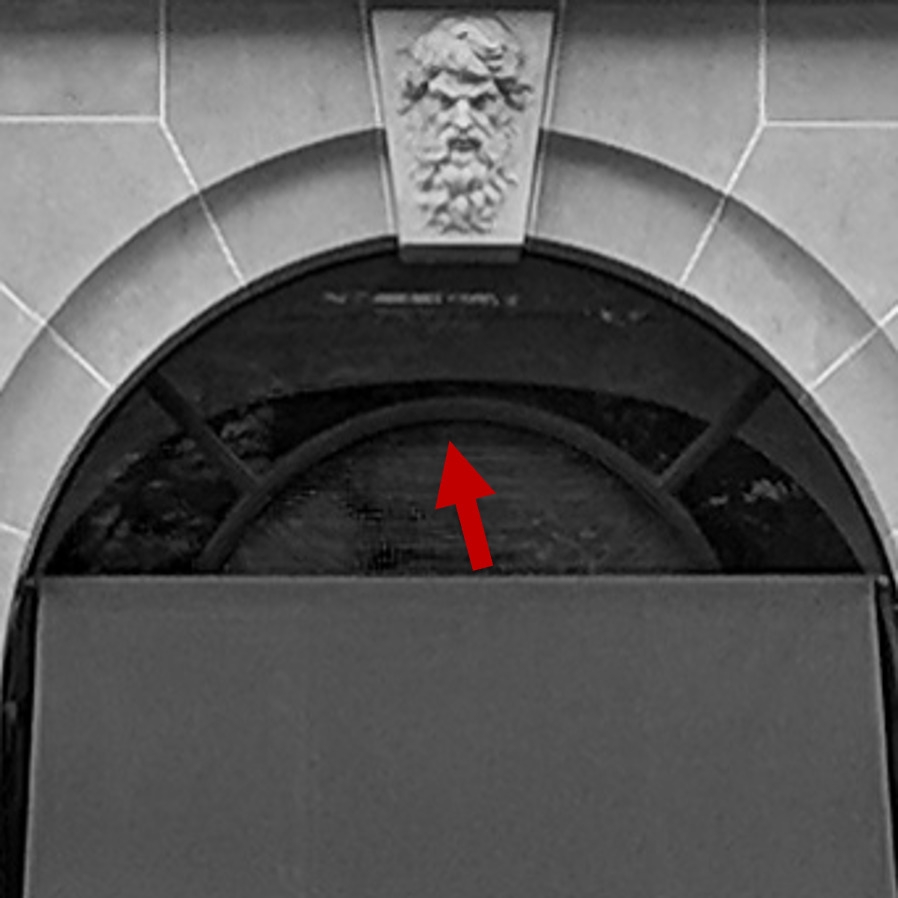}&
\hspace{-2.0ex}\includegraphics[width=0.09\linewidth]{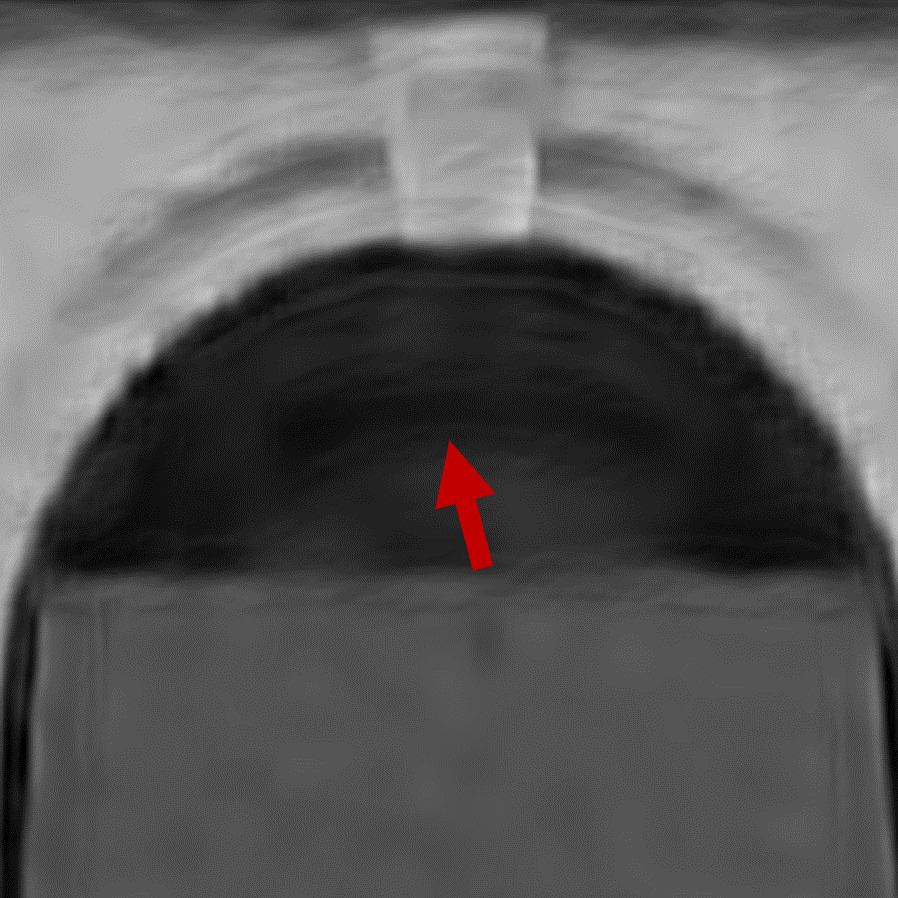}&
\hspace{-2.0ex}\includegraphics[width=0.09\linewidth]{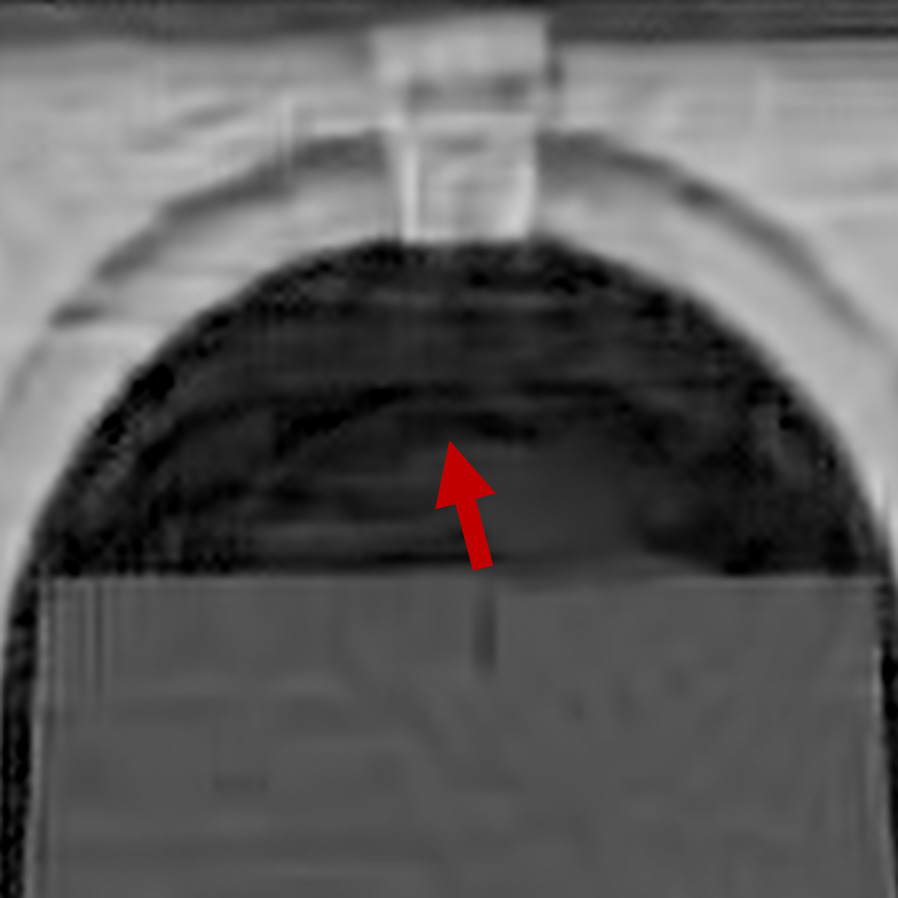}&
\hspace{-2.0ex}\includegraphics[width=0.09\linewidth]{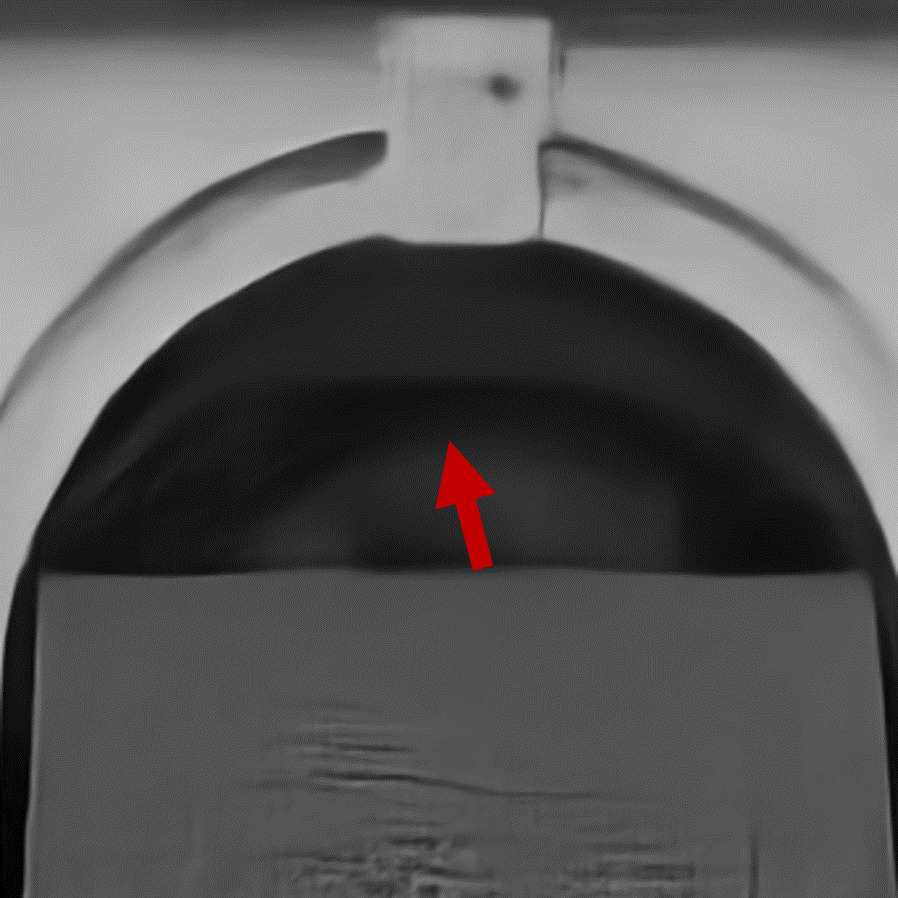}&
\hspace{-2.0ex}\includegraphics[width=0.09\linewidth]{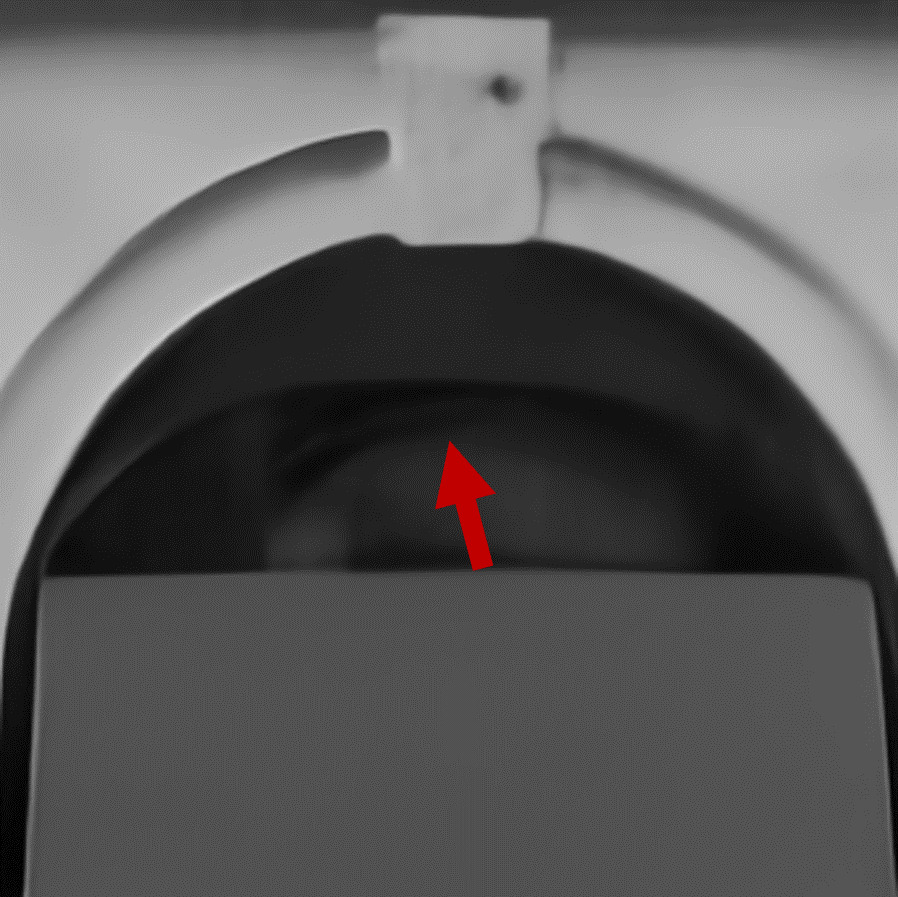}&
\hspace{-2.0ex}\includegraphics[width=0.09\linewidth]{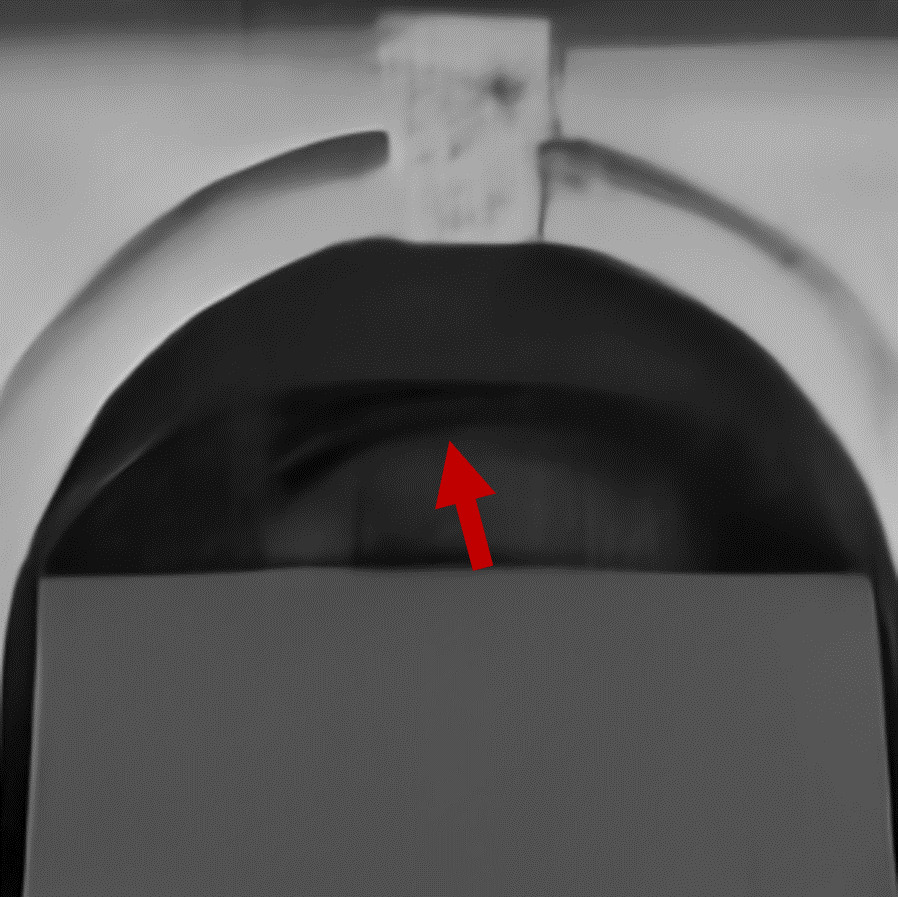}&
\hspace{-2.0ex}\includegraphics[width=0.09\linewidth]{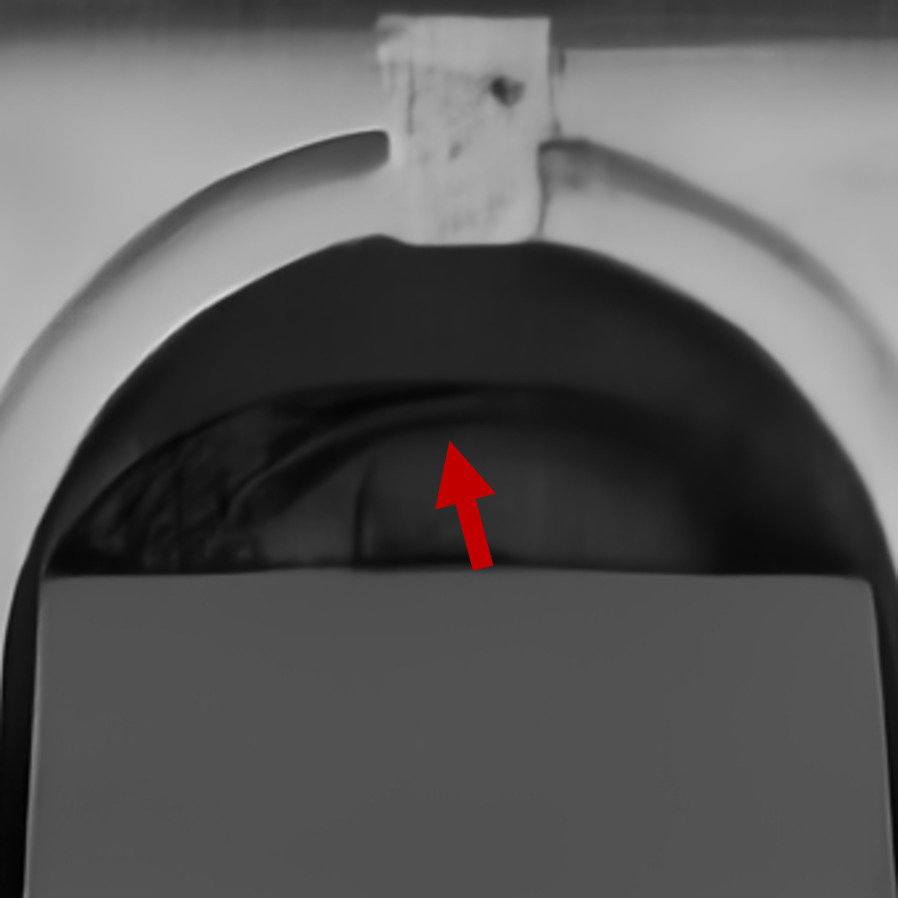}\\
& \scriptsize{Degraded} & \scriptsize{GT} & \scriptsize{PURE-LET~\cite{Li_2017_PURELET}}&\scriptsize{VSTP~\cite{Azzari_2017_VST}}&\scriptsize{DWDN~\cite{Dong_2022_DeepWiener_PAMI}}&\scriptsize{USRNet~\cite{Zhang_2020_USRNet}}&\scriptsize{PhDNet~\cite{Sanghvi_2022_Iterative}}&\scriptsize{Proposed}
\end{tabular}
\caption{Experimental results comparing the proposed FIO-Net with several other Poisson deconvolution methods.}
\label{fig: Synth Expt}
\end{figure*}

In \fref{fig: Synth Expt} we show the visual comparisons. The visual comparisons apparently show another perspective of FIO-Net. If we compare USRNet, PhDNet, and FIO-Net, we see that all three perform similarly. However, as we zoom in to see the details, e.g., the lines on the roof in the first image, the bars on the windows in the second image, and the tail of the alphabet in the third image, we can see the visual improvement of FIO-Net. We remark that all models are trained using the exact same training dataset and tested on the same testing dataset. Therefore, the restored details are due to the network itself rather than data overfitting.

\begin{table*}[!t]
\centering
\setlength\doublerulesep{0.5pt}
\caption{\label{tab:SOTA_comparison}\textbf{Performance of the 5 methods of interest on the test dataset.} PURE-LET \cite{Li_2017_PURELET}, and VSTP \cite{Azzari_2017_VST} are traditional methods and do not use any neural networks. DWDN \cite{Dong_2022_DeepWiener_PAMI} is neural network-based but non-iterative. PhDNet \cite{Sanghvi_2022_TCI} and USRNet \cite{Zhang_2020_USRNet} are unrolled neural network-based solutions. The best-performing method is shown in \textbf{bold} and the second best method is \underline{underlined}.}
\scalebox{0.8}{
\begin{tabular}{ccccccccc}
                Kernel Size & ppp & & PURE-LET~\cite{Li_2017_PURELET}& VSTP~\cite{Azzari_2017_VST}& DWDN~\cite{Dong_2022_DeepWiener_PAMI} & PhDNet~\cite{Sanghvi_2022_TCI} &  USRNet\cite{Zhang_2020_USRNet}&Proposed
        \Bstrut\\
        \hline \hline
         \multirow{6}{*}{Small}&\multirow{2}{*}{10}&PSNR (dB)&22.64&24.93&25.13&25.20&\underline{25.26}&\textbf{25.27}\Tstrut \\
         &&SSIM&0.672&0.733&0.771&\underline{0.775}&\underline{0.775}&\textbf{0.778}\Bstrut\\
         \cline{2-9}
         &\multirow{2}{*}{30}&PSNR (dB)&23.21&25.71&26.11&26.24&26.30&\textbf{26.32}\Tstrut\\
         &&SSIM&0.694&0.770&0.805&0.809&\underline{0.810}&\textbf{0.813}\Bstrut\\
          \cline{2-9}
         &\multirow{2}{*}{50}&PSNR (dB)&23.57&26.58&26.17&26.74&\underline{26.79}&\textbf{26.84}\Tstrut \\
         &&SSIM&0.702&0.779&0.820&0.824&\underline{0.826}&\textbf{0.828}\Bstrut\\
         \hline
         \multirow{6}{*}{Medium}&\multirow{2}{*}{10}&PSNR (dB)&20.91&22.93&23.06&23.24&\underline{23.25}&\textbf{23.28}\Tstrut\\
         &&SSIM&0.621&0.649&0.700&0.704&\underline{0.705}&\textbf{0.707}\Bstrut\\
          \cline{2-9}
         &\multirow{2}{*}{30}&PSNR (dB)&21.71&23.69&24.07&24.30&\underline{24.32}&\textbf{24.36}\Tstrut\\
         &&SSIM&0.644&0.691&0.737&0.744&\underline{0.744}&\textbf{0.747}\Bstrut\\
          \cline{2-9}
         &\multirow{2}{*}{50}&PSNR (dB)&22.14&24.08&24.57&24.86&\underline{24.88}&\textbf{24.92}\Tstrut\\
         &&SSIM&0.661&0.708&0.755&0.763&\underline{0.764}&\textbf{0.767}\Bstrut\\
         \hline
         \multirow{6}{*}{Large}&\multirow{2}{*}{10}&PSNR (dB)&20.53&22.40&22.43&22.60&\underline{22.63}&\textbf{22.65}\Tstrut\\
         &&SSIM&0.588&0.642&0.679&0.683&\underline{0.684}&\textbf{0.685}\Bstrut\\
          \cline{2-9}
         &\multirow{2}{*}{30}&PSNR (dB)&21.22&23.18&23.41&23.65&\underline{23.69}&\textbf{23.71}\Tstrut\\
         &&SSIM&0.613&0.649&0.714&0.721&\underline{0.722}&\textbf{0.723}\Bstrut\\
          \cline{2-9}
         &\multirow{2}{*}{50}&PSNR (dB)&21.60&23.56&23.91&24.19&\underline{24.22}&\textbf{24.25}\Tstrut\\
         &&SSIM&0.625&0.676&0.732&0.740&\underline{0.741}&\textbf{0.742}\Bstrut\\
         \hline
    \end{tabular}}
\end{table*}


\section{Conclusion}
With the growth of photon-limited imaging applications, we recognize the importance of understanding the performance limits of Poisson deconvolution algorithms. To this end, we present a systematic analysis of a large number of existing non-blind Poisson deconvolution methods. Based on this analysis, we deduce five ``secrets'' that are needed for an effective non-blind Poisson deconvolution algorithm design: 
\begin{enumerate}
    \item Use Wiener filter for spatially invariant blur
    \item Use iterative neural networks instead of single forward-pass neural networks
    \item Use feature space deblurring instead of image space deblurring
    \item Do not incorporate Poisson likelihood in the network architecture design
    \item Learn hyperparameters for iterative algorithms in an end-to-end manner.
\end{enumerate}

By combining these five secrets, we obtain a proof-of-concept named the Five-In-One Network (FIO-Net). The results offered by FIO-Net are consistent with the five secrets we presented. Considering that FIO-Net is not a novel design but a combination of five existing ideas, the consistency and the on-par performance with the state-of-the-art result provide additional support to our findings.

\bibliography{references}
\bibliographystyle{ieeetr}

\end{document}


\title{The Secrets of Non-Blind Poisson Deconvolution - Supplementary}

\author{Abhiram~Gnanasambandam,~\IEEEmembership{Member,~IEEE},
and~Yash~Sanghvi,~\IEEEmembership{Student~Member,~IEEE}, and~Stanley~H.~Chan,~\IEEEmembership{Senior~Member,~IEEE}
\thanks{Y. Sanghvi and S. Chan are with the School of Electrical and Computer
Engineering, Purdue University, West Lafayette, IN 47907, USA. Email: {
\{ysanghvi, stanchan\}}@purdue.edu. A. Gnanasambandam is with Samsung Research America, Plano, TX, but this work is completed during his time as a PhD student at Purdue. Email: abhiram.g94@gmail.com. This work is supported, in part, by the National Science Foundation under grant IIS-2133032, ECCS-2030570, and DMS-2134209.} 
}



\maketitle

\begin{abstract}
This supplementary document provides the detailed results of the experiment we conducted. We provide some brief information about each of the methods we are comparing and the performance of these methods on the test dataset is tabulated. 
\end{abstract}

\section{The methods being compared}
These methods can be categorized into two main classes - classical methods and deep learning methods.
\subsection{Classical Methods} 
By classical methods, we mean methods that do not require learning. These methods are typically developed before the deep-learning era. In this paper, we select three representative methods with code publicly available:

\noindent\textbf{1. PURE-LET, by Li and Blu \cite{Li_2017_PURELET}} is a non-iterative deblurring algorithm that uses the Poisson unbiased risk estimator (PURE) as a metric to guide the steps in linear expansion thresholding (LET).

\noindent\textbf{2. VSTP, by Azzari and Foi \cite{Azzari_2017_VST}} uses the variance stabilization transform (VST) to equalize the variance of the Poisson random variable. Then, a deblurring algorithm is applied to handle the blur.

\noindent\textbf{3. Deconvtv, by Chan et al. \cite{Chan_2011_TIP}} is a classical total variation for Gaussian noise removal. Its performance is not necessarily the best compared to other total variation solvers such as \cite{Chowdhury_2020_JMIV, Lingenfelter_2009_Poisson, Lefkimmiatis_2013_TIP, Setzer_2010_Bregman, Harmany_2012_SPIRAL, Getreuer_2012_IPOL, Figueiredo_2010_ADMM}, but its code is readily available for experiments.

\subsection{Deep-Learning Mehods}
\textbf{4. DWDN by Dong et al. \cite{Dong_2020_DeepWiener_NIPS}} is the only single pass deep learning based solution we are comparing. In this method a CNN is used to learn a feature space where the images are deconvolved using Wiener filters and then the deconvolved features are combined using another CNN.  

\textbf{5. SVMAP by Dong et al. \cite{Dong_2021_LearningSpatially}} learns a spatially varying MAP estimate, which could be useful for the spatially varying Poisson noise.

\textbf{6. KerUnc by Nan and Ji \cite{Nan_2020_DeepLearning}} models the error in the estimate of the blur kernel available to us using an error-in-variable (EIV) model of image. The paper then uses unrolled network to solve the problem.

\textbf{7. CPCR by Eboli et al. \cite{Eboli_2020_EndEnd}} is an iterative algorithm based on the HQS algorithm with preconditioned
iterative fixed-point iterations, where the preconditioner and the proximal operator are both learned.

\textbf{8. USRNet by Zhang et al. \cite{Zhang_2020_USRNet}} is an unrolled optimization method which learns to optimize the HQS algorithm.

\textbf{9. PhDNet by Sanghvi et al. \cite{Sanghvi_2022_TCI}} is an unrolled optimization method designed specifically for Poisson noise. 

\textbf{10. RGDN by Gong et al. \cite{Gong_2020_LearningDeep}}  uses neural-networks to solve a gradient descent problem and obtain clean latent image.

\textbf{11. DPIR by Zhang et al. \cite{Zhang_2017_LearningDeep}} uses the plug-and-play (PnP) based ADMM optimization to solve the deconvolution problem.

\textbf{12. DWKF by Pronina et al. \cite{Pronina_2020_Microscopy}} is an iterative method that uses kernel prediction networks for imposing the image priors.

\textbf{13. VEM by Nan et al. \cite{Nan_2020_VariationalEM}} is an iterative method based on variational expectation maximization.

\begin{landscape}
\begin{table}
\centering
\caption{\label{tab:SOTA_comparison_supp}\textbf{Performance of the methods of interest on the test dataset.} }
\begin{tikzpicture}
\node (table) {
\begin{tabular}{ccccccccccccccccc}
        Kernel Size & ppp & & \makecell{PURE-\\ LET \\ \cite{Li_2017_PURELET} } & \makecell{VSTP \\\cite{Azzari_2017_VST} } & \makecell{ Deconv- \\TV \\ \cite{Chan_2011_TIP}}  & \makecell{ DWDN \\ \cite{Dong_2020_DeepWiener_NIPS} } & \makecell{ SVMAP \\  \cite{Dong_2021_LearningSpatially} } & \makecell{ KerUnc \\ \cite{Nan_2020_DeepLearning} } & \makecell{ CPCR \\ \cite{Eboli_2020_EndEnd}} & \makecell{ USRNet \\ \cite{Zhang_2020_USRNet}} &  \makecell{PhDNet \\ \cite{Sanghvi_2022_TCI}} & \makecell{ RGDN \\ \cite{Gong_2020_LearningDeep} } & \makecell{DPIR \\ \cite{Zhang_2017_LearningDeep}} & \makecell{ DWKF \\ \cite{Pronina_2020_Microscopy}}  & \makecell{VEM \\ \cite{Nan_2020_VariationalEM}} & FIONet  \\
        \hline
         \multirow{6}{*}{Small}&\multirow{2}{*}{10}&PSNR (dB)&22.64&24.94&23.66&25.13&23.78&24.27&23.92&\underline{25.26}&25.21&18.51&24.91&23.91&21.77&\textbf{25.27}\Tstrut \\
         &&SSIM&0.500&0.583&0.713&0.771&0.731&0.739&0.736&\underline{0.776}&0.775&0.630&0.764&0.730&0.563&\textbf{0.778}\Bstrut\\
         \cline{2-17}
         &\multirow{2}{*}{30}&PSNR (dB)&23.21&25.72&24.67&26.11&24.15&25.46&24.41&\underline{26.30}&26.24&18.58&25.93&24.49&25.16&\textbf{26.32}\Tstrut\\
         &&SSIM&0.524&0.620&0.786&0.805&0.747&0.780&0.758&\underline{0.810}&0.809&0.637&0.801&0.757&0.760&\textbf{0.812}\Bstrut\\
          \cline{2-17}
         &\multirow{2}{*}{50}&PSNR (dB)&23.57&26.23&25.09&26.58&24.24&25.68&24.58&\underline{26.79}&26.74&18.58&26.43&24.68&25.98&\textbf{26.84}\Tstrut \\
         &&SSIM&0.542&0.649&0.821&0.820&0.751&0.781&0.765&\underline{0.826}&0.824&0.637&0.819&0.767&0.796&\textbf{0.828}\Bstrut\\
         \hline 
         \multirow{6}{*}{Medium}&\multirow{2}{*}{10}&PSNR (dB)&20.92&22.93&21.30&23.07&21.36&22.16&21.10&\underline{23.25}&23.24&17.28&22.46&21.13&19.25&\textbf{23.28}\Tstrut\\
         &&SSIM&0.421&0.500&0.678&0.700&0.634&0.665&0.625&\underline{0.705}&0.704&0.553&0.670&0.631&0.443&\textbf{0.707}\Bstrut\\
          \cline{2-17}
         &\multirow{2}{*}{30}&PSNR (dB)&21.71&23.69&22.13&24.07&21.50&23.48&21.18&\underline{24.33}&24.31&17.27&23.43&21.24&22.79&\textbf{24.36}\Tstrut\\
         &&SSIM&0.444&0.531&\textbf{0.761}&0.737&0.638&0.711&0.626&0.744&0.744&0.550&0.711&0.637&0.660&\underline{0.747}\Bstrut\\
          \cline{2-17}
         &\multirow{2}{*}{50}&PSNR (dB)&22.14&24.08&22.55&24.58&21.52&23.79&21.18&\underline{24.88}&24.86&17.24&23.88&21.27&23.80&\textbf{24.92}\Tstrut\\
         &&SSIM&0.461&0.549&\textbf{0.795}&0.756&0.638&0.715&0.626&0.764&0.764&0.548&0.733&0.638&0.711&\underline{0.767}\Bstrut\\
         \hline 
         \multirow{6}{*}{Large}&\multirow{2}{*}{10}&PSNR (dB)&20.53&22.40&20.56&22.43&20.07&21.41&20.23&\underline{22.63}&22.60&16.74&21.60&20.23&19.20&\textbf{22.65}\Tstrut\\
         &&SSIM&0.419&0.482&0.625&0.680&0.600&0.643&0.605&\underline{0.684}&0.683&0.533&0.645&0.609&0.454&\textbf{0.685}\Bstrut\\
          \cline{2-17}
         &\multirow{2}{*}{30}&PSNR (dB)&21.23&23.18&21.36&23.42&20.11&22.78&20.30&\underline{23.69}&23.66&16.73&22.50&20.32&22.32&\textbf{23.71}\Tstrut\\
         &&SSIM&0.434&0.510&0.689&0.715&0.600&0.688&0.606&\underline{0.722}&0.721&0.529&0.680&0.613&0.647&\textbf{0.723}\Bstrut\\
          \cline{2-17}
         &\multirow{2}{*}{50}&PSNR (dB)&21.60&23.57&21.77&23.92&20.12&23.12&20.30&\underline{24.22}&24.19&16.70&22.93&20.34&23.20&\textbf{24.25}\Tstrut\\
         &&SSIM&0.445&0.526&0.716&0.733&0.601&0.694&0.605&\underline{0.741}&0.740&0.527&0.698&0.614&0.690&\textbf{0.742}\Bstrut\\
    \end{tabular}
};
\end{tikzpicture}
\end{table}
\end{landscape}

\bibliography{references}
\bibliographystyle{ieeetr}